\documentclass[twocolumn]{aastex61}
\pdfoutput=1 
\usepackage{amsmath,amstext}
\usepackage[T1]{fontenc}
\usepackage{apjfonts} 
\usepackage[figure,figure*]{hypcap}
\usepackage{scrextend}
\usepackage{footnote}
\usepackage{hyperref}
\usepackage{subfigure}
\usepackage{array}

\usepackage{epsfig}
\usepackage{natbib}
\usepackage{rotating}

\usepackage{color,soul}

\defcitealias{battersby_cmzoom_2020}{Paper~I}

\shorttitle{CMZoom Catalog}
\shortauthors{CMZoom team}
\begin{document}

\title{CMZoom II: Catalog of Compact Submillimeter Dust Continuum Sources in the Milky Way's Central Molecular Zone}

\correspondingauthor{Hatchfield}
\email{h.hatchfield@uconn.edu}

\author[0000-0003-0946-4365]{H Perry Hatchfield}
\affiliation{University of Connecticut, Department of Physics, 196A Auditorium Road, Unit 3046, Storrs, CT 06269 USA}

\author[0000-0002-6073-9320]{Cara Battersby}
\affiliation{University of Connecticut, Department of Physics, 196A Auditorium Road, Unit 3046, Storrs, CT 06269 USA}
\affiliation{Center for Astrophysics $|$ Harvard \& Smithsonian, MS-78, 60 Garden St., Cambridge, MA 02138 USA}

\author{Eric Keto}
\affiliation{Center for Astrophysics $|$ Harvard \& Smithsonian, MS-78, 60 Garden St., Cambridge, MA 02138 USA}

\author{Daniel Walker}
\affiliation{University of Connecticut, Department of Physics, 196A Auditorium Road, Unit 3046, Storrs, CT 06269 USA}
\affiliation{Joint ALMA Observatory, Alonso de C\'ordova 3107, Vitacura, Santiago, Chile}
\affiliation{National Astronomical Observatory of Japan, 2-21-1 Osawa, Mitaka, Tokyo, 181-8588, Japan}

\author{Ashley Barnes}
\affiliation{Argelander-Institut f\"{u}r Astronomie, Universit\"{a}t Bonn, Auf dem H\"{u}gel 71, 53121, Bonn, DE}

\author{Daniel Callanan}
\affiliation{Astrophysics Research Institute, Liverpool John Moores University, 146 Brownlow Hill, Liverpool L3 5RF, UK}
\affiliation{Center for Astrophysics $|$ Harvard \& Smithsonian, MS-78, 60 Garden St., Cambridge, MA 02138 USA}

\author[0000-0001-6431-9633]{Adam Ginsburg}
\affiliation{University of Florida Department of Astronomy, Bryant Space Science Center, Gainesville, FL, 32611, USA}

\author[0000-0001-9656-7682]{Jonathan D. Henshaw}
\affiliation{Max-Planck-Institute for Astronomy, Koenigstuhl 17, 69117 Heidelberg, Germany}

\author[0000-0002-5094-6393]{Jens Kauffmann}
\affiliation{Haystack Observatory, Massachusetts Institute of Technology, 99 Millstone Road, Westford, MA 01886, USA}

\author[0000-0002-8804-0212]{J. M. Diederik Kruijssen}
\affiliation{Astronomisches Rechen-Institut, Zentrum f{\"u}r Astronomie der Universit{\"a}t Heidelberg, M{\"o}nchhofstra{\ss}e 12-14, D-69120 Heidelberg, Germany}

\author[0000-0001-6353-0170]{Steve N. Longmore}
\affiliation{Astrophysics Research Institute, Liverpool John Moores University, 146 Brownlow Hill, Liverpool L3 5RF, UK}

\author[0000-0003-2619-9305]{Xing Lu}
\affiliation{National Astronomical Observatory of Japan, 2-21-1 Osawa, Mitaka, Tokyo, 181-8588, Japan}

\author[0000-0001-8782-1992]{Elisabeth A. C. Mills}
\affiliation{Department of Physics and Astronomy, University of Kansas, 1251 Wescoe Hall Drive, Lawrence, KS 66045, USA}

\author[0000-0003-2133-4862]{Thushara Pillai}
\affiliation{Boston University Astronomy Department, 725 Commonwealth Avenue, Boston, MA 02215, USA}

\author[0000-0003-2384-6589]{Qizhou Zhang}
\affiliation{Center for Astrophysics $|$ Harvard \& Smithsonian, MS-42, 60 Garden St., Cambridge, MA 02138 USA}

\author[0000-0001-8135-6612]{John Bally}
\affiliation{CASA, University of Colorado, 389-UCB, Boulder, CO 80309}

\author[0000-0002-4013-6469]{Natalie Butterfield}
\affiliation{Green Bank Observatory, 155 Observatory Rd, PO Box 2, Green Bank, WV 24944, USA}

\author[0000-0002-6388-3635]{Yanett A. Contreras}
\affiliation{Leiden Observatory, Leiden University, PO Box 9513, NL 2300 RA Leiden, the Netherlands}

\author[0000-0001-6947-5846]{Luis C. Ho}
\affiliation{Kavli Institute for Astronomy and Astrophysics, Peking University, Beijing 100871, China}
\affiliation{Department of Astronomy, School of Physics, Peking University, Beijing 100871, China}



\author{J{\"u}rgen Ott}
\affiliation{National Radio Astronomy Observatory, 1003 Lopezville Rd., Socorro, NM 87801, USA}

\author{Nimesh Patel}
\affiliation{Center for Astrophysics $|$ Harvard \& Smithsonian, MS-78, 60 Garden St., Cambridge, MA 02138 USA}

\author{Volker Tolls}
\affiliation{Center for Astrophysics $|$ Harvard \& Smithsonian, MS-78, 60 Garden St., Cambridge, MA 02138 USA}


\begin{abstract}
In this paper we present the \textit{CMZoom} Survey's catalog of compact sources (< 10\arcsec, $\sim$0.4pc) within the Central Molecular Zone (CMZ). \textit{CMZoom} is a Submillimeter Array (SMA) large program designed to provide a complete and unbiased map of all high column density gas (N(H$_2$) $\ge$ $10^{23}$ cm$^{-2}$) of the innermost 500pc of the Galaxy in the 1.3mm dust continuum. We generate both a robust catalog designed to reduce spurious source detections, and a second catalog with higher completeness, both generated using a pruned dendrogram. In the robust catalog, we report 285 compact sources, or 816 in the high-completeness catalog. These sources have effective radii between 0.04-0.4 pc, and are the potential progenitors of star clusters. The masses for both catalogs are dominated by the Sagittarius B2 cloud complex, where masses are likely unreliable due to free-free contamination, uncertain dust temperatures, and line-of-sight confusion. Given the survey selection and completeness, we predict that our robust catalog accounts for more than $\sim$99\% of compact substructure capable of forming high-mass stars in the CMZ. This catalog provides a crucial foundation for future studies of high-mass star formation in the Milky Way's Galactic Center.
\end{abstract}


\section{Introduction}
\label{sec:intro}
The Milky Way's innermost ~500 pc, the Central Molecular Zone (CMZ), houses a vast complex of $\sim 3-5\times 10^7$ M$_\odot$ of molecular gas \citep{morris_galactic_1996, Dahmen_Molecular_1998, pierce-price_deep_2000}. The molecular clouds in this environment exist at high densities and temperatures relative to the disc, exhibiting intense pressures, magnetic fields and turbulence (e.g.\ \citealt{mills_detection_2013,rathborne_g0_2014,ginsburg_dense_2016, pillai_magnetic_2015, henshaw_molecular_2016, federrath_link_2016,walker_star_2018}) as well as high cosmic ray ionization rates \citep{goto_h3_2013,harada_chemical_2015,Lepetit_physical_2016, Oka_central_2019} and UV background radiation \citep{lis_quiescent_2001}. Star formation within this complex environment is an area of extensive previous research and current study (e.g.\ \citealt{downes_radio_1966, guesten_new_1982, morris_galactic_1996, yusef-zadeh_Massive_2008, longmore_G0_2012, longmore_variations_2013, kruijssen_what_2014, kauffmann_galactic_2013, kauffmann_galactic_2017,kauffmann_galactic_2017a, lu_deeply_2015, lu_census_2019, lu_star_2019}). The CMZ is a unique laboratory for studying star formation, as these conditions, rarely observed in the rest of the Milky Way, exhibit some similarities to properties of high redshift galaxies, allowing us an indirect glimpse into the cosmic history of star formation for which comparably detailed extragalactic observations are not currently possible \citep{kruijssen_comparing_2013}.

The birthplaces of stars are understood to be the over-densities in giant molecular clouds (GMCs), ``clumps,'' of a characteristic size of \textasciitilde$0.3-3$pc. These clumps fragment further into gravitationally bound ``cores'' of a characteristic size of \textasciitilde $0.03-0.2$pc, which in turn will form individual stellar systems \citep{Bergin_cold_2007,kennicuttjr_star_2012}. The formation of massive stars (M $>$ 8M$_\odot$) is relatively rare due to the fragmentation of dense clumps leading to a greater number of low-mass stars for which the mass distribution is described by an initial mass function \citep[e.g.\ ][]{bastian_universal_2010, Offner_origin_2014}. Studying the nature of transitions between these evolutionary stages, leading from turbulent GMCs on scales $>$3pc down to individual young stellar systems demands large surveys of both star forming and non-star forming structures across the hierarchical continuum of relevant scales (e.g.\ \citealt{kauffmann_galactic_2017,kauffmann_galactic_2017a,walker_star_2018,lu_census_2019,lu_star_2019}). 

There are several indications of recent and active star formation in the Galactic Center. The central few hundred parsecs of the Galaxy house two young massive clusters, the Arches Cluster and the Quintuplet Cluster, suggesting a recent burst of star formation activity in the CMZ within the last \textasciitilde 5 Myr \citep{Figer_massive_2002,habibi_arches_2013,Lu_stellar_2013,hosek_unusual_2019}. At present, there are several known stellar nurseries actively forming stars in the CMZ, most notably the cloud complexes Sagittarius (Sgr B2), Sgr C, and the Dust Ridge \citep{Gordon_anatomy_1993,yusef-zadeh_Massive_2008,yusef-zadeh_Star_2009,schmiedeke_physical_2016,ginsburg_distributed_2018,walker_star_2018,barnes_young_2019, lu_census_2019}. The mean gas densities observed in the CMZ are much higher than those observed in the Galactic disc, leading conventional theories of star formation to predict a correspondingly high Star Formation Rate (SFR) per molecular gas mass relative to the disk \citep{kennicutt_global_1997, lada_star_2012, longmore_variations_2013, barnes_star_2017}. Such a high SFR is not observed, with studies revealing SFRs a factor of 10-100 lower than predicted by current theories of star formation relative to the dense gas content \citep{immer_recent_2012,koepferl_mainsequence_2015, longmore_variations_2013, kruijssen_what_2014, barnes_star_2017}. In particular, \cite{barnes_star_2017} show how the observed low SFR cannot be accounted for by observational bias alone, demonstrating how independent methods of measuring the SFR agree with one another to within a factor of 2. This implies an environmental cause for the deficiency of star formation. It has been proposed that this low SFR and high dense gas content indicate an upcoming starburst in the Galactic Center, possibly part of a dynamical episodic cycle of star formation and quiescence (\citealt{kruijssen_what_2014,krumholz_dynamical_2015,krumholz_dynamical_2017}, see also \citealt{sormani_nuclear_2020}). Recent simulations suggest that variation in the SFR might be driven by variations in the CMZ's total gas mass, perhaps modulated by uneven accretion driven by the Galactic bar \citep{sormani_dynamical_2018,sormani_simulations_2020,tress_simulations_2020}. In order to effectively test theoretical explanations for this present lack of star formation, the \textit{CMZoom} team has carried out a survey of all potential sites of massive star formation at sufficient resolution to resolve and characterize compact substructure. 

In this paper, we present the catalog of compact sources constructed from the \textit{CMZoom} survey's 1.3 mm continuum data (\citealp{battersby_cmzoom_2020}, hereafter \citetalias{battersby_cmzoom_2020}). We refer to the entries in this catalog as either compact sources or (dendrogram) leaves as they span spatial scales between both clump-like emission and core-like emission, between $\sim$ 0.05-0.36 pc (this decision is explained further in Section \ref{sec:catalog_design}). This catalog of compact sources is designed to ultimately help understand the anomalous SFR and hierarchy of dense gas structures in the CMZ by providing a complete survey of all compact sources embedded in gas above a column density of $10^{23}$\,cm$^{-2}$. 

Several of the regions selected in the \textit{CMZoom} survey have already been the topic of more detailed single cloud studies. In particular the cloud G0.253+0.016, colloquially known as "the Brick'' or "the Lima Bean Cloud,'' has been closely studied as a flagship example of the star formation deficiency in the CMZ, where observations detect an excess of dense gas tracers relative to the expected abundance of star formation signatures \citep{lis_star_1994, longmore_G0_2012, kauffmann_galactic_2013, mills_abundant_2015, lu_deeply_2015, mills_abundant_2015, rathborne_g0_2014, henshaw_brick_2019}. There is theoretical framework \citep{kruijssen_dynamical_2015,jeffreson_physical_2018,kruijssen_dynamical_2019,dale_dynamical_2019} and evidence \citep{immer_recent_2012,longmore_candidate_2013,rathborne_g0_2014,walker_comparing_2016,barnes_star_2017,walker_star_2018,barnes_young_2019} suggesting that the clouds composing the Dust Ridge, connecting the Brick to Sgr B2, together exhibit a time-sequence evolution of transition from quiescence to active star formation triggered by dynamical effects. While individual clouds display non-monotonic properties along this stream \citep{kauffmann_galactic_2017,kauffmann_galactic_2017a}, it has been suggested these variations could possibly be explained by using different initial conditions in simulated versions of such a stream \citep{kruijssen_dynamical_2019,dale_dynamical_2019}.

Other notable objects in the region include the massive star forming complexes around Sgr B2 and Sgr A, and Sgr C which have been studied extensively since the discovery of their bright emission in the radio continuum (e.g.\ \citealt{downes_radio_1966, lo_Highresolution_1983}). Many other regions in the \textit{CMZoom} survey have not been the subject of any previous study at comparable resolution and sensitivity. For a more complete description of the source selection and pointings, we refer the reader to \citetalias{battersby_cmzoom_2020}. 

The structure of this paper is as follows. In Section \ref{sec:data_summary} we present a brief overview of the data, summarizing key points from \citetalias{battersby_cmzoom_2020}. In Section \ref{sec:catalog} we provide a description of the cataloging algorithm and the simulated observational procedure used for source recovery experiments and the calculation of our catalogs' completeness. Section \ref{sec:results} describes resultant contents and properties of the ``robust'' and ``high-completeness'' versions of the catalog that we have produced. Section \ref{sec:discussion} highlights the initial analysis of the catalog distributions and the role of Sgr B2 in the catalog, as well as placing limits on the catalog's high-mass stellar precursor completeness and the star formation potential of the CMZ. In Section \ref{sec:summary} we summarize the key points from this work. Appendix \ref{sec:complete_figs} recreates the figures from Section \ref{sec:results} for the high-completeness catalog. Additionally, Appendix \ref{sec:param_study} provides a parameter study for the catalog algorithm, showing the effects of varying the input parameters of the pruned dendrogram algorithm on the final catalog contents and statistical distributions. Appendix \ref{sec:zoomins} is a gallery of zoomed-in regions of the \textit{CMZoom} dust continuum map with the robust catalog leaves over-plotted as contours. Lastly, Appendix \ref{sec:alma_comp} compares several key regions with corresponding data at higher spatial resolutions from the Atacama Large Millimeter / submillimeter Array (ALMA). 
\newpage

\section{Description of the Data}
\label{sec:data_summary}

The \textit{CMZoom} survey is a 550 hour project using the Submillimeter Array\footnote{The Submillimeter Array is a joint project between the Smithsonian
Astrophysical Observatory and the Academia Sinica Institute of Astronomy and
Astrophysics, and is funded by the Smithsonian Institution and the Academia
Sinica.} (SMA), mapping the highest column density gas in the innermost $5.0 ^\circ$ in longitude and $0.5 ^\circ$ in latitude of the Galaxy at 1.3\,mm, as well as selected additional regions detailed in \citetalias{battersby_cmzoom_2020}. The survey is designed to include completely all CMZ clouds with column densities above $10^{23}$\,cm$^{-2}$ as derived from the {\it Herschel} cold dust continuum (by the spectral energy distribution fitting procedure outlined in \citealt{battersby_characterizing_2011}, explored in \citealt{mills_origins_2017}), with the exception of the cloud to the southeast of Sgr B2 and isolated bright pixels in the {\it Herschel} continuum. The \textit{CMZoom} pointings also contain some lower column density regions of interest across the CMZ, such as far-side cloud candidates, the circumnuclear disk, the Arches filaments and a bridge of emission that, in projection, appears to connect the dust ridge and 50 km s$^{-1}$ clouds (see figure 1 in \citetalias{battersby_cmzoom_2020}). 

The survey uses the SMA in both its compact and subcompact configurations in order to effectively probe the range of spatial scales between 3$\arcsec$ and about 45$\arcsec$ (corresponding to a physical range between 0.12-1.8 pc at a Galactic Center distance of 8.178 $\pm$ 0.013 (stat.) $\pm$ 0.022 (sys.) kpc \citep{thegravitycollaboration_geometric_2019}. With this setup, we achieve a typical spatial resolution of $~3.2$" (0.13 pc) in the 230 GHz dust continuum wideband (8+ GHz) from which these catalogs are constructed. Single dish data from the Bolocam Galactic Plane Survey have been feathered with the SMA data in order to supplement our sensitivity to larger scale structure. For the catalogs created in this work, we used the SMA-only data, as we are interested only in the most compact emission structures. All data were calibrated in \textsc{mir idl} using standard SMA calibration procedures, and imaging and deconvolution were completed in a combination of \textsc{mir idl} and \textsc{casa}. All the \textit{CMZoom} pointings, SMA configurations, and imaging pipeline details are described in more detail in \citetalias{battersby_cmzoom_2020} and the data products have been made available on \texttt{Dataverse} at \hyperlink{https://dataverse.harvard.edu/dataverse/cmzoom}{https://dataverse.harvard.edu/dataverse/cmzoom}.


 
\section{Catalog Design and Methodology}

\label{sec:catalog}

\subsection{Pruned Dendrogram Algorithm}
\label{sec:catalog_design}

The 1.3 mm dust continuum emission observed in the \textit{CMZoom} survey contains structures widely varying in shape, spatial scale, intensity, and local noise (see \citetalias{battersby_cmzoom_2020} for more details). We developed a cataloging algorithm in order to consistently catalog the compact emission in the \textit{CMZoom} continuum data despite this diversity. To this end, the catalog was developed using an implementation of the dendrogram algorithm, with additional pruning for local noise. 

\begin{figure}
\begin{center}
\includegraphics[trim = 0mm 10mm 0mm 0mm, width = .45 \textwidth]{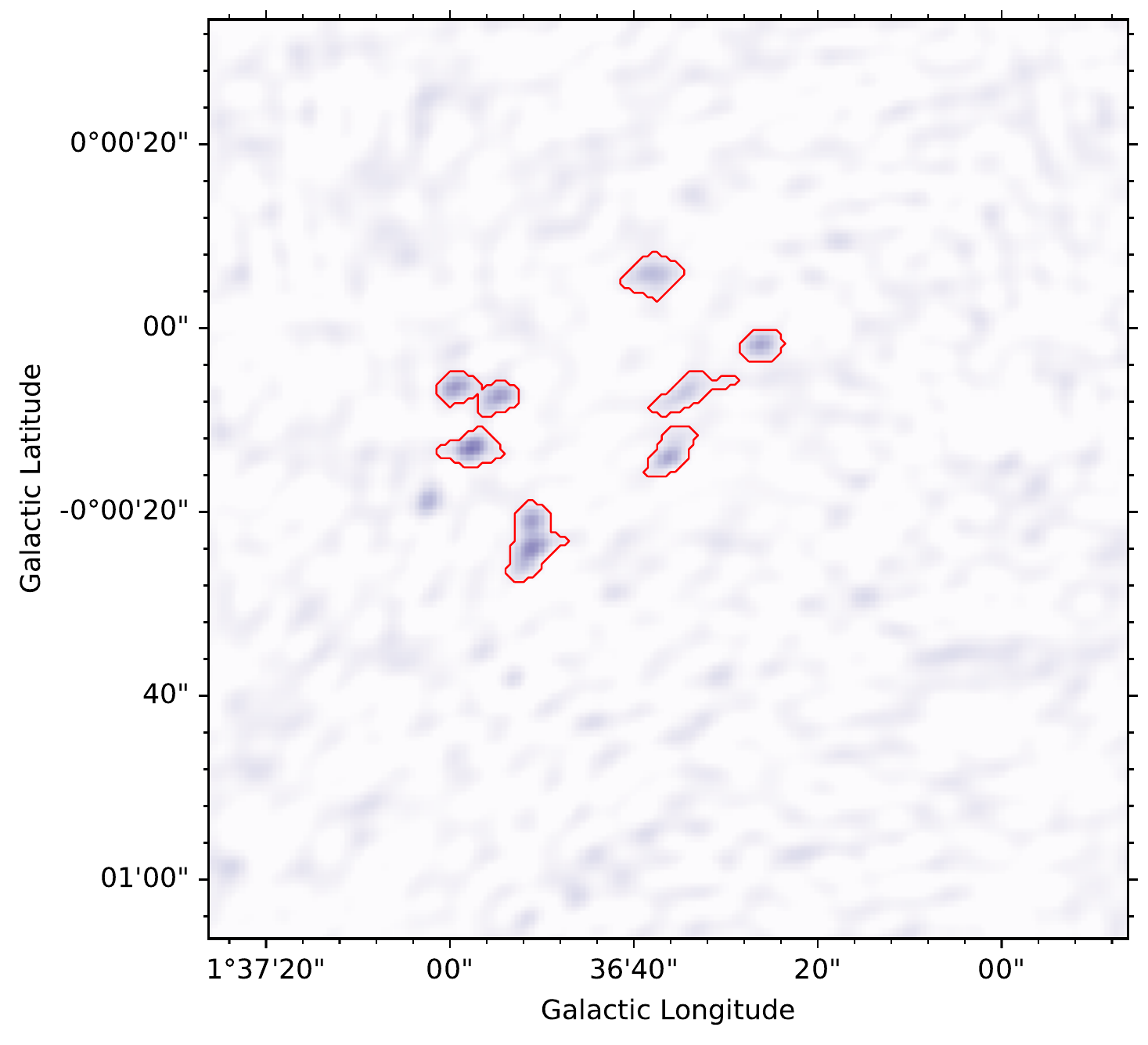}
\end{center}
\caption{A typical simulated observation used to calculate the completeness of the \textit{CMZoom} catalog algorithm. For this example, the skymap was populated with ten point-sources. The red contours are the leaf outlines or the simulated catalog, generated by applying the same version of the pruned dendrogram algorithm to the simulated image. In this case, nine of the ten point-sources are recovered, two nearby point-sources are grouped together in one leaf, and there are no false positive source detections.}
\label{fig:simobs_example}
\end{figure}

\begin{figure}
\begin{center}
\includegraphics[trim = 0mm 0mm 0mm 0mm, clip, width=0.4 \textwidth]{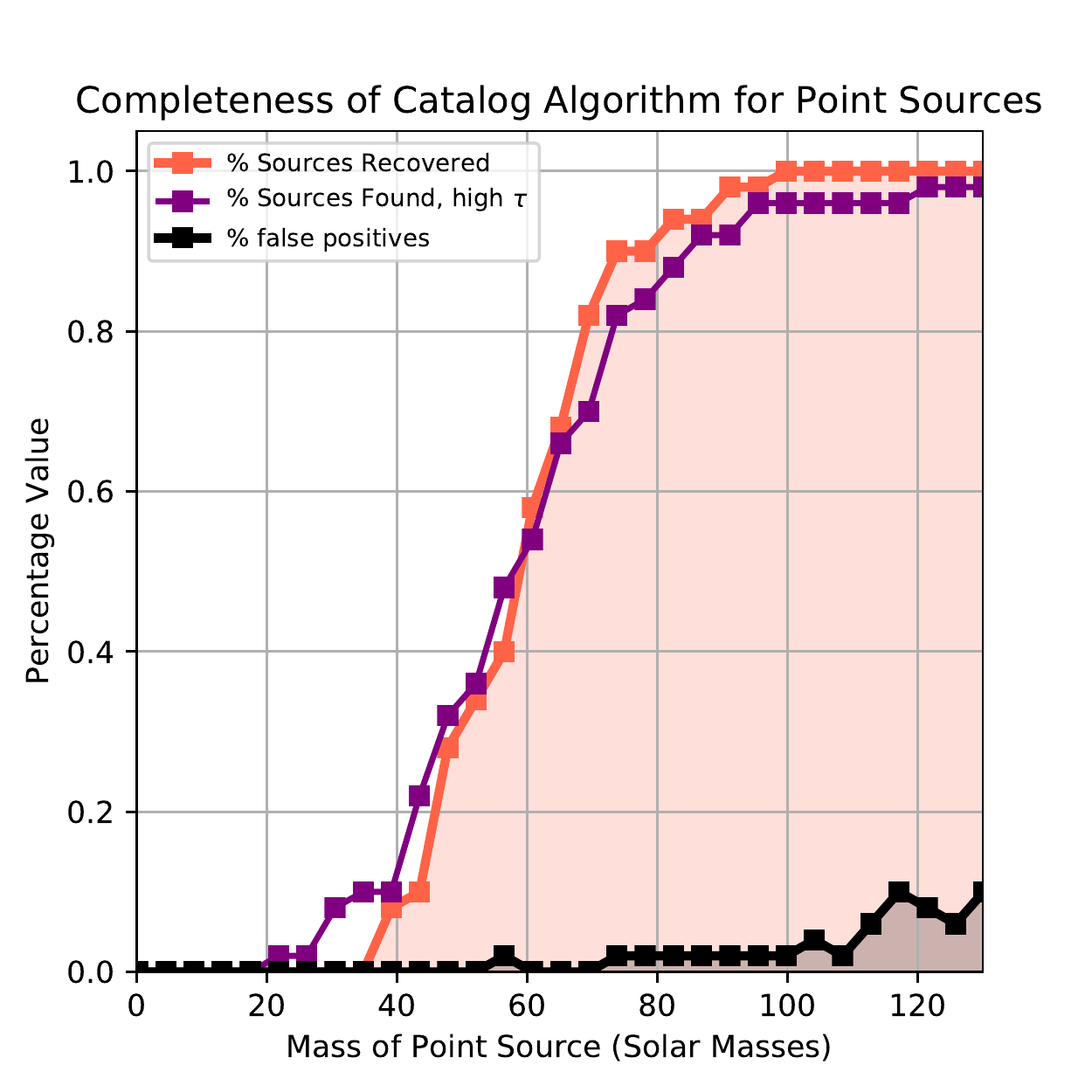}
\end{center}
\caption{The completeness percentage for source recovery using the procedure outlined in section 3.2. The red line shows the number of point-sources recovered from the artificial skymap for varying point-source masses (assuming a dust temperature $T_d = 20$K), while the purple line shows a sample recovery for weather conditions with a high zenith opacity ($\tau_0 = 0.2$). Our cataloging algorithm recovers 95\% of objects with a mass of $\sim$ 80 M$_\odot$ for an assumed dust temperature of T$_d$=20K. This completeness mass is very sensitive to the assumed dust temperature. For example, when we instead assume a dust temperature T$_d$=50K, we find a much lower 95\% completeness mass of $\sim$ 25 M$_\odot$. Higher dust temperatures are possible on scales smaller than those resolved by {\it Herschel} (36\arcsec) in areas of active star formation such as Sgr B2 and Dust Ridge cloud C.}
\label{fig:simobs_percent}
\end{figure}

The dendrogram algorithm is a hierarchical clustering algorithm, representing the significant structures in a data set as the branches and leaves of a tree-like data structure. The highest level structures in the cluster hierarchy are called ``leaves,'' which represent local maxima of the flux. These leaves are, therefore, the highest column density and most compact sources in the \textit{CMZoom} maps. The parent ``branches'' of these leaves represent the less dense regions immediately surrounding the compact sources. The parent branches of those branches represent lower and lower level emission, and so on down to some flux floor. 

The dendrogram tree for a given dataset is uniquely determined by the choice of three parameters: a minimum structure value, a minimum number of pixels, and a minimum significance value. The minimum structure value is the lowest peak pixel value that will be considered as a structure in the tree. The minimum number of pixels determines the minimum size of a structure included in the tree. Finally, the minimum significance parameter is a measure of how high the pixel value of a new peak must be relative to nearby structures to be considered independent and added to the catalog, typically chosen relative to the noise (e.g.\ \citealt{rosolowsky_structural_2008,henshaw_investigating_2016}).\footnote{A more detailed description and the documentation for the dendrogram algorithm implementation can be found at \hyperlink{https://dendrograms.readthedocs.io/}{https://dendrograms.readthedocs.io/}.}

The continuum map was translated to units of megajanskys per steradian to allow comparisons between data sets despite variations in the beam size across the survey (\citetalias{battersby_cmzoom_2020}, section 4.2 provides the complete description of the beam size variation). The initial dendrogram parameters were chosen relative to the lowest RMS noise estimates for the entire surveyed region, 3 MJy sr$^{-1}$. The minimum pixel value is chosen to be a factor of three greater than this global noise estimate. The minimum pixel number was selected to be roughly half of the average beam size over the survey. The minimum significance parameter was chosen to be equivalent to the global low noise estimate, which allows detection of substructure within low flux regions while not significantly altering the distribution of cataloged structure in higher flux regions (this is justified in appendices \ref{sec:complete_figs} and \ref{sec:param_study}).

This initial dendrogram is highly populated due to the fact that the algorithm considers the significance of leaves as compared the low global noise estimate without any regard for local noise, which can be more important in regions with irregular noise properties. We have chosen to narrow the catalog by ``pruning'' the list of dendrogram leaves relative to their local RMS noise. This pruning process amounts to the removal of each dendrogram leaf that does not meet a chosen set of requirements on both the peak leaf flux and the mean leaf flux. In choosing these parameters, we have opted to construct two catalogs: the former prioritizing only the most robust leaves with a peak flux at least 6$\sigma$ above the local RMS estimate and the latter prioritizing a higher catalog completeness, which includes leaves with a peak flux at least 4$\sigma$ above the local RMS estimate. Both versions of the catalog require a mean flux of at least 2$\sigma$ above the local RMS. Such high thresholds are required since the u-v coverage is not perfect, and residual non-Gaussian imaging artifacts can strongly effect source selection unless sufficiently strict local noise restrictions are implemented. 

A map of the local RMS noise estimates across the entire survey was constructed from the residual images generated in the cleaning process described in detail in \citetalias{battersby_cmzoom_2020}. In order to generate an estimate of the RMS for each pixel in the survey, we find the spatially averaged standard deviation of these residual maps, calculated as 
\begin{equation}
        \sigma_{x,y} =\sqrt{ \left(R_{x,y} - (R \ast G)_{x,y}\right)^2 \ast G_{x,y}}
\end{equation}
where $R_{x,y}$ corresponds to the residual value at pixel coordinates $(x,y)$, and $\ast G$ denotes convolution with a Gaussian kernel with $\sigma=14$ pixels, or $\sim$ 7\arcsec. This effectively recovers an RMS value for each pixel by taking the difference between the pixel's value in a smoothed and un-smoothed version of the residuals. We find a median RMS over all surveyed regions is 13 MJy sr$^{-1}$. For more information about the RMS calculation and properties, we refer the reader to Section 4.2 of \citetalias{battersby_cmzoom_2020}. The smoothing kernel size of 14 pixels is justified with a parameter study in Appendix \ref{sec:param_study}. The choices of the pruning thresholds for local noise pruning and the resulting completeness for each catalog are explored using the simulated source recovery experiment described in the following section.

\subsection{Source Recovery and Simulated Observations}
\label{sec:simobs}
There are many potential sources of error in the cataloging process including, but not limited to the emergence of imaging artifacts in the cleaning process that might erroneously be identified as independent leaves, variations in measurement conditions such as the changes of antenna configuration or weather conditions, as well the inherent biases of the dendrogram algorithm \citep{rosolowsky_structural_2008}. To probe the effects of these sources of error, we estimate a completeness percentage for the catalog, derived from an algorithm built around the NRAO \texttt{CASA} simulated observation functionality \citep{mcmullin_casa_2007}. 

The scripts to generate this completeness percentage have been made available on Github\footnote{\hyperlink{https://github.com/CMZoom/core_catalogue}{https://github.com/CMZoom/core\_catalogue}} in a generalized form such that it can be used to test the accuracy of any cataloging algorithm, with particular ease if the data were observed using one of the observatories compatible with the current version of \texttt{CASA}, though other interferometers can be included with relative ease. The procedure for determining our completeness using simulated observations is as follows: 

Step 1) An artificial image is randomly populated with ten point-sources at a fixed intensity representing the emission of a field of unresolved core-like objects.

Step 2) A simulated observation of the artificial skymap is generated using the \texttt{CASA simobserve} function. This function allows us to mimic the true SMA observations by generating visibilities from a sample antenna configuration from the \textit{CMZoom} observations, and a typical length of observation used for pointings in the survey. These fake data can be combined with noise in order to simulate the effects of thermal noise and atmospheric attenuation using the \texttt{AATM} package\footnote{\hyperlink{https://www.mrao.cam.ac.uk/~bn204/alma/atmomodel.html}{https://www.mrao.cam.ac.uk/~bn204/alma/atmomodel.html}}, an atmospheric modeling package designed for use with ALMA \citep{pardo_atmospheric_2001}. The completeness results for our catalogs constructed from the simulated observations is resilient to variations typical in atmospheric conditions for \textit{CMZoom} observations.

Step 3) The uv data generated by the simulated observations are then imaged using the \texttt{CASA tclean} task. This procedure is identical to the imaging process for a typical low RMS noise region in the \textit{CMZoom} survey. The \texttt{tclean} task was called with multi-scale parameter set to scales [0, 3, 9, and 27]. Briggs weighting with a robust parameter of 0.5 was used throughout. This choice is described in more detail in \citetalias{battersby_cmzoom_2020}. We used the \texttt{auto-multithresh}\footnote{\hyperlink{https://casa.nrao.edu/casadocs/casa-5.3.0/synthesis-imaging/masks-for-deconvolution}{https://casa.nrao.edu/casadocs/casa-5.3.0/synthesis-imaging/masks-for-deconvolution}} feature (with identical input parameters to those detailed in \citetalias{battersby_cmzoom_2020}), and a continuum threshold of 1 mJy beam${-1}$. An example of the resulting simulated image is shown in Figure \ref{fig:simobs_example}.

Step 4) The residual images from \texttt{tclean} are used to produce an RMS map, just as described in Section \ref{sec:catalog_design}. An initial dendrogram catalog is run on the images resulting from the \texttt{tclean} process, using the same dendrogram parameters as in the catalog procedure, scaled by a global noise estimate corresponding to a low value noise estimate in the artificial image. The leaves of this dendrogram are then pruned using a noise map created identically to the catalog procedure in \ref{sec:catalog_design} so that the only remaining leaves have a peak flux above at least 6$\sigma$ and average flux above at least 2$\sigma$ average for the robust cataloging, or a peak flux above at least 4$\sigma$ and average flux above at least 2$\sigma$ for the high-completeness catalog.

Step 5) Finally, the pruned dendrogram leaves are compared to the locations of the injected point-sources. A given point-source is considered to be recovered if its coordinates are located anywhere within a pruned leaf.

\vspace{2mm}

A plot of this completeness percentage for clumps as a function of leaf mass is depicted in Figure \ref{fig:simobs_percent}. We found that point-sources above 80 solar masses, under the typical observational conditions for the \textit{CMZoom} survey, are detected with more than 95\% completeness by the cataloging algorithm. The detection rate rapidly drops off, leaving only $\sim$80\% of structures detected at a $\sim$70 solar mass threshold, and about 50\% at $\sim$60 solar masses. To calculate these masses from the fluxes in the images, we assume a constant dust temperature of 20K, a typical value from {\it Herschel} temperature estimates in the CMZ \citep{mills_origins_2017}. These mass completeness estimates account only for the observational biases in observing point-source-like emission, and do not take into account the unrelated biases arising from more extended high-level emission structures. The bright, dense objects included in our catalog represent only the highest levels of this complex structure, and can be modeled by point-source-like emission for the purposes of this simplistic completeness estimate. We work on the assumption that the point-source analysis is sufficient for modeling the bright emission that we are prioritizing in our cataloging procedure. We emphasize that some of the cataloged emission is in more extended sources and does not closely match this point-source-like emission model, and we advise caution in interpreting these completeness estimates for more spatially intricate sources. We expect that some catalog leaves may include many unresolved or extremely nearby point-sources, particularly in extreme regions like the Sgr B2 complex. 

A similar test was constructed to provide a simplistic measure of the rate of false positive identifications, by measuring the number of leaves not associated with any of the injected point-sources. The results of this completeness test are also displayed in Figure \ref{fig:simobs_percent}. This is an idealized measure of the false positive value for objects in the catalog, since our technique considers only point-source emission, and is not necessarily applicable to the component of extended emission detected in the \textit{CMZoom} survey. While the real interstellar medium (ISM) is indisputably more complex than our simple model, this analysis provides a basic measure of the completeness of the survey, based on the effects of weather, imaging artifacts, and the cataloging scheme.

\section{Catalog Results}\label{sec:results}
Here we present the initial results of the catalog, initially focusing on the robust version of the catalog. Corresponding figures for the high-completeness catalog can be found in Appendix \ref{sec:complete_figs}. A complete gallery of all surveyed regions with cataloged leaves shown as contours is provided in Appendix \ref{sec:zoomins}. We wish to emphasize two notes before continuing.
\begin{itemize}
    \item The nature of the objects cataloged in this work occupy a range of spatial sizes that do not correspond unanimously to either cores (\textasciitilde 0.03-0.2 pc) or clumps (\textasciitilde 0.3-3 pc). Some of sources appear more filamentary and extended, while other sources are more compact and resemble large cores that likely are fragmented below our resolution limit. We briefly compare by eye the nature of the SMA emission to that of higher spatial resolution maps from ALMA in several key regions in Appendix \ref{sec:alma_comp}. The mass-radius relationship of our cataloged sources is compared to previous studies of the Galactic Center and disk in Section \ref{sec:comparison}.
    \item We refer to the contents of our catalog as ``dendrogram leaves" and ``compact sources" interchangeably. Typically we use ``leaves" when discussing the nature of the objects in the context of the cataloging algorithm, and ``compact sources" when describing the physical nature or properties of these objects. They are effectively interchangeable in either instance. 
\end{itemize}

\subsection{Leaf Distribution in l-b Space}

\begin{figure*}
\begin{center}
\includegraphics[trim = 0mm 0mm 25mm 0mm, clip, width = 0.63\textwidth, angle=270]{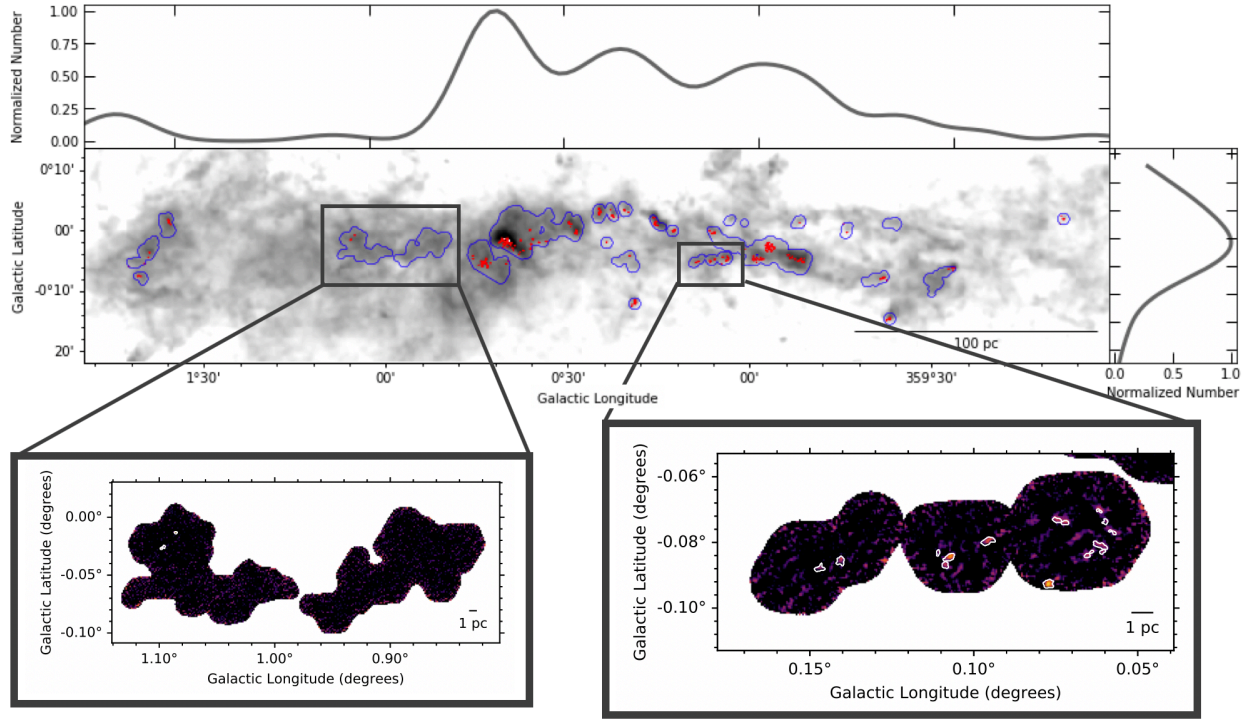}
\end{center}
\caption{The kernel density estimate of the normalized number of compact sources from the robust catalog as a function of Galactic longitude (top) and latitude (right) with leaf contours over a map of {\it Herschel}  (center).  The highest concentrations of detected compact sources are in Sgr B2, the Dust Ridge, and the 50 and 20 km s$^{-1}$ clouds. Zoom-ins are shown of several regions with white catalog leaf contours over the SMA 1.3mm dust continuum. The regions shown are G1.085-0.027, G1.038-0.074, G0.891-0.048 on the bottom left, and G0.068-0.075, G0.106-0.082, and G0.145-0.086 (the Three Little Pigs) on the bottom right.}
\label{fig:lb_hist}
\end{figure*}

\label{sec:lb_distro}
The robust catalog contains 285 leaves, all located within the Galactic latitude range of -0.25$^\circ$ to 0.07$^\circ$ and the Galactic longitude range -0.87$^\circ$ to 1.69$^\circ$. Their distribution in l-b space is shown in Figure \ref{fig:lb_hist}. There is a high density of leaves associated with the massive star forming complex Sgr B2, many of which are the most massive objects contained in the catalog. See Appendix \ref{sec:alma_comp} for a direct comparison between the Sgr B2 leaves in this work and the dust continuum emission observed at 3mm using ALMA by \cite{ginsburg_distributed_2018}. Outside the Sgr B2 cloud complex, the leaf distribution peaks in the dust ridge and around Sgr A* / circumnuclear disk (CND). The CND is a highly time variable source of synchrotron emission in the submillimeter regime (e.g.\ \citealt{serabyn_highfrequency_1997}), and the flux detected in its immediate surroundings appears to be dominated by imaging artifacts. For this reason we have chosen to exclude all leaves in this region from the following analysis, though they are included in the version of the catalog released with this paper.

\begin{figure*}
\begin{center}
\includegraphics[trim = 0mm 0mm 0mm 0mm, clip, width = 1.0\textwidth]{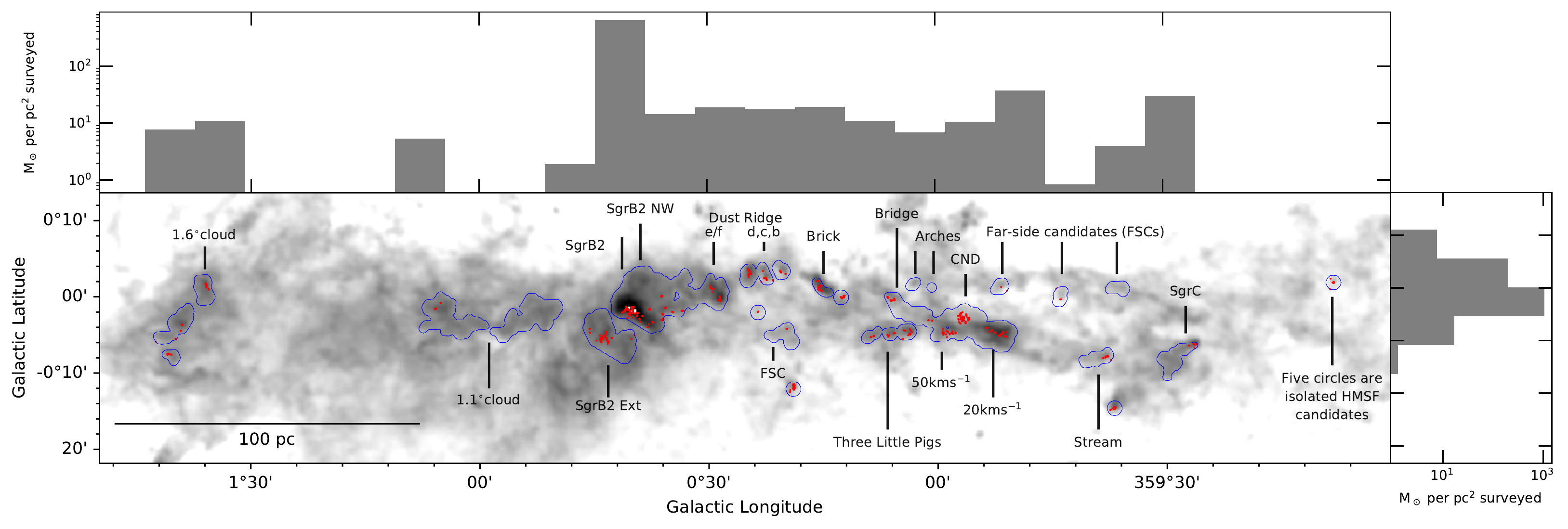}
\end{center}
\caption{The average mass of compact sources per unit area surveyed from the robust catalog as a function of Galactic longitude (top) and latitude (right) with leaf contours (in red) and the surveyed area footprint (blue) over the {\it Herschel} N(H$_2$) map from Battersby et al (in prep) (bottom). Clouds/regions are also labeled by their colloquial names. We detect high concentrations of mass in compact sources per area surveyed in Sgr B2, Sgr C, the Dust Ridge, the 20 and 50 km s$^{-1}$ clouds, the Brick, and the three little pigs, while other regions of high column density lacking a similar degree of compact substructure. This is quantified in \citetalias{battersby_cmzoom_2020} (Section 5.4) as a compact dense gas fraction, calculated as both SMA flux divided by Bolocam Galactic Plane Survey flux or SMA mass divided by {\it Herschel} mass. For an in-depth look at the compact dense gas fraction of these clouds, we refer the reader to \citetalias{battersby_cmzoom_2020}.}
\label{fig:lb_weighted_robust}
\end{figure*}

The distribution of leaves in Galactic longitude and latitude is displayed in figures \ref{fig:lb_hist} and \ref{fig:lb_weighted_robust}. Figure \ref{fig:lb_weighted_robust} shows the mass of sources per unit area surveyed as a function of latitude and longitude as well. While this distribution is highly influenced by the mapping strategy, it still highlights some of the high column density regions that lack compact substructure, such as the 1.1 degree cloud complex. The leaf distribution is largely dominated by the Sgr B2 cloud complex, the Dust Ridge, and the 20 and 50 km s$^{-1}$ clouds. The lack of compact substructure in many of the other surveyed clouds, such as the 1.1 degree cloud complex, is analyzed in detail in section 5 of \citetalias{battersby_cmzoom_2020}. 

\subsection{Comparison with Cloud Scale Column Densities}
\label{sec:measurements}
To supplement this catalog, we measure the local column density and temperature of the SMA compact sources and their surroundings from the {\it Herschel} property maps created using the dust SED fitting technique detailed in \cite{battersby_characterizing_2011}, to be presented in Battersby et al (in preparation).
The values are extracted from these maps at the position of the centroid of the catalog sources. These are cloud scale column density measurements at a resolution of 36\arcsec (1.5 pc). 

\begin{figure}
\includegraphics[trim = 0mm 0mm 0mm 0mm, clip, width = 0.45\textwidth]{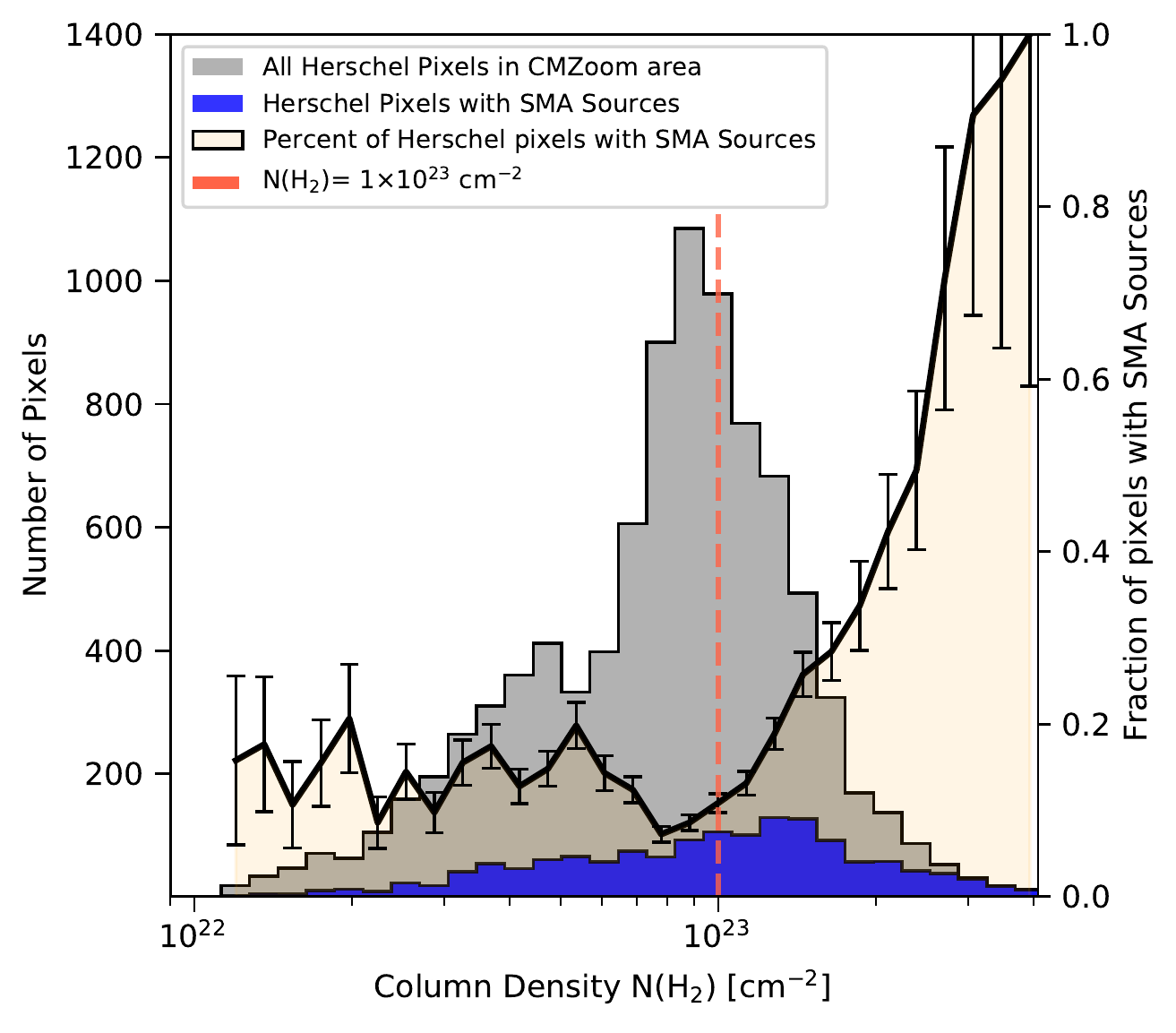}
\caption{ A histogram of the Herschel column density (derived using the procedure in \cite{battersby_characterizing_2011}, from Battersby et al in prep) associated with every pixel in the \textit{CMZoom} survey. The gray histogram represents the column density associated with every pixel covered by the SMA survey, while the blue represents only the pixels within one Herschel beam (36\arcsec) of a catalog leaf centroid in the robust version of the catalog. The ratio of pixels within one Herschel beam of a source in the SMA catalog to the total number of pixels in the SMA's map for a given column density is shown as a solid line, which experiences a sharp uptick around a column density of 1-2$\times 10^{23}$ cm$^{-2}$ (highlighted with the red dotted line at $10^{23}$ cm$^{-2}$. The error bars represent the Poisson uncertainty from the number of Herschel pixels in each column density bin. Pixels in the Sgr B2 region have been excluded from this analysis due to their much higher column densities resulting from potentially unreliable fluxes (see section \ref{sec:sgrb2}). A similar analysis is available for Sgr B2 sources in Figure 15 of \cite{ginsburg_distributed_2018}, looking instead at the fraction of material at a given column density that is associated with young stellar objects.}
\label{fig:herschel_hist}
\end{figure}

Figure \ref{fig:herschel_hist} shows the histogram of Herschel column density for pixels within the area observed by \textit{CMZoom}. Overlaid on this is a subset histogram of {\it Herschel} column density for pixels associated with our cataloged SMA compact sources. We determine pixels as associated if they lie within one {\it Herschel} beam FWHM (i.e., 36\arcsec) of a SMA catalog leaf centroid. The percentage of {\it Herschel} pixels that contain an SMA catalog source appears to sharply increase around a {\it Herschel} column density of $1-2\times 10^{23}$ cm$^{-2}$. The amount of compact substructure as measured by dendrogram leaf number seems highly dependent on the measured cloud scale column density. 

While the catalog presented in this work does not include information about the star forming properties of the leaves, the sharp increase in leaf occurrence at this threshold and its relationship to the star forming properties of leaves will be further investigated in upcoming work. We note that the \emph{Herschel}-derived column densities saturate in some pixels toward the center of Sgr B2 and may not be representative of true column densities in that region, and so all pixels and leaves in the Sgr B2 region have been removed from this analysis, and are not included in the histogram in Figure \ref{fig:herschel_hist}. Because these leaves are all highly concentrated at high column densities, they do not significantly influence the trend described in this section.
\pagebreak

\subsection{Physical properties}\label{Section:analysis_phys}
\label{sec:phys_prop}

\begin{figure}
\includegraphics[trim = 0mm 0mm 0mm 0mm, clip, width = 0.45\textwidth]{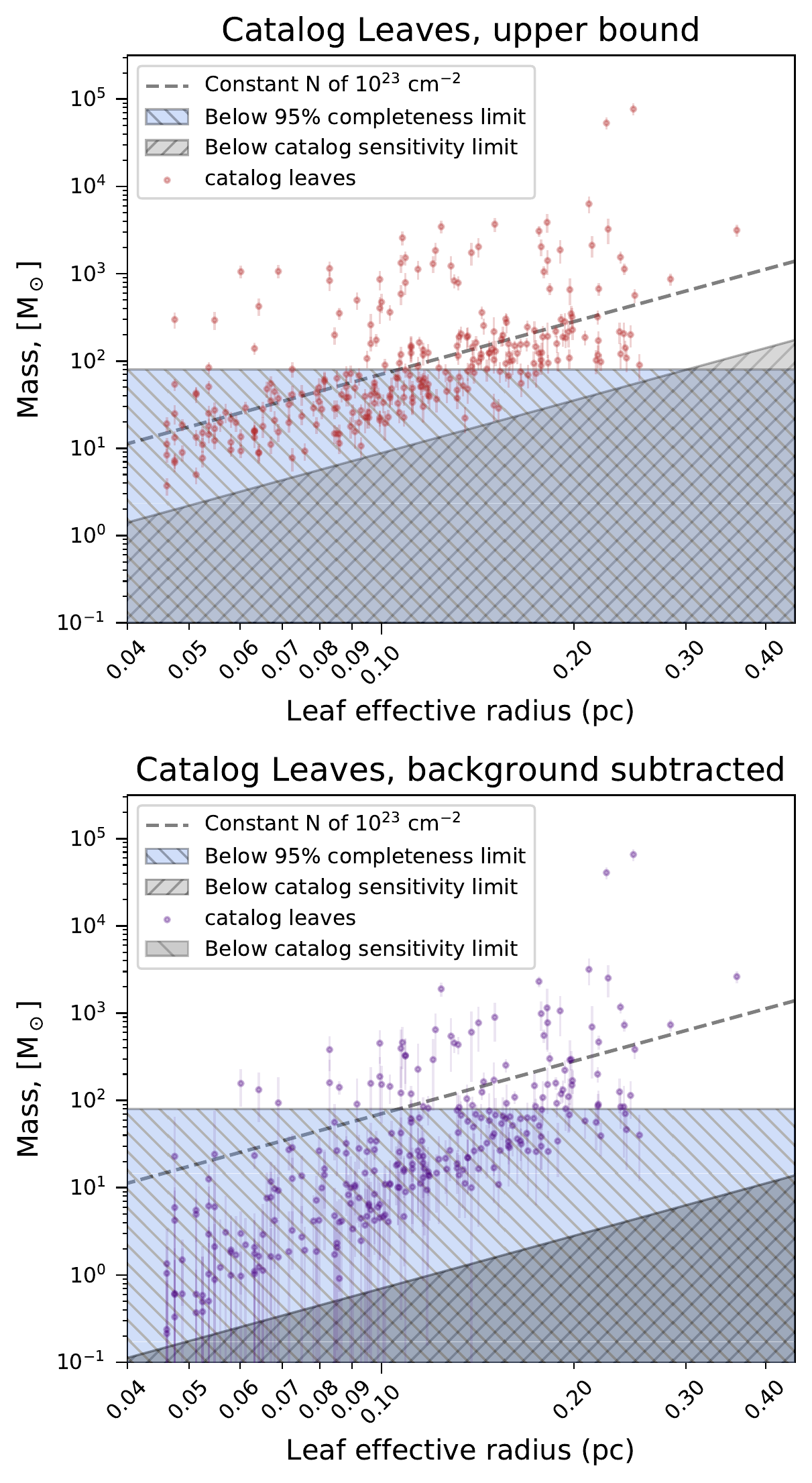}
\caption{Two representations of the mass ($M_\odot$) vs. radius (parsecs) for each leaf in the robust version of the catalog. The upper panel shows the mass calculated using the flux achieved by integrating over the entire leaf structure, acting as an upper bound for the true structure mass. The bottom panel displays the background subtracted leaf masses, calculated by subtracting the minimum value of the leaf and integrated over the entire leaf area. This background subtracted mass is likely an underestimate for the true structure mass. Both panels show a line of constant column density N = $10^{23}$ cm$^{-2}$ for reference. The range of masses below 95\% completeness is hatched right and shaded blue, and the region below minimum flux possible for a cataloged leaf is hatched left and shaded gray.}
\label{fig:mr2fig}
\end{figure}

The robust catalog is presented in tables \ref{tab:base}, \ref{tab:prop} and \ref{tab:unc}. We extract a number of physical properties of our cataloged leaves directly from the dendrogram structure using the \texttt{PP\_Statistic} package in the \texttt{Astrodendro} package, including the minimum flux value, maximum flux value, integrated flux, structure area, and center position. The structure area is converted into an effective projected radius for each leaf using $R_{\rm eff}\equiv(N_{\rm pix}A_{\rm pix}/\pi)^{1/2}$ (where $N_{\rm pix}$ is the number of pixels associated with a given dendrogram leaf and $A_{\rm pix}$ is the area of a single pixel). For each leaf, \texttt{Astrodendro} also extracts an effective ellipse major and minor axis, and position angle using moment analysis. For more information on the exact implementation of how these features are extracted, please refer to the \texttt{Astrodendro} documentation\footnote{\hyperlink{https://dendrograms.readthedocs.io/en/stable/}{https://dendrograms.readthedocs.io/en/stable/}}. These approximate ellipse values are included in the catalog, but we use the effective leaf radius calculated above instead for the following analysis.

Assuming optically thin dust continuum emission, the beam-averaged column density at the location of peak emission, $N_{\rm H_{2}}$ is estimated for each dendrogram leaf using
\begin{equation}\label{eq:rms}
N_{\rm H_{2}}=\frac{F^{\rm peak}_{\rm \nu}R_{\rm gd}}{\mu_{\rm H_{2}}m_{\rm H}\kappa_{\nu}B_{\rm \nu}(T_{\rm d})},
\end{equation}
where $F^{\rm peak}_{\rm \nu}$ is the peak flux density over the leaf (in Jy beam$^{-1}$), $R_{\rm gd}$ is the gas to dust mass ratio (for which we assume a value of 100, e.g.\ \citealt{battersby_characterizing_2011}), mean atomic weight $\mu_{\rm H_{2}}=2.8$ (e.g.\ \citealt{kauffmann_mambo_2008}), and $m_{\rm H}$ is the mass of atomic hydrogen, $\kappa_{\nu}$ is the dust opacity per unit mass at a frequency $\nu$, and $B_{\nu}(T_{\rm d})$ is the Planck function at a dust temperature, $T_{\rm d}$.


The dust opacity per unit mass is determined from $\kappa_{\nu}=~\kappa_{0}(\nu/\nu_0)^{\beta}$, assuming a dust emissivity index, $\beta$, where $\kappa_{0}$ is based on the moderately coagulated thin ice mantle dust model of \cite{ossenkopf_dust_1994} at a frequency, $\nu_{0}$. At a frequency of $\sim226$\,GHz we adopt a value of $\kappa_{\nu}\approx0.867$\,cm$^{2}$g$^{-1}$ (extrapolating from $\kappa_{0}=0.899$\,cm$^{2}$g$^{-1}$ at $\nu_{0}=230$\,GHz with $\beta=1.75$; e.g.\ \citealp{battersby_characterizing_2011}). While we do not have a measure of dust temperature  ($T_{\rm d}$) at the resolution of our SMA observations, we use corresponding dust temperatures derived from {\it Herschel} measurements. This temperature map and the {\it Herschel} column density map were generated using the dust SED fitting procedure outlined in \citealt{battersby_characterizing_2011}, and further detailed in \cite{mills_origins_2017}. These local temperature estimates have a resolution of 36\arcsec. While higher resolution temperature maps derived using the PPMAP procedure \citep{Marsh_multitemperature_2017} differ only by a few degrees K for the set of pixels considered in this work. This small change in temperature on smaller scales affects the derived physical properties significantly less than the systematic errors described below.

The SMA derived column densities range from $2.9~\times~10^{22}$  ${\rm cm^{-2}}<~N_{\rm H_{2}}<9.0~\times10^{25}$ ${\rm cm^{-2}}$, with a median value of $N_{\rm H_{2}}\sim 1.2\times10^{23}\,{\rm cm^{-2}}$. The high end of this distribution is dominated by two leaves with column densities $6.2\times10^{25} {\rm cm^{-2}}$ and $9.0\times10^{25} {\rm cm^{-2}}$. These extreme sources are within the Sgr B2 complex, and are colloquially known as Sgr B2 main (M) and north (N) respectively. 

\begin{figure}
\begin{center}
\includegraphics[trim = 0mm 0mm 0mm 0mm, clip, width = 0.45\textwidth]{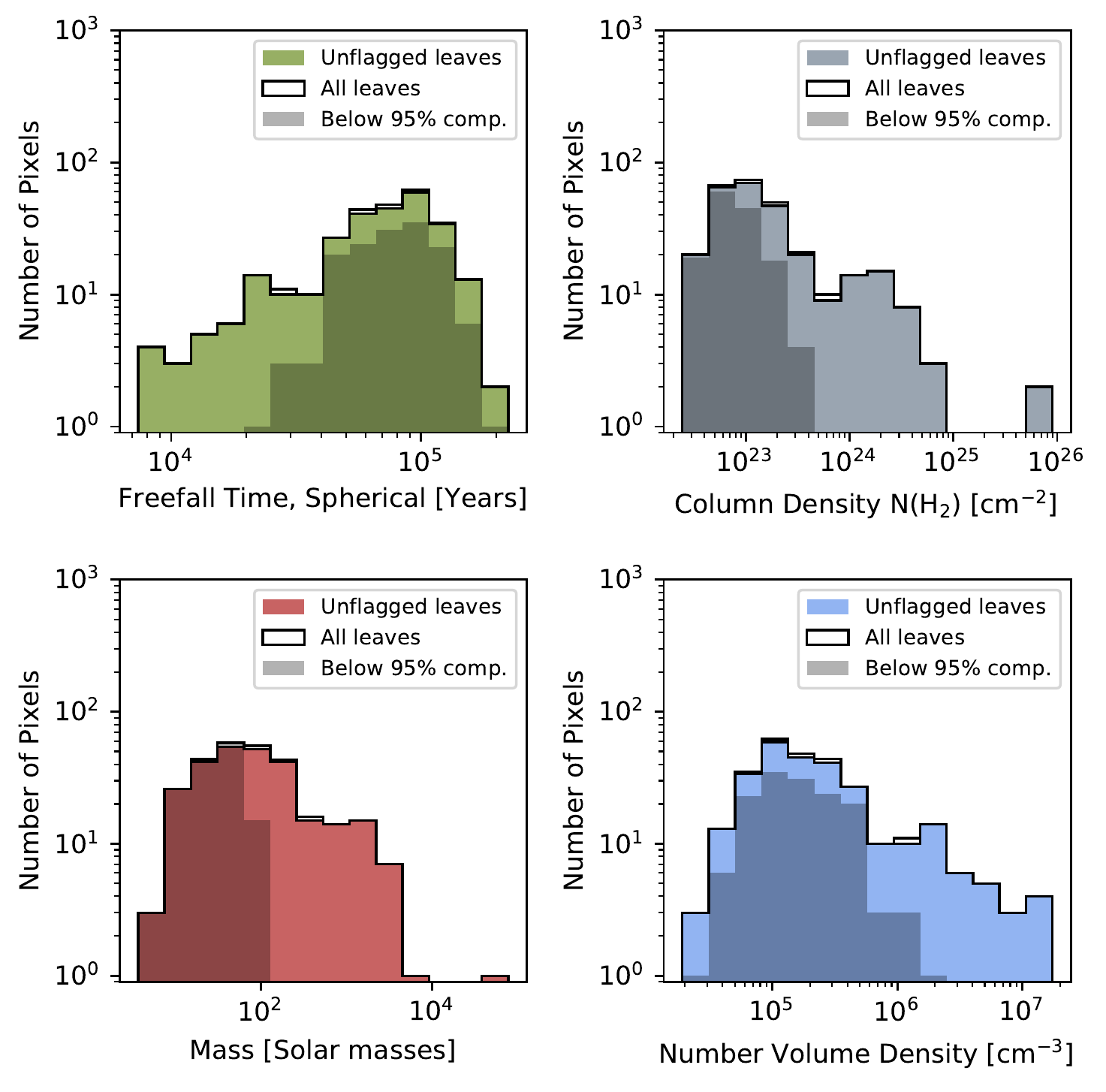}
\end{center}
\caption{Histograms of the physical properties for the leaves contained in the high reliability version of the catalog. These properties are calculated according to the methods described in section \ref{sec:phys_prop}. The upper left panel shows the distribution of freefall times calculated by assuming a spherical distribution of gas with the leaf's effective radius. The upper left panel shows the distribution of column densities for each leaf. The distribution of leaf masses is shown in the bottom left panel, and the volume density distribution is shown in the bottom right panel which is calculated again by assuming a spherical volume distribution according to the leaf's effective radius. The darker shaded regions correspond to the portion of leaves below the catalog's 95\% mass completeness limit. Leaves flagged as suspicious for being within 15 pixels of the map's edge are included as a non-filled histogram, though they largely follow the same trends as the more robust leaves. A roughly bimodal distribution is apparent, and its origin is explored in section \ref{sec:sgrb2}.}
\label{fig:physprop_histo}
\end{figure}

Following the method described in \cite{battersby_infrared_2010}, we find an isothermal estimate for the mass of each leaf using
\begin{equation}
M_{\rm leaf}=\frac{d^{2}S_{\nu}R_{\rm gd}}{\kappa_{\nu}B_{\nu}(T_{\rm d})}, 
\label{eq:mass}
\end{equation}

where T$_d$ is the local dust temperature estimate, $d$ is the distance to the source ($\sim8.178$\,kpc; \citealt{thegravitycollaboration_geometric_2019}) and $S_{\nu}$ is the integrated leaf flux (in Jy), $\kappa_{\nu}$ is the dust opacity per unit mass discussed above, and $B_{\nu}(T_{\rm d})$ is the Planck function at the local dust temperature, $T_{\rm d}$. The resultant estimates for leaf masses range from $\sim 4.1\,{\rm M_{\odot}}<M_{\rm leaf}<\sim 7.6\times 10^4 \,{\rm M_{\odot}}$, with a median value of $\sim 86 {\rm M_{\odot}}$. As with the column densities, the two most massive leaves in the catalog, Sgr B2 north and main, lie well apart from the rest of the mass distribution of leaves (see Figure \ref{fig:mr2fig}). Histograms of the physical properties are shown in Figure \ref{fig:physprop_histo}.

Two estimates for the leaf mass are included in the catalog: the upper bound leaf mass (reported above) and the background subtracted leaf mass. The upper bound mass is calculated using the total integrated flux from the pixel values for the whole leaf. This mass estimate assumes that all of the flux recovered in the leaf belongs to one, spatially cohesive clump, ignoring the more diffuse envelope surrounding the clump. Our alternative, lower mass estimate are calculated by subtracting away  the lowest flux value in the leaf from each pixel in the leaf,  removing all flux associated with the envelope (similar to \citealt{henshaw_investigating_2016}). This ``background subtracted" mass assumes that each leaf's lowest value (the outer contour of the leaf) is a good approximation for the average flux for the background emission. As a consequence, this background subtracted mass ignores any contribution from the lower dendrogram structures to the source mass, and therefore is likely to be an underestimate of the total mass. 

It is also important to note that both masses calculated in this way are subject to a systematic uncertainty of a factor of approximately two due to uncertainty in the dust opacity as described in detail in \cite{battersby_infrared_2010}.  The uncertainties associated with each of the leaf properties (see Table \ref{tab:unc}) account for this dominant systematic uncertainty as well as fluctuations in the {\it Herschel} dust temperatures, uncertainty in the Galactic Center distance assumption and the local noise estimates in the 1.3\,mm continuum flux. We assume an uncertainty in the distance to any given catalog source be $\sim \pm 240$pc, as this corresponds to the maximum longitudinal extent of sources identified in the survey, as the line-of-sight distance is not well known for many of the clouds hosting these sources.

The equivalent particle number density averaged over a leaf with radius $R_{\rm eff}$, and mass, $M$, can be estimated using
\begin{equation}
n_{\rm H_{2}}=\frac{M}{\frac{4}{3}\pi R_{\rm eq}^{3}\mu_{\rm H_{2}}m_{\rm H}},
\end{equation}
assuming the line-of-sight radius is equivalent to the radius on the plane of the sky. The range in particle number density is $2.3\times10^{4}\,{\rm cm^{-3}}<n_{\rm H_{2}}<1.7\times10^{7}\,{\rm cm^{-3}}$, with a median value of $n_{\rm H_{2}}=2.0\times10^{5}\,{\rm cm^{-3}}$. The corresponding range in the local freefall time is calculated by,
\begin{equation} \label{eq:tff}
t_{\rm ff}=\bigg(\frac{3\pi}{32G\mu_{\rm H_{2}}m_{\rm H} n_{\rm H_{2}}}\bigg)^{1/2},
\end{equation}
where $G$ is the gravitational constant and remaining constants are the same as defined above. For the robust catalog, dendrogram leaves range between $7.4\times10^{3}\,{\rm yr}<t_{\rm ff}<2.0\times10^{5}\,{\rm yr}$, with a mean value of $t_{\rm ff}=6.9\times10^{4}\,{\rm yr}$. Each of the above properties is displayed for a subset of leaves in Table \ref{tab:prop}. Uncertainties derived from propagating the dust opacity, assumed Galactic Center distance, typical fluctuations in {\it Herschel} dust temperature, and the local noise estimate for our 1.3mm dust continuum flux measurements are presented in Table \ref{tab:unc}. 
\pagebreak
\begin{table*}
\centering
\caption{Small subset of the leaf properties table, available in full \href{https://doi.org/10.7910/DVN/RDE1CH}{here}.}
\begin{tabular}{lllllllll}
\hline\hline
 Leaf ID & Area & $l$ & $b$ & R$_{\rm eff}$ & Integrated Flux & Peak Cont. Flux & Mean Cont. Flux & RMS \\
 & (as$^2)$ & (deg.) & (deg.) & (pc) & (Jy) & (Jy sr$^{-1}$) & (Jy sr$^{-1}$) & (Jy sr$^{-1}$) \\
\hline
G359.615-0.243a & 259.25 & -0.39 & -0.24 & 0.36 & 2.09e+00 & 1.73e+09 & 3.43e+08 & 9.88e+06 \\
G359.615-0.243b & 34.75 & -0.38 & -0.24 & 0.13 & 7.52e-02 & 1.70e+08 & 9.20e+07 & 1.25e+07 \\
G359.615-0.243c & 45.00 & -0.38 & -0.25 & 0.15 & 6.71e-02 & 1.39e+08 & 6.35e+07 & 1.68e+07 \\
G0.316-0.201a & 115.25 & 0.32 & -0.20 & 0.24 & 9.12e-01 & 1.09e+09 & 3.37e+08 & 9.56e+06 \\
G0.316-0.201b & 75.50 & 0.32 & -0.19 & 0.19 & 1.41e-01 & 1.47e+08 & 7.97e+07 & 1.43e+07 \\
G0.316-0.201c & 38.75 & 0.32 & -0.21 & 0.14 & 7.12e-02 & 1.28e+08 & 7.82e+07 & 1.91e+07 \\
G0.316-0.201d & 60.00 & 0.31 & -0.19 & 0.17 & 7.86e-02 & 1.23e+08 & 5.57e+07 & 1.92e+07 \\
G0.316-0.201e & 19.00 & 0.32 & -0.20 & 0.10 & 3.88e-02 & 1.22e+08 & 8.68e+07 & 1.67e+07 \\
G0.316-0.201f & 28.25 & 0.31 & -0.19 & 0.12 & 3.28e-02 & 7.71e+07 & 4.94e+07 & 1.16e+07 \\
G0.316-0.201g & 15.50 & 0.31 & -0.20 & 0.09 & 3.13e-02 & 1.27e+08 & 8.60e+07 & 1.35e+07 \\
... & ... & ... & ... & ... & ... & ... & ... & ...\\
\hline\hline
\end{tabular}
\label{tab:base}
\end{table*}

\subsection{Mass-Radius Comparison to Previous Studies of the CMZ and Galactic Disk}\label{sec:comparison}

\begin{figure*}
\begin{center}
\includegraphics[trim = 0mm 0mm 0mm 0mm, clip, width = 0.9\textwidth]{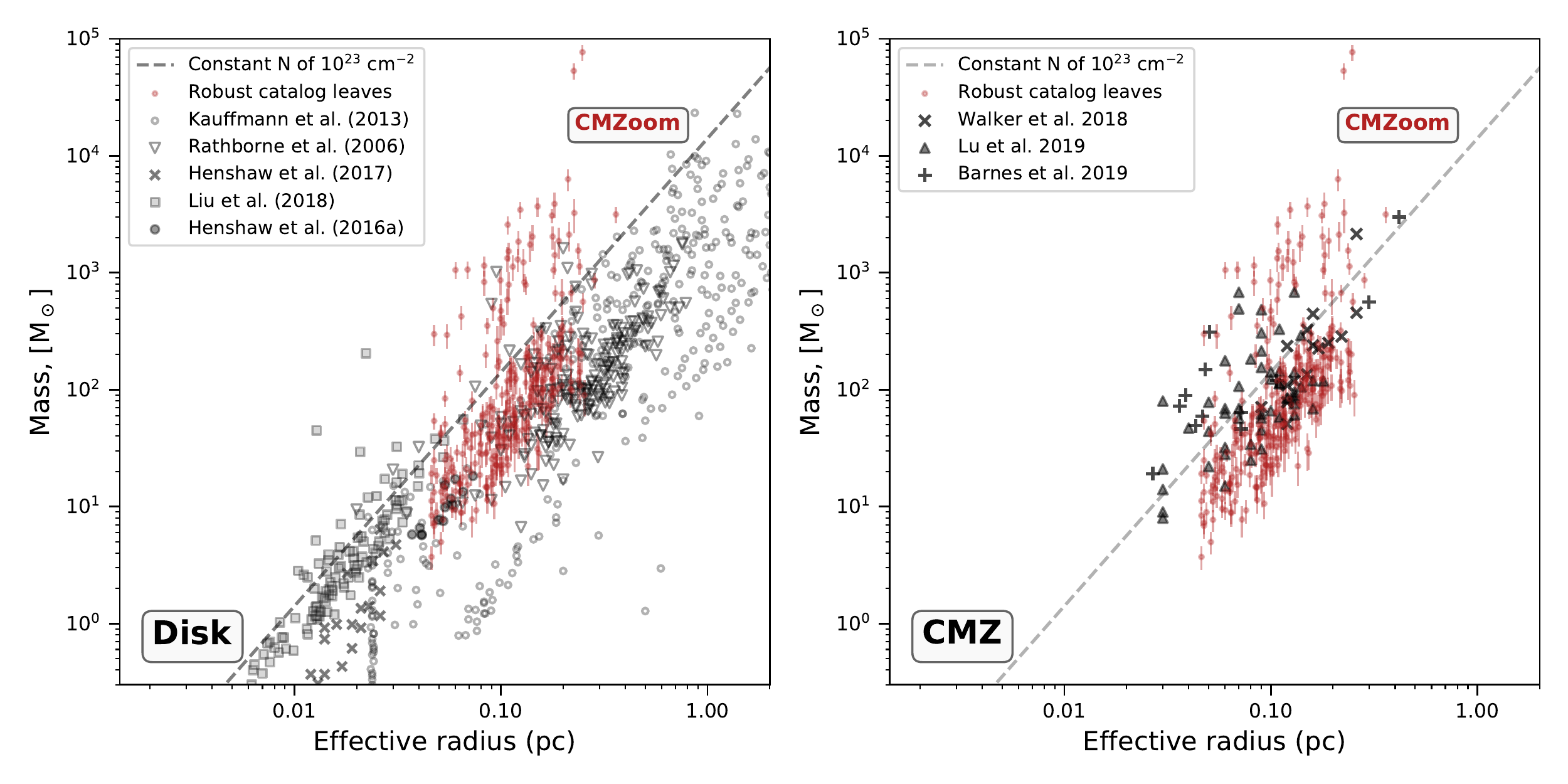}
\end{center}
\caption{A mass-radius comparison of our catalog leaves (from the robust catalog, shown as red dots with mass uncertainties) with cataloged objects from previous work studying cores in the Milky Way's disk (left panel) and in the Galactic Center (right panel). A line of constant column density N=$10^{23}$ cm$^{-2}$ is shown plotted over the data for reference. For the Galactic disk, data are plotted from \cite{rathborne_infrared_2006} \cite{kauffmann_low_2013}, \cite{henshaw_investigating_2016}, \cite{Henshaw_unveiling_2017}, and \cite{liu_mm_2018}. For the Galactic Center, data are plotted from \cite{walker_star_2018}, \cite{barnes_young_2019}, and \cite{lu_census_2019}.}
\label{fig:core_comparison}
\end{figure*}

As previously mentioned, our cataloged sources occupy a spatial scale ranging between $\sim$ 0.04-0.4 pc, meaning they largely constitute objects intermediate to clumps and cores. We briefly compare the mass-radius distribution of our robust catalog sources to clumps and cores identified by previous observational catalogs in Figure \ref{fig:core_comparison}. Our sources largely agree with the mass-radius distribution of previous studies of the CMZ. Many of these previous studies (e.g.\ \citealt{walker_star_2018,lu_census_2019}) aim to identify star forming objects and their properties, and CMZ sources with known star formation sites seem to have a higher mass per radius than the bulk of our catalog. This trend will be investigated further in upcoming work (Hatchfield et al. in preparation). 
\pagebreak

\subsection{High-Completeness Catalog}

As an alternative to the primary, robust catalog released with this paper, we also present a more lenient catalog that prioritizes greater completeness over a low false-positive rate. We use the same cataloging algorithm described in section \ref{sec:catalog}, while adjusting the parameters to suit the new design philosophy. Specifically we lower the pruning parameters from the high-reliability version,
$$ \text{Flux}_{\text{1.3mm}}^{\text{peak}} > 6\sigma_{RMS},  $$
where $\text{Flux}_{\text{1.3mm}}^{\text{peak}}$ is the peak pixel flux, and $\sigma_{RMS}$ is the local RMS noise estimate, to
$$ \text{Flux}_{\text{1.3mm}}^{\text{peak}} > 4\sigma_{RMS}.  $$
The requirement that each leaf's mean flux be $> 2\sigma$ above the local RMS applied to both catalogs. This reduction in the maximum required per pixel flux results in a significant increase in the number cataloged leaves in the high-completeness catalog compared to the high-reliability catalog. The change produces a catalog of 816 leaves as compared to the robust catalog with its 285 leaves. Despite this large change the number of leaves, most of the leaves exclusive to the high-completeness catalog have relatively low-mass and together constitute an additional 2.8$\times 10^4$M$_\odot$, which is only 13\% of the robust catalog's total mass. The vast majority of mass is cataloged in both versions. 

The more diffuse nature of emission in the \textit{CMZoom} field as well as non-Gaussian components in the noise resulting from primary beam correction and cleaning artifacts together make it difficult to measure the trade off of completeness vs. false positive rate using the same simulated observation procedure outlined in Section \ref{sec:simobs}. The assumptions inherent in this procedure, namely the cataloging of a random spatial distribution of uniform mass point-sources, might be ill suited for characterizing the lower flux component of emission in the \textit{CMZoom} field. The simulated observations result in a steep increase in the number of false positives and a small increase in completeness for changes in the pruning, with a 95\% completeness of $\sim 50$ M$_\odot$ (assuming T$_d$=20K).

\begin{table*}
\caption{Small subset of the leaf properties table (continued).}
\centering
\begin{tabular}{llllllll}
\hline\hline
Leaf ID & N$_{\rm Herschel}$ & N$_{\rm SMA}$ & Mass & Mass (bg. sub.) & n & $\rho$ & t$_{\rm ff}$ \\
 & (cm$^{-2})$ & (cm$^{-2})$ & (M$_\odot)$ & (M$_\odot$) & (cm$^{-3})$ & (g cm$^{-3})$ & (yr) \\
\hline
G359.615-0.243a & 1.46e+23 & 1.73e+24& 3.15e+03 & 2.62e+03 & 2.33e+05& 1.09e-15 & 6.37e+04 \\
G359.615-0.243b & 8.59e+22 & 1.73e+23& 1.14e+02 & 4.28e+01 & 1.72e+05& 8.07e-16 & 7.41e+04 \\
G359.615-0.243c & 6.81e+22 & 1.42e+23& 1.03e+02 & 4.16e+01 & 1.05e+05& 4.92e-16 & 9.49e+04 \\
G0.316-0.201a & 5.74e+22 & 9.00e+23& 1.13e+03 & 7.31e+02 & 2.82e+05& 1.32e-15 & 5.79e+04 \\
G0.316-0.201b & 5.97e+22 & 1.28e+23& 1.84e+02 & 8.04e+01 & 8.66e+04& 4.06e-16 & 1.05e+05 \\
G0.316-0.201c & 3.86e+22 & 1.24e+23& 1.04e+02 & 4.97e+01 & 1.32e+05& 6.21e-16 & 8.45e+04 \\
G0.316-0.201d & 4.94e+22 & 9.76e+22& 9.39e+01 & 6.23e+01 & 6.23e+04& 2.92e-16 & 1.23e+05 \\
G0.316-0.201e & 4.65e+22 & 1.18e+23& 5.62e+01 & 1.57e+01 & 2.09e+05& 9.80e-16 & 6.73e+04 \\
G0.316-0.201f & 5.94e+22 & 6.15e+22& 3.92e+01 & 1.36e+01 & 8.05e+04& 3.77e-16 & 1.08e+05 \\
G0.316-0.201g & 4.01e+22 & 1.03e+23& 3.82e+01 & 1.10e+01 & 1.93e+05& 9.03e-16 & 7.01e+04 \\
... & ... & ... & ... & ... & ... & ... & ... \\
\hline\hline
\end{tabular}
\label{tab:prop}
\end{table*}

Additionally, a large number of new leaves are included in the catalog which are in immediate proximity to the edge of the \textit{CMZoom} maps, which is suspicious due to more extreme fluctuations in the noise at the map edges resulting from the primary beam correction process. To mitigate this, we have flagged all leaves with centroids within 15 pixels of the map's edge, and confirmed that the removal of these leaves does not seem to change the overall distribution in the physical property statistics. We include recreated versions of the physical property figures in Appendix \ref{sec:complete_figs}).  

\begin{table*}
\caption{Small subset of leaf property uncertainties.}
\centering
\begin{tabular}{llllll}
\hline\hline
Leaf ID & Mass Unc. & N$_{\rm SMA}$ unc. & n unc. & $\rho$ unc. & t$_{\rm ff}$ unc. \\
 & (M$_\odot)$ & (cm$^{-2})$ & (cm$^{-3})$ & (g cm$^{-3})$ (yr) \\
\hline
G359.615-0.243a & 4.92e+02 & 3.48e+24& 7.85e+04 & 3.68e-16 & 2.15e+04 \\
G359.615-0.243b & 2.34e+01 & 3.47e+23& 6.20e+04 & 2.90e-16 & 2.67e+04 \\
G359.615-0.243c & 3.15e+01 & 2.85e+23& 4.47e+04 & 2.09e-16 & 4.04e+04 \\
G0.316-0.201a & 1.72e+02 & 1.80e+24& 1.15e+05 & 5.41e-16 & 2.37e+04 \\
G0.316-0.201b & 4.31e+01 & 2.57e+23& 3.84e+04 & 1.80e-16 & 4.63e+04 \\
G0.316-0.201c & 2.99e+01 & 2.49e+23& 5.76e+04 & 2.70e-16 & 3.67e+04 \\
G0.316-0.201d & 3.52e+01 & 1.96e+23& 3.84e+04 & 1.80e-16 & 7.59e+04 \\
G0.316-0.201e & 1.38e+01 & 2.37e+23& 8.48e+04 & 3.97e-16 & 2.73e+04 \\
G0.316-0.201f & 1.09e+01 & 1.24e+23& 4.20e+04 & 1.97e-16 & 5.66e+04 \\
G0.316-0.201g & 8.26e+00 & 2.06e+23& 8.90e+04 & 4.17e-16 & 3.23e+04 \\
... & ... & ... & ... & ... & ... \\
\hline\hline
\end{tabular}
 \label{tab:unc}
\end{table*}

Qualitatively, the distribution of leaves' physical properties is similar in both catalogs, with the catalog design changes mainly permitting the detection of a larger number of low-mass leaves.

\section{Discussion}
\label{sec:discussion}
\subsection{The Elephant in the Room: Sagittarius B2}
\label{sec:sgrb2}
\begin{figure}
\begin{center}
\includegraphics[trim = 0mm 0mm 0mm 0mm, clip, width = 0.45\textwidth]{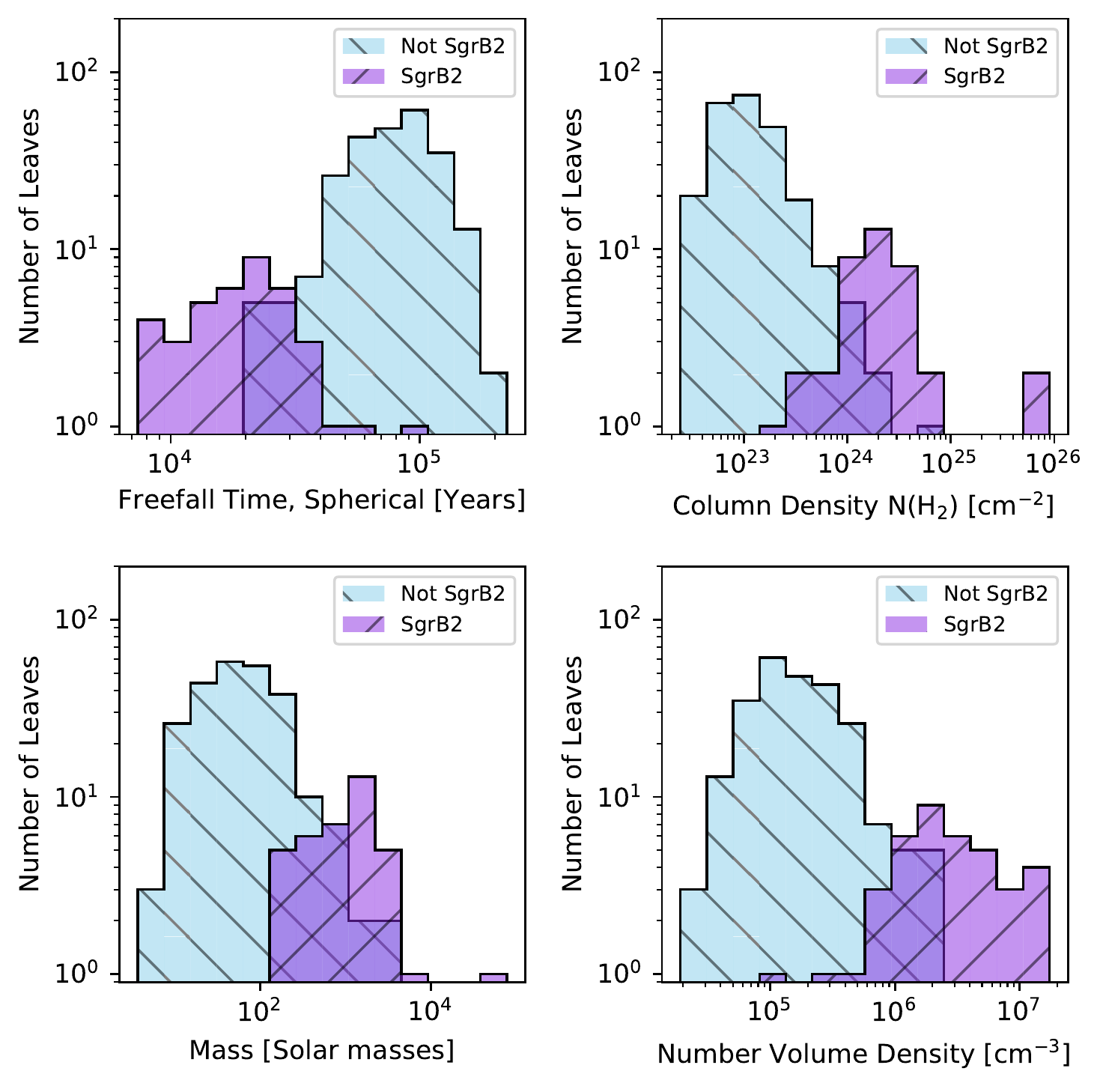}
\end{center}
\caption{Separated histograms of the physical properties for the leaves within and apart from Sgr B2, for the robust version of catalog. These properties are calculated according to the methods described in section \ref{sec:phys_prop}, and separated according to the mask released with this work at \hyperlink{https://dataverse.harvard.edu/dataverse/cmzoom}{https://dataverse.harvard.edu/dataverse/cmzoom}.}
\label{fig:sgrb2_phys_prop}
\end{figure}

The distribution of leaves in Figure \ref{fig:mr2fig} reveals a striking characteristic of the mass-radius relation of the catalog. There appears to be a population set apart from the majority in the mass-radius distribution of the leaves. This feature is resilient to changes in the catalog algorithm parameters, as detailed in Appendix \ref{sec:param_study}. Almost all of the points in the "higher mass" mode are located in the region surrounding Sgr B2, the most intense site of star formation in the Galactic Center. In Figure \ref{fig:sgrb2_phys_prop} we compare the catalog divided into leaves associated with Sgr B2 and all others. It is clear from the histograms of physical properties that objects in Sgr B2 are detected with higher flux per leaf area than elsewhere in the surveyed region, leading to higher masses, column and volume densities, and shorter freefall times. As measured, sources associated with Sgr B2 have more mass than the rest of the CMZ combined by nearly a factor of 10, assuming the validity of the {\it Herschel} dust temperatures used in the mass calculation from Section \ref{sec:phys_prop}. 

There are a few possible explanations for this extreme population. Firstly, it is possible that there is significant contamination in the 1.3mm flux from non-dust emission. Of the possible sources of non-dust continuum emission in Sgr B2, the most likely culprit is free-free emission coming from HII regions. Previous observations have identified numerous HII regions distributed throughout the Sgr B2 complex \citep{Mehringer_very_1995,depree_sagittarius_1996,depree_evidence_2015,ginsburg_distributed_2018,lu_star_2019}. \citet{ginsburg_distributed_2018} find the Sgr B2 complex's 3mm continuum emission to be well populated by both extended and compact HII regions. In particular, Sgr B2 Main (identified in our catalog as leaf ID G0.699-0.028b, the second most massive source in the catalog) is dominated by HII region continuum flux at 3mm (see figure 3 and table 4 of \citealt{ginsburg_distributed_2018}). While we expect free-free emission to be somewhat weaker at 1.3mm, it is still likely the free-free emission accounts for some of the observed flux these regions.

\begin{figure*}
\begin{center}
\includegraphics[trim = 0mm 0mm 0mm 0mm, clip, width = 0.5\textwidth]{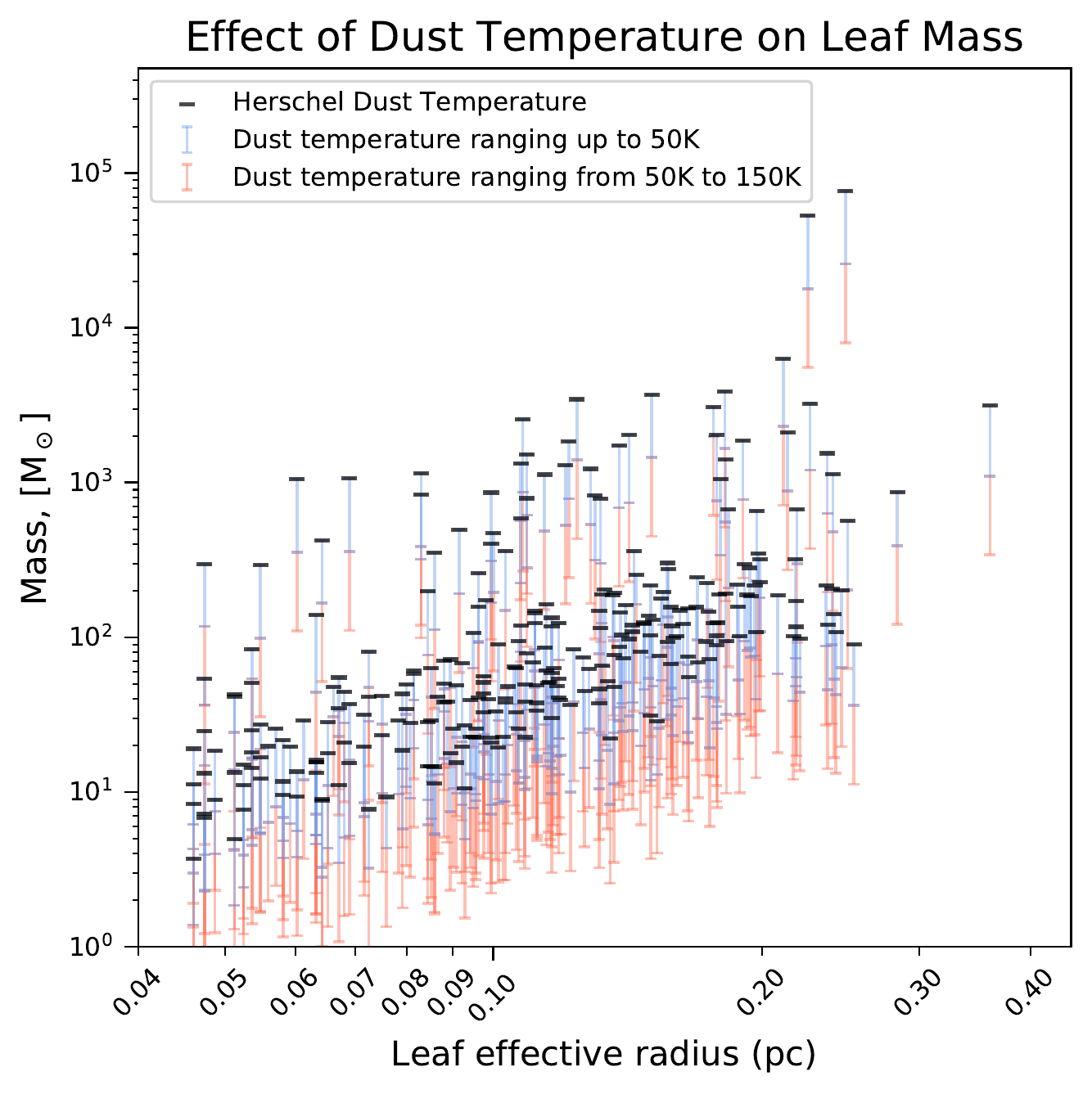}
\end{center}
\caption{The mass ($M_\odot$) vs. radius (parsecs) for each leaf in the high reliability version of the catalog for variations in the assumed dust temperatures. The black cap represents leaf mass for an assumed temperature of 20K. The blue bar shows how much the mass of each leaf decreases for an assumed dust temperature of up to 50K, and the red bar shows how much the mass decreases for a dust temperature of 150K, which is higher than typically expected for dust in the CMZ (\citealp{pierce-price_deep_2000, etxaluze_herschel_2013}). The temperature assumed would have to be close to 150K to explain the separation of the Sgr B2 leaves' distribution in the mass-radius relation. This is unlikely to as the multi-wavelength modeling from \cite{schmiedeke_physical_2016} derived considerably lower dust temperatures for most of the Sgr B2 complex. }
\label{fig:mr_temp}
\end{figure*}

Secondly, temperature has a strong effect on the inferred mass of our objects, so it is possible that the Sgr B2 area hosts higher dust temperatures than the {\it Herschel} estimates used in this work. In calculating the mass for each leaf we assume the validity of the cloud scale dust temperature, so the difference between the two modes could be explained by a systematic difference in dust temperature on smaller scales. In Figure \ref{fig:mr_temp} we demonstrate how the mass of each leaf would vary with a temperature between the observed {\it Herschel} dust temperature (close to 20K for most leaves) and 50K, and then from 50K to 150K. If we assume T$_{\rm dust}=\, \sim120 - 150$K, the Sgr B2 source masses become consistent with the broader CMZ leaf population. The {\it Herschel} temperatures reported in this catalog agree with previous measurements of CMZ dust temperatures, typically range between 20 and 30 K \citep{pierce-price_deep_2000, etxaluze_herschel_2013}. The SMA data presented in this work probe smaller spatial scales than the {\it Herschel} maps (smallest beam FWHM = 5.9$^{\prime\prime}$), so it is possible that different dust temperatures would be resolved on smaller spatial scales. \citet{Marsh_multitemperature_2017} uses the PPMAP procedure to derive temperature maps at a spatial resolution of 12\arcsec, but these temperatures differ from the lower resolution data by only a few degrees K for pixels in the \textit{CMZoom} survey footprint. While it is possible that even smaller spatial scales could reveal higher dust temperatures in areas of active or recent star formation such as Sgr B2, it is unlikely to have a widespread effect on our catalogs. It is also worth noting the \textit{Herschel} temperatures in certain pixels in the Sgr B2 region are likely to be unreliable due to saturation and line contamination.

\begin{figure}
\includegraphics[trim = 0mm 0mm 0mm 0mm, clip, width = 0.45\textwidth]{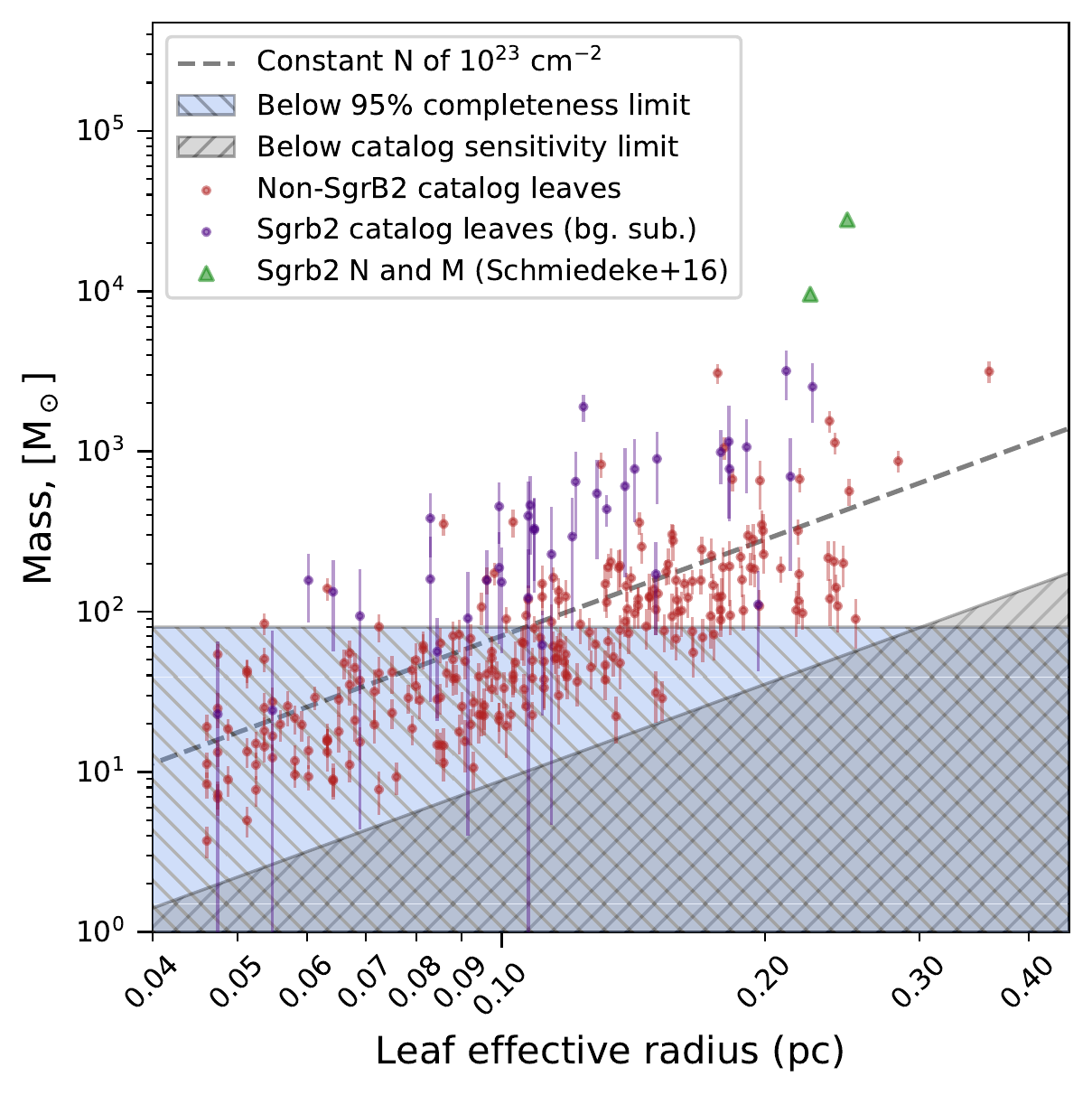}
\caption{The mass ($M_\odot$) vs. radius (parsecs) for each leaf in the robust version of the catalog, using the background subtracted mass values for leaves in the Sgr B2 complex (Sgr B2 leaves' background subtracted masses are shown in purple, the modified masses for Sgr B2 N and M from \citealt{schmiedeke_physical_2016} are shown as green triangles, while the rest of the catalog is shown in red). Using this background subtracted mass mostly resolves the separation between the bimodal distribution of leaves in Figure \ref{fig:mr2fig}, suggesting that the line-of-sight overlapping envelopes might be significantly affecting the mass of these leaves in particular (see section \ref{sec:sgrb2}). For comparison with Figure \ref{fig:mr2fig}, the same line of constant column density N = $10^{23}$ cm$^{-2}$ is shown. Again, the range of masses below 95\% completeness is hatched right and shaded blue, and the region below minimum flux possible for a cataloged leaf is hatched left and shaded gray.}
\label{fig:sgrb2_mr}
\end{figure}

Lastly, it is possible that the measured masses and column densities are due to line-of-sight confusion.
It has been suggested that the line-of-sight structure of the Sgr B2 complex could lead to complications in determining the mass of individual structures, with some evidence suggesting Sgr B2 north and main might have overlapping, elongated envelopes \citep{Goldsmith_high_1990,schmiedeke_physical_2016}. In this case, the background subtracted masses calculated in Section \ref{sec:phys_prop} would be more appropriate for comparisons with the less complicated regions in our catalog. These background subtracted masses are in much better agreement with the rest of the catalog's contents, shown in Figure \ref{fig:sgrb2_mr}. This may be suggesting that the line-of-sight envelope effects contribute significantly to our reported masses in this region.

In reality, it is likely that a combination of these effects would impact the masses calculated in this work. Flux from HII regions certainly accounts for some of the flux in the two most massive leaves (leaf IDs G0.699-0.028a and G0.699-0.028b), but \citet{ginsburg_distributed_2018} finds only 10 compact HII regions associated with sources other than Sgr B2 main, meaning HII regions are likely not solely responsible for the  difference in mass-radius distribution, but may be more likely to explain the extreme mass we calculate for Sgr B2 north and main. As described in Figure \ref{fig:mr_temp}, a dust temperature greater than 150K would be required to fully explain the separation of these sources. While such high dust temperatures could exist for sources with extreme, ongoing star formation activity like Sgr B2 north and main, they are unlikely to be so high in this region unanimously.  A combination of high local dust temperatures, line-of-sight envelope overlap, and non-continuum contamination together can readily explain the apparent divergence of these leaves from the more general distribution of CMZ compact structures in mass-radius space. 

\subsection{Maximum Star Formation Potential of the CMZ}\label{sec:sf_limit}
The high-completeness catalog presented in this work constitutes more than 95\% of compact sources with masses greater than 80 M$_\odot$ (for the robust catalog, or 50 M$_\odot$ for the high-completeness catalog) embedded in high column density gas ($\geq 10^{23} $cm$^{-2}$) in CMZ clouds. Determining the star formation potential of any one of these objects is challenging due to the complexity of the environmental conditions in CMZ clouds. some of these structures are well known to be actively star forming, while some others show tenuous signatures of star formation activity and many seem entirely quiescent. The connections between source properties and star formation signatures are investigated in upcoming work.

Because \textit{CMZoom} covers all high column density material in the CMZ, we suspect these catalogs contain all possible sites of star cluster formation over a relevant timescale, up to our completeness limits. While it is not necessary that each of these compact sources will collapse down on a freefall time to form stars, we can determine an upper limit on current star formation potential of the CMZ. 

This star formation potential is the maximum possible SFR for the CMZ, calculated by assuming each compact source in our catalog collapses on a freefall time and forms stars with an assumed star formation efficiency. Here this is done for the robust catalog, though the SFR estimated using the high-completeness catalog is similar, albeit with larger uncertainty. We choose a range of star formation efficiencies accounting for the wide range of proposed efficiencies on scales of clumps ($ 0.1 \leq \epsilon_{\rm clump} \leq 0.3$) to the possible efficiencies of individual cores ($ 0.25 \leq \epsilon_{\rm core} \leq 0.75$) since our compact sources are intermediate to these two hierarchical categories \citep{lada_embedded_2003,mckee_theory_2007}. We calculate our SFR limit for both the upper and lower bounds of this efficiency range, and report both. This choice of SFE is consistent with the results of \cite{lu_star_2019}, in which star formation efficiencies for gravitationally bound cores in Galactic Center clouds are measured to be $\sim$ 0.3, with a systematic error of about a factor of 3. We calculate the SFR in two ways. Firstly, we find an "individual" SFR for each compact source in the catalog using
\begin{equation}
    \text{SFR}_{\rm indiv.} = \sum_{\rm i} \frac{\epsilon \, m_{\rm i}}{t_{\rm ff,i}},
\end{equation}
where the sum is over each source in the catalog, $\epsilon$ is the star formation efficiency, $m_i$ is the mass of the source, and $t_{\rm ff,i}$ is the freefall time of the source, calculated using equation \ref{eq:tff}. Alternatively, we calculate the "mean" SFR for all sources over the mean freefall time, using 
\begin{equation}
    \text{SFR}_{\rm mean} = \frac{\epsilon }{t_{\rm ff,mean}}\sum_{\rm i} m_{\rm i},
\end{equation}
where $t_{\rm ff,mean}$ is the mean freefall time of sources in the catalog (for the robust catalog, this value is 7.48$\times 10^4$ Myr).

To arrive at a more accurate upper value for the star formation potential, we use the background-subtracted values for masses of objects in the Sgr B2 complex. Additionally, we replace the extreme masses measured in this work (described in Section \ref{sec:phys_prop}) for Sgr B2 North and Main with more reliable gas masses derived in \cite{schmiedeke_physical_2016} to arrive at the above estimates for the star formation potential. Without these modifications, Sgr B2 would completely dominate the SFR potential, exceeding the combined SFR of the entire Milky Way, which is not plausible. For an assumed efficiency of $\epsilon=0.1$, we find 
\[
  \text{SFR}_{\rm indiv.}=
  \begin{cases}
                                   0.29\text{M}_\odot \text{yr}^{-1} & \text{for $\epsilon=0.1$} \\
                                   2.2\text{M}_\odot \text{yr}^{-1} & \text{for $\epsilon=0.75$}
  \end{cases}
\]
or using the mean freefall time,
\[
  \text{SFR}_{\rm mean}=
  \begin{cases}
                                   0.08\text{M}_\odot \text{yr}^{-1} & \text{for $\epsilon=0.1$} \\
                                   0.64\text{M}_\odot \text{yr}^{-1} & \text{for $\epsilon=0.75$}
  \end{cases}
\]

The vast majority of mass generating this SFR potential is associated with leaves in the Sgr B2 complex, where we suspect our masses to be most unreliable for reasons discussed in Section \ref{sec:sgrb2}.

If instead we consider the star formation potential of all cataloged sources outside the Sgr B2 complex, we find
\[
  \text{SFR}_{\rm indiv.}=
  \begin{cases}
                                   0.06\text{M}_\odot \text{yr}^{-1} & \text{for $\epsilon=0.1$} \\
                                   0.47\text{M}_\odot \text{yr}^{-1} & \text{for $\epsilon=0.75$}
  \end{cases}
\]
or using the mean freefall time,
\[
  \text{SFR}_{\rm mean}=
  \begin{cases}
                                   0.04\text{M}_\odot \text{yr}^{-1} & \text{for $\epsilon=0.1$} \\
                                   0.30\text{M}_\odot \text{yr}^{-1} & \text{for $\epsilon=0.75$}
  \end{cases}
\]

We conclude that the maximum star formation potential of the CMZ is between 0.08 - 2.20M$_\odot$ yr$^{-1}$, or, excluding Sgr B2, between 0.04 - 0.47M$_\odot$ yr$^{-1}$. We wish to emphasize that these SFR potentials may not be accurate predictions of the CMZ's future SFR, as some of these objects may not ultimately collapse to form stars. Because the CMZ appears to be a highly complex environment, the longevity and boundedness of any given compact structure might be compromised by turbulence, dynamical interactions like inflow and shear, as well as other disrupting mechanisms. Therefore, these SFR potential estimates should be understood as an upper limit on the possible future SFR for the assumed star formation efficiency. Of the entire SFR of the Milky Way, estimated to be around 1.9 $\pm$ 0.4 M$_\odot$ yr$^{-1}$ \citep{chomiuk_Unification_2011}, present tracers of star formation in the CMZ indicates that 0.06 M$_\odot$ yr$^{-1}$ of this star formation activity occurs within the central 500pc of the Galaxy (e.g.\ \citealp{longmore_variations_2013, barnes_star_2017}). The CMZ might roughly maintain the current SFR if we assume a low SFE for these objects, or alternatively if the dense structures required longer timescales than their freefall times to collapse and form stars (e.g.\ \citealt{lu_star_2019}), and if some fraction of compact substructure remaining unbound and never collapsing to form stars.

\subsection{High-Mass Star Precursor Completeness}\label{sec:star_completeness}
One of the key goals of the \textit{CMZoom} survey is to compile potential sites of massive star formation as completely as possible. There is substantial evidence suggesting that massive stars largely form in high column density material (e.g.\ \citealt{Kauffmann_how_2010,lada_star_2010}), and \textit{CMZoom} covers all of the CMZ's high column density material. The simulated observations performed in section \ref{sec:simobs} are used to estimate how many leaves in the \textit{CMZoom} survey area might be excluded from the catalog as a function of the leaf mass (see Figure \ref{fig:simobs_percent}). If we can estimate the number of missing leaves as a function of mass, and make the assumption that each of these leaves is destined to form stars according to some IMF, we can estimate an upper limit on the missing number of high-mass star precursor objects missed by the catalog, thus deriving our completeness for high-mass stellar precursors in the CMZ.

The curve in Figure \ref{fig:simobs_percent} gives an estimate for the detected percentage of leaves of a given mass. The fraction of leaves missing at any given mass is equal to $1-P(m)$ where $P(m)$ is the completeness as a function of leaf mass $m$. Because $P(m)$ is defined discretely for a subset of leaf masses, we bin the catalog leaf masses together with the nearest defined value for $P(m)$. This is a poor approximation for small leaf masses, where the completeness percentage changes rapidly as a function of mass. However, these lower mass objects are also unlikely to produce many high-mass stars, so this approximation does not significantly affect our results for the high-mass stellar precursor completeness.


By sampling the Kroupa IMF (from equation 2 of \citealt{Kroupa_variation_2001}), we can convert these suspected missing leaves counts to an upper limit on the number of missed massive star precursors. We calculate the number of missing massive stellar precursors, denoted $N_{\rm miss}$, as
\begin{equation}\label{missing_massive}
    N_{\rm miss} = \sum_m \big(1-P(m)\big)N_{\rm leaves}(m)N_{\geq8\rm M_\odot}(\epsilon_{\rm SF} m),
\end{equation}
where $P(m)$ is the completeness percentage for catalog leaves with mass $m$, $N_{\rm leaves}(m)$ is the number of leaves binned to the nearest mass $m$ with a defined value of $P(m)$, and $N_{\geq8\rm M_\odot}(\epsilon_{\rm SF} m)$ is the number of massive stars in a cluster with a final star mass defined by assumed star formation efficiency $\epsilon_{\rm SF}$ and leaf mass $m$. To determine $N_{\geq8\rm M_\odot}(\epsilon_{\rm SF} m)$, we draw samples from the Kroupa IMF distribution until we reach the target cluster mass, and then either keep or discard the final draw to produce a total mass closest to the target cluster mass (this is identical to the "STOP\_NEAREST" method in Appendix A of \citealt{krumholz_slug_2015}). We then average the number of massive ($\geq 8$ M$_\odot$) stars for 1000 iterations of each leaf mass bin. This is performed for mass bins corresponding to each of the masses used in the simulated observation procedure (i.e. the masses in Figure \ref{fig:simobs_percent}), and a star formation efficiencies ranging between 0.1-0.75, as in Section \ref{sec:sf_limit}.

We choose the same range of star formation efficiencies as section \ref{sec:sf_limit}. Multiplying this average by the estimated number of missing leaves at each mass bin, we arrive at an estimated number of missing massive stellar precursors, which we find to be $\lessapprox1$\% (if we assume a star formation efficiency of 0.75) or $\approx$0\% (if we assume a star formation efficiency of 0.1). In summary, we estimate that \textit{CMZoom} is complete to more than 99\% of sites capable of forming massive stars embedded in high column density gas within the innermost \textasciitilde 500pc of the Galaxy.

\section{Summary}
\label{sec:summary}
In this work, we produce the most complete catalogs to date of compact sources in the cold dust continuum across the Milky Way's innermost 500pc. These compact sources are identified in the 1.3mm \textit{CMZoom} survey as the leaves produced by a dendrogram algorithm, pruned according to local noise conditions. We present two catalogs:
\begin{itemize}
    \item a ``robust" catalog designed for high accuracy, with sources required to have a peak 1.3mm dust continuum flux  6$\sigma$ above the local RMS noise estimate. The robust catalog is $>$95\% complete for compact sources with mass $\geq$ 80 M$_\odot$.
    \item a ``high-completeness" catalog, designed to include more of the lower flux compact structures in the \textit{CMZoom} field, which demands a larger tolerance for false positive detections. The high-completeness catalog requires a peak 1.3mm dust continuum flux 4$\sigma$ above the same local RMS noise estimate, and is $>$95\% complete for compact sources with mass $\geq$ 50 M$_\odot$. 
\end{itemize}
The above mass completeness estimates assume a dust temperature that we take to be a typical value for {\it Herschel} derived dust temperatures, 20K. Since temperatures on smaller scales could be greater, these 95\% mass completeness estimates should be understood as upper limits on our true completeness masses. By treating each of these leaves as an eventual site of star formation (an intentionally generous assumption), along with an IMF for the resulting stars, we place an upper limit on the number of possible high-mass star precursors missing from the robust catalog, which we find to be $\lessapprox1$\% . 


For objects included in both the robust and complete catalogs we find a bimodal distribution in a number of physical properties, separating the population of sources associated with Sgr B2 from the sources elsewhere in the CMZ. This difference is also resilient to changes in the cataloging procedure (e.g.\ Figure \ref{fig:mr_full_parameterstudy}). This separation is likely caused by a combination of several factors:  increased dust temperatures on scales smaller than the {\it Herschel} measurements used in this work, overlapping envelopes and line-of-sight degeneracy of Sgr B2 clouds, and non-dust continuum contamination from extended and compact HII regions distributed throughout the cloud complex. 

We calculate a limit on the maximum possible star formation potential of the Milky Way's CMZ, assuming that stars are not formed in this environment in isolation or in gas with a column density lower than 10$^{23}$ N(H$_2$) cm$^{-2}$. Accounting for a range of possible star formation efficiencies for our cataloged sources, we find a maximum star formation potential of 0.08 - 2.2 M$_\odot$ yr$^{-1}$. This SFR is dominated by the Sgr B2 complex, where more reliable SFR estimates are available (e.g.\ \citealt{schmiedeke_physical_2016,ginsburg_distributed_2018}), and if we exclude Sgr B2, we find the maximum SFR potential for the rest of the CMZ is 0.04-0.47M$_\odot$ yr$^{-1}$. This SFR potential ranges from near current estimates for active star formation in the CMZ to much higher values rivaling the entire Milky Way's SFR, if a high star formation efficiency is expected for these compact structures. 

Many of these sources are known to be sites of active star formation, and exploring the relationship between star formation tracer activity and the line emission properties for these catalogs will be explored in upcoming work (Hatchfield et al in prep.). Future studies detailing the nature of these sources, their star formation activity, chemistry, outflow properties, gas temperatures and kinematics will serve as an important stepping stone toward understanding the process of star formation, from the Milky Way GMCs to the great diversity of extragalactic environments.

\acknowledgments
The authors wish to recognize and acknowledge the very significant cultural role and reverence that the summit of Maunakea has always had within the indigenous Hawaiian community. We are most fortunate to have the opportunity to conduct observations from this mountain. We acknowledge and thank the staff of the SMA for their assistance, and we thank Ray Blundell for his ongoing support of the \textit{CMZoom} project. We thank Ken Young (Taco), Charlie Qi, and Mark Gurwell for their help working with the SMA data. We thank Glen Petitpas and Ryan Howie for scheduling these observations and Shelbi Holster and the SMA operators for making them happen. We thank Margaret Simonini and Anna Huang for their logistical assistance with this project and associated travel. We thank Jimmy Casta\~no, Liz Gehret, Mark Graham, Elizabeth Guti\'errez, Dennis Lee, and Irene Vargas-Salazar for their early investigations into these data which helped to guide its progress and improve its quality. This research has made use of NASA's Astrophysics Data System. The National Radio Astronomy Observatory is a facility of the National Science Foundation operated under cooperative agreement by Associated Universities, Inc.This paper makes use of the following ALMA data: ADS/JAO. ALMA\#2013.1.00617.S, ALMA\#2013.1.00269., ALMA\#2017.1.00687.S, and ALMA\#2018.1.00850.S.  ALMA is a partnership of ESO (representing its member states), NSF (USA) and NINS (Japan), together with NRC (Canada), MOST and ASIAA (Taiwan), and KASI (Republic of Korea), in cooperation with the Republic of Chile. The Joint ALMA Observatory is operated by ESO, AUI/NRAO and NAOJ.

HPH thanks the anonymous referees for their constructive and valuable feedback. HPH gratefully acknowledges support from the National Science Foundation under Award Nos. 1816715. HPH thanks the LSSTC Data Science Fellowship Program, which is funded by LSSTC, NSF Cybertraining Grant \#1829740, the Brinson Foundation, and the Moore Foundation; his participation in the program has benefited this work. HPH also acknowledges many helpful conversations and valuable advice from Imad Pasha and Matthew Phelps. CB gratefully acknowledges support from the National Science Foundation under Award Nos. 1602583 and 1816715. ATB would like to acknowledge funding from the European Research Council (ERC) under the European Union's Horizon 2020 research and innovation programme (grant agreement No.726384/Empire). A.G. acknowledges support from the NSF under grant 2008101. JMDK gratefully acknowledges funding from the Deutsche Forschungsgemeinschaft (DFG, German Research Foundation) through an Emmy Noether Research Group (grant number KR4801/1-1) and the DFG Sachbeihilfe (grant number KR4801/2-1), as well as from the European Research Council (ERC) under the European Union's Horizon 2020 research and innovation programme via the ERC Starting Grant MUSTANG (grant agreement number 714907). XL acknowledges support by JSPS KAKENHI grants No.\ 18K13589 \& 20K14528. EACM gratefully acknowledges support by the National Science Foundation under grant No. AST-1813765. LH acknowledges support by the National Science Foundation of China (11721303, 11991052) and the National Key R\&D Program of China (2016YFA0400702).

\facilities{SMA, CSO} 
\software{This research made use of \texttt{APLpy}, an open-source plotting package for Python \citep{aplpy2012,aplpy2019}, \texttt{astropy}, a community-developed core Python package for astronomy \citep{astropy:2013,astropy:2018}, \texttt{glue-viz} \citep{robitaille_glueviz_2017, beaumont_glue}, \texttt{CASA} \citep{mcmullin_casa_2007}, \texttt{MIR}\footnote{\href{https://www.cfa.harvard.edu/~cqi/mircook.html}{https://www.cfa.harvard.edu/~cqi/mircook.html}} , \texttt{NumPy} \citep{van2011numpy} and \texttt{Miriad} \citep{miriad95}.}
\bibliographystyle{yahapj}
\bibliography{cmzoom_lib_8_2020}

\begin{thebibliography}{}
\providecommand\natexlab[1]{#1}
\providecommand\JournalTitle[1]{#1}

\bibitem[{{Astropy Collaboration} {et~al.}(2013){Astropy Collaboration},
  {Robitaille}, {Tollerud}, {Greenfield}, {Droettboom}, {Bray}, {Aldcroft},
  {Davis}, {Ginsburg}, {Price-Whelan}, {Kerzendorf}, {Conley}, {Crighton},
  {Barbary}, {Muna}, {Ferguson}, {Grollier}, {Parikh}, {Nair}, {Unther},
  {Deil}, {Woillez}, {Conseil}, {Kramer}, {Turner}, {Singer}, {Fox}, {Weaver},
  {Zabalza}, {Edwards}, {Azalee Bostroem}, {Burke}, {Casey}, {Crawford},
  {Dencheva}, {Ely}, {Jenness}, {Labrie}, {Lim}, {Pierfederici}, {Pontzen},
  {Ptak}, {Refsdal}, {Servillat}, \& {Streicher}}]{astropy:2013}
{Astropy Collaboration}, {Robitaille}, T.~P., {Tollerud}, E.~J., {et~al.} 2013,
  \href{http://dx.doi.org/10.1051/0004-6361/201322068}{\JournalTitle{\aap},
  558, A33}

\bibitem[{{Astropy Collaboration} {et~al.}(2018){Astropy Collaboration},
  {Sip{\H{o}}cz}, {G{\"u}nther}, {Lim}, {Crawford}, {Conseil}, {Shupe},
  {Craig}, {Dencheva}, {Ginsburg}, {VanderPlas}, {Bradley},
  {P{\'e}rez-Su{\'a}rez}, {de Val-Borro}, {Paper Contributors}, {Aldcroft},
  {Cruz}, {Robitaille}, {Tollerud}, {Coordination Committee}, {Ardelean},
  {Babej}, {Bach}, {Bachetti}, {Bakanov}, {Bamford}, {Barentsen}, {Barmby},
  {Baumbach}, {Berry}, {Biscani}, {Boquien}, {Bostroem}, {Bouma}, {Brammer},
  {Bray}, {Breytenbach}, {Buddelmeijer}, {Burke}, {Calderone}, {Cano
  Rodr{\'\i}guez}, {Cara}, {Cardoso}, {Cheedella}, {Copin}, {Corrales},
  {Crichton}, {D{\textquoteright}Avella}, {Deil}, {Depagne}, {Dietrich},
  {Donath}, {Droettboom}, {Earl}, {Erben}, {Fabbro}, {Ferreira}, {Finethy},
  {Fox}, {Garrison}, {Gibbons}, {Goldstein}, {Gommers}, {Greco}, {Greenfield},
  {Groener}, {Grollier}, {Hagen}, {Hirst}, {Homeier}, {Horton}, {Hosseinzadeh},
  {Hu}, {Hunkeler}, {Ivezi{\'c}}, {Jain}, {Jenness}, {Kanarek}, {Kendrew},
  {Kern}, {Kerzendorf}, {Khvalko}, {King}, {Kirkby}, {Kulkarni}, {Kumar},
  {Lee}, {Lenz}, {Littlefair}, {Ma}, {Macleod}, {Mastropietro}, {McCully},
  {Montagnac}, {Morris}, {Mueller}, {Mumford}, {Muna}, {Murphy}, {Nelson},
  {Nguyen}, {Ninan}, {N{\"o}the}, {Ogaz}, {Oh}, {Parejko}, {Parley}, {Pascual},
  {Patil}, {Patil}, {Plunkett}, {Prochaska}, {Rastogi}, {Reddy Janga},
  {Sabater}, {Sakurikar}, {Seifert}, {Sherbert}, {Sherwood-Taylor}, {Shih},
  {Sick}, {Silbiger}, {Singanamalla}, {Singer}, {Sladen}, {Sooley},
  {Sornarajah}, {Streicher}, {Teuben}, {Thomas}, {Tremblay}, {Turner},
  {Terr{\'o}n}, {van Kerkwijk}, {de la Vega}, {Watkins}, {Weaver}, {Whitmore},
  {Woillez}, {Zabalza}, \& {Contributors}}]{astropy:2018}
{Astropy Collaboration}, {Price-Whelan}, A.~M., {Sip{\H{o}}cz}, B.~M.,
  {G{\"u}nther}, H.~M., {et~al.} 2018,
  \href{http://dx.doi.org/10.3847/1538-3881/aabc4f}{\JournalTitle{\aj}, 156,
  123}

\bibitem[{Barnes {et~al.}(2017)Barnes, Longmore, Battersby, Bally, Kruijssen,
  Henshaw, \& Walker}]{barnes_star_2017}
Barnes, A.~T., Longmore, S.~N., Battersby, C., {et~al.} 2017,
  \href{http://dx.doi.org/10.1093/mnras/stx941}{\JournalTitle{MNRAS}, 469,
  2263}

\bibitem[{Barnes {et~al.}(2019)Barnes, Longmore, Avison, Contreras, Ginsburg,
  Henshaw, Rathborne, Walker, Alves, Bally, Battersby, Beltr{\'a}n, Beuther,
  Garay, Gomez, Jackson, Kainulainen, Kruijssen, Lu, Mills, Ott, \&
  Peters}]{barnes_young_2019}
Barnes, A.~T., Longmore, S.~N., Avison, A., {et~al.} 2019,
  \href{http://dx.doi.org/10.1093/mnras/stz796}{\JournalTitle{MNRAS}, 486, 283}

\bibitem[{Bastian {et~al.}(2010)Bastian, Covey, \&
  Meyer}]{bastian_universal_2010}
Bastian, N., Covey, K.~R., \& Meyer, M.~R. 2010,
  \href{http://dx.doi.org/10.1146/annurev-astro-082708-101642}{\JournalTitle{ARA\&A},
  48, 339}

\bibitem[{Battersby {et~al.}(2010)Battersby, Bally, Jackson, Ginsburg, Shirley,
  Schlingman, \& Glenn}]{battersby_infrared_2010}
Battersby, C., Bally, J., Jackson, J.~M., {et~al.} 2010,
  \href{http://dx.doi.org/10.1088/0004-637X/721/1/222}{\JournalTitle{ApJ}, 721,
  222}

\bibitem[{Battersby {et~al.}(2011)Battersby, Bally, Ginsburg, Bernard, Brunt,
  Fuller, Martin, Molinari, Mottram, Peretto, Testi, \&
  Thompson}]{battersby_characterizing_2011}
Battersby, C., Bally, J., Ginsburg, A., {et~al.} 2011,
  \href{http://dx.doi.org/10.1051/0004-6361/201116559}{\JournalTitle{A\&A},
  535, A128}

\bibitem[{Battersby {et~al.}(2020)Battersby, Keto, Walker, Barnes, Callanan,
  Ginsburg, Hatchfield, Henshaw, Kauffmann, Kruijssen, Longmore, Lu, Mills,
  Pillai, Zhang, Bally, Butterfield, Contreras, Ho, Ott, Patel, \&
  Tolls}]{battersby_cmzoom_2020}
Battersby, C., Keto, E., Walker, D., {et~al.} 2020,
  \href{http://dx.doi.org/10.3847/1538-4365/aba18e}{\JournalTitle{ApJS}, 249,
  35}

\bibitem[{{Beaumont} {et~al.}(2015){Beaumont}, {Goodman}, \&
  {Greenfield}}]{beaumont_glue}
{Beaumont}, C., {Goodman}, A., \& {Greenfield}, P. 2015, in Astronomical
  Society of the Pacific Conference Series, Vol. 495, Astronomical Data
  Analysis Software an Systems XXIV (ADASS XXIV), ed. A.~R. {Taylor} \&
  E.~{Rosolowsky}, 101

\bibitem[{Bergin \& Tafalla(2007)}]{Bergin_cold_2007}
Bergin, E.~A., \& Tafalla, M. 2007,
  \href{http://dx.doi.org/10.1146/annurev.astro.45.071206.100404}{\JournalTitle{ARA\&A},
  45, 339}

\bibitem[{Chomiuk \& Povich(2011)}]{chomiuk_Unification_2011}
Chomiuk, L., \& Povich, M.~S. 2011,
  \href{http://dx.doi.org/10.1088/0004-6256/142/6/197}{\JournalTitle{ApJ}, 142,
  197}

\bibitem[{Dahmen {et~al.}(1998)Dahmen, Huttemeister, Wilson, \&
  Mauersberger}]{Dahmen_Molecular_1998}
Dahmen, G., Huttemeister, S., Wilson, T.~L., \& Mauersberger, R. 1998,
  \JournalTitle{A\&A}, 331, 959

\bibitem[{Dale {et~al.}(2019)Dale, Kruijssen, \&
  Longmore}]{dale_dynamical_2019}
Dale, J.~E., Kruijssen, J. M.~D., \& Longmore, S.~N. 2019,
  \href{http://dx.doi.org/10.1093/mnras/stz888}{\JournalTitle{MNRAS}, 486,
  3307}

\bibitem[{De~Pree {et~al.}(1996)De~Pree, Gaume, Goss, \&
  Claussen}]{depree_sagittarius_1996}
De~Pree, C.~G., Gaume, R.~A., Goss, W.~M., \& Claussen, M.~J. 1996,
  \href{http://dx.doi.org/10.1086/177364}{\JournalTitle{ApJ}, 464, 788}

\bibitem[{De~Pree {et~al.}(2015)De~Pree, Peters, Mac~Low, Wilner, Goss,
  {Galv{\'a}n-Madrid}, Keto, Klessen, \& Monsrud}]{depree_evidence_2015}
De~Pree, C.~G., Peters, T., Mac~Low, M.~M., {et~al.} 2015,
  \href{http://dx.doi.org/10.1088/0004-637X/815/2/123}{\JournalTitle{ApJ}, 815,
  123}

\bibitem[{Downes \& Maxwell(1966)}]{downes_radio_1966}
Downes, D., \& Maxwell, A. 1966,
  \href{http://dx.doi.org/10.1086/148943}{\JournalTitle{ApJ}, 146, 653}

\bibitem[{Etxaluze {et~al.}(2013)Etxaluze, Goicoechea, Cernicharo, Polehampton,
  {Noriega-Crespo}, Molinari, Swinyard, Wu, \& Bally}]{etxaluze_herschel_2013}
Etxaluze, M., Goicoechea, J.~R., Cernicharo, J., {et~al.} 2013,
  \href{http://dx.doi.org/10.1051/0004-6361/201321258}{\JournalTitle{A\&A},
  556, A137}

\bibitem[{Federrath {et~al.}(2016)Federrath, Rathborne, Longmore, Kruijssen,
  Bally, Contreras, Crocker, Garay, Jackson, Testi, \&
  Walsh}]{federrath_link_2016}
Federrath, C., Rathborne, J.~M., Longmore, S.~N., {et~al.} 2016,
  \href{http://dx.doi.org/10.3847/0004-637X/832/2/143}{\JournalTitle{ApJ}, 832,
  143}

\bibitem[{Figer {et~al.}(2002)Figer, Najarro, Gilmore, Morris, Kim, Serabyn,
  McLean, Gilbert, Graham, Larkin, Levenson, \& Teplitz}]{Figer_massive_2002}
Figer, D.~F., Najarro, F., Gilmore, D., {et~al.} 2002,
  \href{http://dx.doi.org/10.1086/344154}{\JournalTitle{ApJ}, 581, 258}

\bibitem[{Ginsburg {et~al.}(2016)Ginsburg, Henkel, Ao, Riquelme, Kauffmann,
  Pillai, Mills, {Requena-Torres}, Immer, Testi, Ott, Bally, Battersby,
  Darling, Aalto, Stanke, Kendrew, Kruijssen, Longmore, Dale, Guesten, \&
  Menten}]{ginsburg_dense_2016}
Ginsburg, A., Henkel, C., Ao, Y., {et~al.} 2016,
  \href{http://dx.doi.org/10.1051/0004-6361/201526100}{\JournalTitle{A\&A},
  586, A50}

\bibitem[{Ginsburg {et~al.}(2018)Ginsburg, Bally, Barnes, Bastian, Battersby,
  Beuther, Brogan, Contreras, Corby, Darling, De~Pree, {Galv{\'a}n-Madrid},
  Garay, Henshaw, Hunter, Kruijssen, Longmore, Lu, Meng, Mills, Ott, Pineda,
  {S{\'a}nchez-Monge}, Schilke, Schmiedeke, Walker, \&
  Wilner}]{ginsburg_distributed_2018}
Ginsburg, A., Bally, J., Barnes, A., {et~al.} 2018,
  \href{http://dx.doi.org/10.3847/1538-4357/aaa6d4}{\JournalTitle{ApJ}, 853,
  171}

\bibitem[{Goldsmith {et~al.}(1990)Goldsmith, Lis, Hills, \&
  Lasenby}]{Goldsmith_high_1990}
Goldsmith, P.~F., Lis, D.~C., Hills, R., \& Lasenby, J. 1990,
  \href{http://dx.doi.org/10.1086/168372}{\JournalTitle{ApJ}, 350, 186}

\bibitem[{Gordon {et~al.}(1993)Gordon, Berkermann, Mezger, Zylka, Haslam,
  Kreysa, Sievers, \& Lemke}]{Gordon_anatomy_1993}
Gordon, M.~A., Berkermann, U., Mezger, P.~G., {et~al.} 1993,
  \JournalTitle{A\&A}, 280, 208

\bibitem[{Goto {et~al.}(2013)Goto, Indriolo, Geballe, \& Usuda}]{goto_h3_2013}
Goto, M., Indriolo, N., Geballe, T.~R., \& Usuda, T. 2013,
  \href{http://dx.doi.org/10.1021/jp400017s}{\JournalTitle{JPCA}, 117, 9919}

\bibitem[{Guesten \& Downes(1982)}]{guesten_new_1982}
Guesten, R., \& Downes, D. 1982, \JournalTitle{A\&A}, 117, 343

\bibitem[{Habibi {et~al.}(2013)Habibi, Stolte, Brandner, Hu{\ss}mann, \&
  Motohara}]{habibi_arches_2013}
Habibi, M., Stolte, A., Brandner, W., Hu{\ss}mann, B., \& Motohara, K. 2013,
  \href{http://dx.doi.org/10.1051/0004-6361/201220556}{\JournalTitle{A\&A},
  556, A26}

\bibitem[{Harada {et~al.}(2015)Harada, Riquelme, Viti, {Jim{\'e}nez-Serra},
  {Requena-Torres}, Menten, Mart{\'i}n, Aladro, {Martin-Pintado}, \&
  Hochg{\"u}rtel}]{harada_chemical_2015}
Harada, N., Riquelme, D., Viti, S., {et~al.} 2015,
  \href{http://dx.doi.org/10.1051/0004-6361/201526994}{\JournalTitle{A\&A},
  584, A102}

\bibitem[{Henshaw {et~al.}(2016{\natexlab{a}})Henshaw, Caselli, Fontani,
  {Jimenez-Serra}, Tan, Longmore, Pineda, Parker, \&
  Barnes}]{henshaw_investigating_2016}
Henshaw, J.~D., Caselli, P., Fontani, F., {et~al.} 2016{\natexlab{a}},
  \href{http://dx.doi.org/10.1093/mnras/stw1794}{\JournalTitle{MNRAS}, 463,
  146}

\bibitem[{Henshaw {et~al.}(2016{\natexlab{b}})Henshaw, Longmore, Kruijssen,
  Davies, Bally, Barnes, Battersby, Burton, Cunningham, Dale, Ginsburg, Immer,
  Jones, Kendrew, Mills, Molinari, Moore, Ott, Pillai, Rathborne, Schilke,
  Schmiedeke, Testi, Walker, Walsh, \& Zhang}]{henshaw_molecular_2016}
Henshaw, J.~D., Longmore, S.~N., Kruijssen, J. M.~D., {et~al.}
  2016{\natexlab{b}},
  \href{http://dx.doi.org/10.1093/mnras/stw121}{\JournalTitle{MNRAS}, 457,
  2675}

\bibitem[{Henshaw {et~al.}(2017)Henshaw, {Jim{\'e}nez-Serra}, Longmore,
  Caselli, Pineda, Avison, Barnes, Tan, \& Fontani}]{Henshaw_unveiling_2017}
Henshaw, J.~D., {Jim{\'e}nez-Serra}, I., Longmore, S.~N., {et~al.} 2017,
  \href{http://dx.doi.org/10.1093/mnrasl/slw154}{\JournalTitle{MNRAS}, 464,
  L31}

\bibitem[{Henshaw {et~al.}(2019)Henshaw, Ginsburg, Haworth, Longmore,
  Kruijssen, Mills, Sokolov, Walker, Barnes, Contreras, Bally, Battersby,
  Beuther, Butterfield, Dale, Henning, Jackson, Kauffmann, Pillai, Ragan,
  Riener, \& Zhang}]{henshaw_brick_2019}
Henshaw, J.~D., Ginsburg, A., Haworth, T.~J., {et~al.} 2019,
  \href{http://dx.doi.org/10.1093/mnras/stz471}{\JournalTitle{MNRAS}, 485,
  2457}

\bibitem[{Hosek {et~al.}(2019)Hosek, Lu, Anderson, Najarro, Ghez, Morris,
  Clarkson, \& Albers}]{hosek_unusual_2019}
Hosek, Jr., M.~W., Lu, J.~R., Anderson, J., {et~al.} 2019,
  \href{http://dx.doi.org/10.3847/1538-4357/aaef90}{\JournalTitle{ApJ}, 870,
  44}

\bibitem[{Immer {et~al.}(2012)Immer, Schuller, Omont, \&
  Menten}]{immer_recent_2012}
Immer, K., Schuller, F., Omont, A., \& Menten, K.~M. 2012,
  \href{http://dx.doi.org/10.1051/0004-6361/201117857}{\JournalTitle{A\&A},
  537, A121}

\bibitem[{Jeffreson {et~al.}(2018)Jeffreson, Kruijssen, Krumholz, \&
  Longmore}]{jeffreson_physical_2018}
Jeffreson, S. M.~R., Kruijssen, J. M.~D., Krumholz, M.~R., \& Longmore, S.~N.
  2018, \href{http://dx.doi.org/10.1093/mnras/sty1154}{\JournalTitle{MNRAS},
  478, 3380}

\bibitem[{Kauffmann {et~al.}(2008)Kauffmann, Bertoldi, Bourke, Evans, \&
  Lee}]{kauffmann_mambo_2008}
Kauffmann, J., Bertoldi, F., Bourke, T.~L., Evans, II, N.~J., \& Lee, C.~W.
  2008,
  \href{http://dx.doi.org/10.1051/0004-6361:200809481}{\JournalTitle{A\&A},
  487, 993}

\bibitem[{Kauffmann \& Pillai(2010)}]{Kauffmann_how_2010}
Kauffmann, J., \& Pillai, T. 2010,
  \href{http://dx.doi.org/10.1088/2041-8205/723/1/L7}{\JournalTitle{ApJL}, 723,
  L7}

\bibitem[{Kauffmann {et~al.}(2013{\natexlab{a}})Kauffmann, Pillai, \&
  Goldsmith}]{kauffmann_low_2013}
Kauffmann, J., Pillai, T., \& Goldsmith, P.~F. 2013{\natexlab{a}},
  \href{http://dx.doi.org/10.1088/0004-637X/779/2/185}{\JournalTitle{ApJ}, 779,
  185}

\bibitem[{Kauffmann {et~al.}(2013{\natexlab{b}})Kauffmann, Pillai, \&
  Zhang}]{kauffmann_galactic_2013}
Kauffmann, J., Pillai, T., \& Zhang, Q. 2013{\natexlab{b}},
  \href{http://dx.doi.org/10.1088/2041-8205/765/2/L35}{\JournalTitle{ApJL},
  765, L35}

\bibitem[{Kauffmann {et~al.}(2017{\natexlab{a}})Kauffmann, Pillai, Zhang,
  Menten, Goldsmith, Lu, \& Guzm{\'a}n}]{kauffmann_galactic_2017}
Kauffmann, J., Pillai, T., Zhang, Q., {et~al.} 2017{\natexlab{a}},
  \href{http://dx.doi.org/10.1051/0004-6361/201628088}{\JournalTitle{A\&A},
  603, A89}

\bibitem[{Kauffmann {et~al.}(2017{\natexlab{b}})Kauffmann, Pillai, Zhang,
  Menten, Goldsmith, Lu, Guzm{\'a}n, \& Schmiedeke}]{kauffmann_galactic_2017a}
---. 2017{\natexlab{b}},
  \href{http://dx.doi.org/10.1051/0004-6361/201628089}{\JournalTitle{A\&A},
  603, A90}

\bibitem[{Kennicutt(1997)}]{kennicutt_global_1997}
Kennicutt, J. \href{http://dx.doi.org/10.1086/305588}{1997}

\bibitem[{Kennicutt~Jr \& Evans~II(2012)}]{kennicuttjr_star_2012}
Kennicutt~Jr, R.~C., \& Evans~II, N.~J. 2012,
  \href{http://dx.doi.org/10.1146/annurev-astro-081811-125610}{\JournalTitle{ARA\&A},
  50, 531}

\bibitem[{Koepferl {et~al.}(2015)Koepferl, Robitaille, Morales, \&
  Johnston}]{koepferl_mainsequence_2015}
Koepferl, C.~M., Robitaille, T.~P., Morales, E. F.~E., \& Johnston, K.~G. 2015,
  \href{http://dx.doi.org/10.1088/0004-637X/799/1/53}{\JournalTitle{ApJ}, 799,
  53}

\bibitem[{Kroupa(2001)}]{Kroupa_variation_2001}
Kroupa, P. 2001,
  \href{http://dx.doi.org/10.1046/j.1365-8711.2001.04022.x}{\JournalTitle{MNRAS},
  322, 231}

\bibitem[{Kruijssen {et~al.}(2015)Kruijssen, Dale, \&
  Longmore}]{kruijssen_dynamical_2015}
Kruijssen, J. M.~D., Dale, J.~E., \& Longmore, S.~N. 2015,
  \href{http://dx.doi.org/10.1093/mnras/stu2526}{\JournalTitle{MNRAS}, 447,
  1059}

\bibitem[{Kruijssen \& Longmore(2013)}]{kruijssen_comparing_2013}
Kruijssen, J. M.~D., \& Longmore, S.~N. 2013,
  \href{http://dx.doi.org/10.1093/mnras/stt1634}{\JournalTitle{MNRAS}, 435,
  2598}

\bibitem[{Kruijssen {et~al.}(2014)Kruijssen, Longmore, Elmegreen, Murray,
  Bally, Testi, \& Kennicutt}]{kruijssen_what_2014}
Kruijssen, J. M.~D., Longmore, S.~N., Elmegreen, B.~G., {et~al.} 2014,
  \href{http://dx.doi.org/10.1093/mnras/stu494}{\JournalTitle{MNRAS}, 440,
  3370}

\bibitem[{Kruijssen {et~al.}(2019)Kruijssen, Dale, Longmore, Walker, Henshaw,
  Jeffreson, Petkova, Ginsburg, Barnes, Battersby, Immer, Jackson, Keto,
  Krieger, Mills, {S{\'a}nchez-Monge}, Schmiedeke, Suri, \&
  Zhang}]{kruijssen_dynamical_2019}
Kruijssen, J. M.~D., Dale, J.~E., Longmore, S.~N., {et~al.} 2019,
  \href{http://dx.doi.org/10.1093/mnras/stz381}{\JournalTitle{MNRAS}, 484,
  5734}

\bibitem[{Krumholz {et~al.}(2015)Krumholz, Fumagalli, {da Silva}, Rendahl, \&
  Parra}]{krumholz_slug_2015}
Krumholz, M.~R., Fumagalli, M., {da Silva}, R.~L., Rendahl, T., \& Parra, J.
  2015, \href{http://dx.doi.org/10.1093/mnras/stv1374}{\JournalTitle{MNRAS},
  452, 1447}

\bibitem[{Krumholz \& Kruijssen(2015)}]{krumholz_dynamical_2015}
Krumholz, M.~R., \& Kruijssen, J. M.~D. 2015,
  \href{http://dx.doi.org/10.1093/mnras/stv1670}{\JournalTitle{MNRAS}, 453,
  739}

\bibitem[{Krumholz. {et~al.}(2017)Krumholz., Kruijssen, \&
  Crocker}]{krumholz_dynamical_2017}
Krumholz., M.~R., Kruijssen, J. M.~D., \& Crocker, R.~M. 2017,
  \href{http://dx.doi.org/10.1093/mnras/stw3195}{\JournalTitle{MNRAS}, 466,
  1213}

\bibitem[{Lada {et~al.}(2012)Lada, Forbrich, Lombardi, \&
  Alves}]{lada_star_2012}
Lada, C.~J., Forbrich, J., Lombardi, M., \& Alves, J.~F. 2012,
  \href{http://dx.doi.org/10.1088/0004-637X/745/2/190}{\JournalTitle{ApJ}, 745,
  190}

\bibitem[{Lada \& Lada(2003)}]{lada_embedded_2003}
Lada, C.~J., \& Lada, E.~A. 2003,
  \href{http://dx.doi.org/10.1146/annurev.astro.41.011802.094844}{\JournalTitle{ARA\&A},
  41, 57}

\bibitem[{Lada {et~al.}(2010)Lada, Lombardi, \& Alves}]{lada_star_2010}
Lada, C.~J., Lombardi, M., \& Alves, J.~F. 2010,
  \href{http://dx.doi.org/10.1088/0004-637X/724/1/687}{\JournalTitle{ApJ}, 724,
  687}

\bibitem[{Le~Petit {et~al.}(2016)Le~Petit, Ruaud, Bron, Godard, Roueff,
  Languignon, \& Le~Bourlot}]{Lepetit_physical_2016}
Le~Petit, F., Ruaud, M., Bron, E., {et~al.} 2016,
  \href{http://dx.doi.org/10.1051/0004-6361/201526658}{\JournalTitle{A\&A},
  585, A105}

\bibitem[{Lis {et~al.}(1994)Lis, Menten, Serabyn, \& Zylka}]{lis_star_1994}
Lis, D.~C., Menten, K.~M., Serabyn, E., \& Zylka, R. 1994,
  \href{http://dx.doi.org/10.1086/187230}{\JournalTitle{ApJ}, 423, L39}

\bibitem[{Lis {et~al.}(2001)Lis, Serabyn, Zylka, \& Li}]{lis_quiescent_2001}
Lis, D.~C., Serabyn, E., Zylka, R., \& Li, Y. 2001,
  \href{http://dx.doi.org/10.1086/319815}{\JournalTitle{ApJ}, 550, 761}

\bibitem[{Liu {et~al.}(2018)Liu, Dunham, Pascucci, Bourke, Hirano, Longmore,
  Andrews, {Carrasco-Gonz{\'a}lez}, Forbrich, {Galv{\'a}n-Madrid}, Girart,
  Green, Ju{\'a}rez, K{\'o}sp{\'a}l, Manara, Palau, Takami, Testi, \&
  Vorobyov}]{liu_mm_2018}
Liu, H.~B., Dunham, M.~M., Pascucci, I., {et~al.} 2018,
  \href{http://dx.doi.org/10.1051/0004-6361/201731951}{\JournalTitle{A\&A},
  612, A54}

\bibitem[{Lo \& Claussen(1983)}]{lo_Highresolution_1983}
Lo, K.~Y., \& Claussen, M.~J. 1983,
  \href{http://dx.doi.org/10.1038/306647a0}{\JournalTitle{Natur}, 306, 647}

\bibitem[{Longmore {et~al.}(2012)Longmore, Rathborne, Bastian, Alves, Ascenso,
  Bally, Testi, Longmore, Battersby, Bressert, Purcell, Walsh, Jackson, Foster,
  Molinari, Meingast, Amorim, Lima, Marques, Moitinho, Pinhao, Rebordao, \&
  Santos}]{longmore_G0_2012}
Longmore, S.~N., Rathborne, J., Bastian, N., {et~al.} 2012,
  \href{http://dx.doi.org/10.1088/0004-637X/746/2/117}{\JournalTitle{ApJ}, 746,
  117}

\bibitem[{Longmore {et~al.}(2013{\natexlab{a}})Longmore, Kruijssen, Bally, Ott,
  Testi, Rathborne, Bastian, Bressert, Molinari, Battersby, \&
  Walsh}]{longmore_candidate_2013}
Longmore, S.~N., Kruijssen, J. M.~D., Bally, J., {et~al.} 2013{\natexlab{a}},
  \href{http://dx.doi.org/10.1093/mnrasl/slt048}{\JournalTitle{MNRASL}, 433,
  L15}

\bibitem[{Longmore {et~al.}(2013{\natexlab{b}})Longmore, Bally, Testi, Purcell,
  Walsh, Bressert, Pestalozzi, Molinari, Ott, Cortese, Battersby, Murray, Lee,
  Kruijssen, Schisano, \& Elia}]{longmore_variations_2013}
Longmore, S.~N., Bally, J., Testi, L., {et~al.} 2013{\natexlab{b}},
  \href{http://dx.doi.org/10.1093/mnras/sts376}{\JournalTitle{MNRAS}, 429, 987}

\bibitem[{Lu {et~al.}(2013)Lu, Do, Ghez, Morris, Yelda, \&
  Matthews}]{Lu_stellar_2013}
Lu, J.~R., Do, T., Ghez, A.~M., {et~al.} 2013,
  \href{http://dx.doi.org/10.1088/0004-637X/764/2/155}{\JournalTitle{ApJ}, 764,
  155}

\bibitem[{Lu {et~al.}(2015)Lu, Zhang, Kauffmann, Pillai, Longmore, Kruijssen,
  Battersby, \& Gu}]{lu_deeply_2015}
Lu, X., Zhang, Q., Kauffmann, J., {et~al.} 2015,
  \href{http://dx.doi.org/10.1088/2041-8205/814/2/L18}{\JournalTitle{ApJL},
  814, L18}

\bibitem[{Lu {et~al.}(2019{\natexlab{a}})Lu, Mills, Ginsburg, Walker, Barnes,
  Butterfield, Henshaw, Battersby, Kruijssen, Longmore, Zhang, Bally,
  Kauffmann, Ott, Rickert, \& Wang}]{lu_census_2019}
Lu, X., Mills, E. A.~C., Ginsburg, A., {et~al.} 2019{\natexlab{a}},
  \href{http://dx.doi.org/10.3847/1538-4365/ab4258}{\JournalTitle{ApJS}, 244,
  35}

\bibitem[{Lu {et~al.}(2019{\natexlab{b}})Lu, Zhang, Kauffmann, Pillai,
  Ginsburg, Mills, Kruijssen, Longmore, Battersby, Liu, \& Gu}]{lu_star_2019}
Lu, X., Zhang, Q., Kauffmann, J., {et~al.} 2019{\natexlab{b}},
  \href{http://dx.doi.org/10.3847/1538-4357/ab017d}{\JournalTitle{ApJ}, 872,
  171}

\bibitem[{Marsh {et~al.}(2017)Marsh, Whitworth, Lomax, Ragan, Becciani,
  Cambr{\'e}sy, Di~Giorgio, Eden, Elia, Kacsuk, Molinari, Palmeirim, Pezzuto,
  Schneider, Sciacca, \& Vitello}]{Marsh_multitemperature_2017}
Marsh, K.~A., Whitworth, A.~P., Lomax, O., {et~al.} 2017,
  \href{http://dx.doi.org/10.1093/mnras/stx1723}{\JournalTitle{MNRAS}, 471,
  2730}

\bibitem[{McKee \& Ostriker(2007)}]{mckee_theory_2007}
McKee, C.~F., \& Ostriker, E.~C. 2007,
  \href{http://dx.doi.org/10.1146/annurev.astro.45.051806.110602}{\JournalTitle{ARA\&A},
  45, 565}

\bibitem[{McMullin {et~al.}(2007)McMullin, Waters, Schiebel, Young, \&
  Golap}]{mcmullin_casa_2007}
McMullin, J.~P., Waters, B., Schiebel, D., Young, W., \& Golap, K. 2007, in
  Astronomical {{Data Analysis Software}} and {{Systems XVI}}, Vol. 376, 127

\bibitem[{Mehringer {et~al.}(1995)Mehringer, {de Pree}, Gaume, Goss, \&
  Claussen}]{Mehringer_very_1995}
Mehringer, D.~M., {de Pree}, C.~G., Gaume, R.~A., Goss, W.~M., \& Claussen,
  M.~J. 1995, \href{http://dx.doi.org/10.1086/187808}{\JournalTitle{ApJL}, 442,
  L29}

\bibitem[{Mills \& Battersby(2017)}]{mills_origins_2017}
Mills, E. A.~C., \& Battersby, C. 2017,
  \href{http://dx.doi.org/10.3847/1538-4357/835/1/76}{\JournalTitle{ApJ}, 835,
  76}

\bibitem[{Mills {et~al.}(2015)Mills, Butterfield, Ludovici, Lang, Ott, Morris,
  \& Schmitz}]{mills_abundant_2015}
Mills, E. A.~C., Butterfield, N., Ludovici, D.~A., {et~al.} 2015,
  \href{http://dx.doi.org/10.1088/0004-637X/805/1/72}{\JournalTitle{ApJ}, 805,
  72}

\bibitem[{Mills \& Morris(2013)}]{mills_detection_2013}
Mills, E. A.~C., \& Morris, M.~R. 2013,
  \href{http://dx.doi.org/10.1088/0004-637X/772/2/105}{\JournalTitle{ApJ}, 772,
  105}

\bibitem[{Morris \& Serabyn(1996)}]{morris_galactic_1996}
Morris, M., \& Serabyn, E. 1996,
  \href{http://dx.doi.org/10.1146/annurev.astro.34.1.645}{\JournalTitle{ARA\&A},
  34, 645}

\bibitem[{Offner {et~al.}(2014)Offner, Clark, Hennebelle, Bastian, Bate,
  Hopkins, Moraux, \& Whitworth}]{Offner_origin_2014}
Offner, S. S.~R., Clark, P.~C., Hennebelle, P., {et~al.} 2014,
  \href{http://dx.doi.org/10.2458/azu_uapress_9780816531240-ch003}{\JournalTitle{2014prpl
  conf}, 53}

\bibitem[{Oka {et~al.}(2019)Oka, Geballe, Goto, Usuda, {Benjamin}, McCall, \&
  Indriolo}]{Oka_central_2019}
Oka, T., Geballe, T.~R., Goto, M., {et~al.} 2019,
  \href{http://dx.doi.org/10.3847/1538-4357/ab3647}{\JournalTitle{ApJ}, 883,
  54}

\bibitem[{Ossenkopf \& Henning(1994)}]{ossenkopf_dust_1994}
Ossenkopf, V., \& Henning, T. 1994, \JournalTitle{A\&A}, 291, 943

\bibitem[{Pardo {et~al.}(2001)Pardo, Cernicharo, \&
  Serabyn}]{pardo_atmospheric_2001}
Pardo, J.~R., Cernicharo, J., \& Serabyn, E. 2001,
  \href{http://dx.doi.org/10.1109/8.982447}{\JournalTitle{ITAP}, 49, 1683}

\bibitem[{{Pierce-Price} {et~al.}(2000){Pierce-Price}, Richer, Greaves,
  Holland, Jenness, Lasenby, White, Matthews, {Ward-Thompson}, Dent, Zylka,
  Mezger, Hasegawa, Oka, Omont, \& Gilmore}]{pierce-price_deep_2000}
{Pierce-Price}, D., Richer, J.~S., Greaves, J.~S., {et~al.} 2000,
  \href{http://dx.doi.org/10.1086/317884}{\JournalTitle{ApJ}, 545, L121}

\bibitem[{Pillai {et~al.}(2015)Pillai, Kauffmann, Tan, Goldsmith, Carey, \&
  Menten}]{pillai_magnetic_2015}
Pillai, T., Kauffmann, J., Tan, J.~C., {et~al.} 2015,
  \href{http://dx.doi.org/10.1088/0004-637X/799/1/74}{\JournalTitle{ApJ}, 799,
  74}

\bibitem[{Rathborne {et~al.}(2006)Rathborne, Jackson, \&
  Simon}]{rathborne_infrared_2006}
Rathborne, J.~M., Jackson, J.~M., \& Simon, R. 2006,
  \href{http://dx.doi.org/10.1086/500423}{\JournalTitle{ApJ}, 641, 389}

\bibitem[{Rathborne {et~al.}(2014)Rathborne, Longmore, Jackson, Foster,
  Contreras, Garay, Testi, Alves, Bally, Bastian, Kruijssen, \&
  Bressert}]{rathborne_g0_2014}
Rathborne, J.~M., Longmore, S.~N., Jackson, J.~M., {et~al.} 2014,
  \href{http://dx.doi.org/10.1088/0004-637X/786/2/140}{\JournalTitle{ApJ}, 786,
  140}

\bibitem[{Robitaille(2019)}]{aplpy2019}
Robitaille, T. 2019, {APLpy v2.0: The Astronomical Plotting Library in Python}

\bibitem[{Robitaille {et~al.}(2017)Robitaille, Beaumont, Qian, Borkin, \&
  Goodman}]{robitaille_glueviz_2017}
Robitaille, T., Beaumont, C., Qian, P., Borkin, M., \& Goodman, A. 2017,
  {glueviz v0.13.1: multidimensional data exploration (Zenodo)}

\bibitem[{{Robitaille} \& {Bressert}(2012)}]{aplpy2012}
{Robitaille}, T., \& {Bressert}, E. 2012, {APLpy: Astronomical Plotting Library
  in Python}, Astrophysics Source Code Library,
  \href{http://arxiv.org/abs/1208.017}{{\sffamily ascl:1208.017}}

\bibitem[{Rosolowsky {et~al.}(2008)Rosolowsky, Pineda, Kauffmann, \&
  Goodman}]{rosolowsky_structural_2008}
Rosolowsky, E.~W., Pineda, J.~E., Kauffmann, J., \& Goodman, A.~A. 2008,
  \href{http://dx.doi.org/10.1086/587685}{\JournalTitle{ApJ}, 679, 1338}

\bibitem[{{Sault} {et~al.}(1995){Sault}, {Teuben}, \& {Wright}}]{miriad95}
{Sault}, R.~J., {Teuben}, P.~J., \& {Wright}, M.~C.~H. 1995, in Astronomical
  Society of the Pacific Conference Series, Vol.~77, Astronomical Data Analysis
  Software and Systems IV, ed. R.~A. {Shaw}, H.~E. {Payne}, \& J.~J.~E.
  {Hayes}, 433

\bibitem[{Schmiedeke {et~al.}(2016)Schmiedeke, Schilke, M{\"o}ller,
  {S{\'a}nchez-Monge}, Bergin, Comito, Csengeri, Lis, Molinari, Qin, \&
  Rolffs}]{schmiedeke_physical_2016}
Schmiedeke, A., Schilke, P., M{\"o}ller, T., {et~al.} 2016,
  \href{http://dx.doi.org/10.1051/0004-6361/201527311}{\JournalTitle{A\&A},
  588, A143}

\bibitem[{Serabyn {et~al.}(1997)Serabyn, Carlstrom, Lay, Lis, Hunter, Lacy, \&
  Hills}]{serabyn_highfrequency_1997}
Serabyn, E., Carlstrom, J., Lay, O., {et~al.} 1997,
  \href{http://dx.doi.org/10.1086/311010}{\JournalTitle{ApJL}, 490, L77}

\bibitem[{Sormani \& Li(2020)}]{sormani_nuclear_2020}
Sormani, M.~C., \& Li, Z. 2020,
  \href{http://dx.doi.org/10.1093/mnras/staa1139}{\JournalTitle{MNRAS}, 494,
  6030}

\bibitem[{Sormani {et~al.}(2018)Sormani, Sobacchi, Fragkoudi, Ridley, Tre{\ss},
  Glover, \& Klessen}]{sormani_dynamical_2018}
Sormani, M.~C., Sobacchi, E., Fragkoudi, F., {et~al.} 2018,
  \href{http://dx.doi.org/10.1093/mnras/sty2246}{\JournalTitle{MNRAS}, 481, 2}

\bibitem[{Sormani {et~al.}(2020)Sormani, Tress, Glover, Klessen, Battersby,
  Clark, Hatchfield, \& Smith}]{sormani_simulations_2020}
Sormani, M.~C., Tress, R.~G., Glover, S. C.~O., {et~al.} 2020,
  \href{http://dx.doi.org/10.1093/mnras/staa1999}{\JournalTitle{MNRAS}, 497,
  5024}

\bibitem[{The
  GRAVITY~Collaboration(2019)}]{thegravitycollaboration_geometric_2019}
The GRAVITY~Collaboration, A. 2019,
  \href{http://dx.doi.org/10.1051/0004-6361/201935656}{\JournalTitle{A\&A},
  625, L10}

\bibitem[{Tress {et~al.}(2020)Tress, Sormani, Glover, Klessen, Battersby,
  Clark, Hatchfield, \& Smith}]{tress_simulations_2020}
Tress, R.~G., Sormani, M.~C., Glover, S. C.~O., {et~al.} 2020,
  \JournalTitle{arXiv:2004.06724 [astro-ph]},
  \href{http://arxiv.org/abs/2004.06724}{{\sffamily arXiv:2004.06724
  [astro-ph]}}

\bibitem[{Van Der~Walt {et~al.}(2011)Van Der~Walt, Colbert, \&
  Varoquaux}]{van2011numpy}
Van Der~Walt, S., Colbert, S.~C., \& Varoquaux, G. 2011,
  \JournalTitle{Computing in Science \& Engineering}, 13, 22

\bibitem[{Walker {et~al.}(2016)Walker, Longmore, Bastian, Kruijssen, Rathborne,
  {Galv{\'a}n-Madrid}, \& Liu}]{walker_comparing_2016}
Walker, D.~L., Longmore, S.~N., Bastian, N., {et~al.} 2016,
  \href{http://dx.doi.org/10.1093/mnras/stw313}{\JournalTitle{MNRAS}, 457,
  4536}

\bibitem[{Walker {et~al.}(2018)Walker, Longmore, Zhang, Battersby, Keto,
  Kruijssen, Ginsburg, Lu, Henshaw, Kauffmann, Pillai, Mills, Walsh, Bally, Ho,
  Immer, \& Johnston}]{walker_star_2018}
Walker, D.~L., Longmore, S.~N., Zhang, Q., {et~al.} 2018,
  \href{http://dx.doi.org/10.1093/mnras/stx2898}{\JournalTitle{MNRAS}, 474,
  2373}

\bibitem[{{Yusef-Zadeh} \& Wardle(2008)}]{yusef-zadeh_Massive_2008}
{Yusef-Zadeh}, F., \& Wardle, M. 2008, \JournalTitle{Massive Star Formation:
  Observations Confront Theory}, 387, 361

\bibitem[{{Yusef-Zadeh} {et~al.}(2009){Yusef-Zadeh}, Hewitt, Arendt, Whitney,
  Rieke, Wardle, Hinz, Stolovy, Lang, Burton, \&
  Ramirez}]{yusef-zadeh_Star_2009}
{Yusef-Zadeh}, F., Hewitt, J.~W., Arendt, R.~G., {et~al.} 2009,
  \href{http://dx.doi.org/10.1088/0004-637X/702/1/178}{\JournalTitle{ApJ}, 702,
  178}

\end{thebibliography}

\appendix
\section{High-Completeness Catalog figures and Tables}\label{sec:complete_figs}
\begin{figure}
\begin{center}
\includegraphics[trim = 0mm 0mm 0mm 0mm, clip, width = 0.45\textwidth]{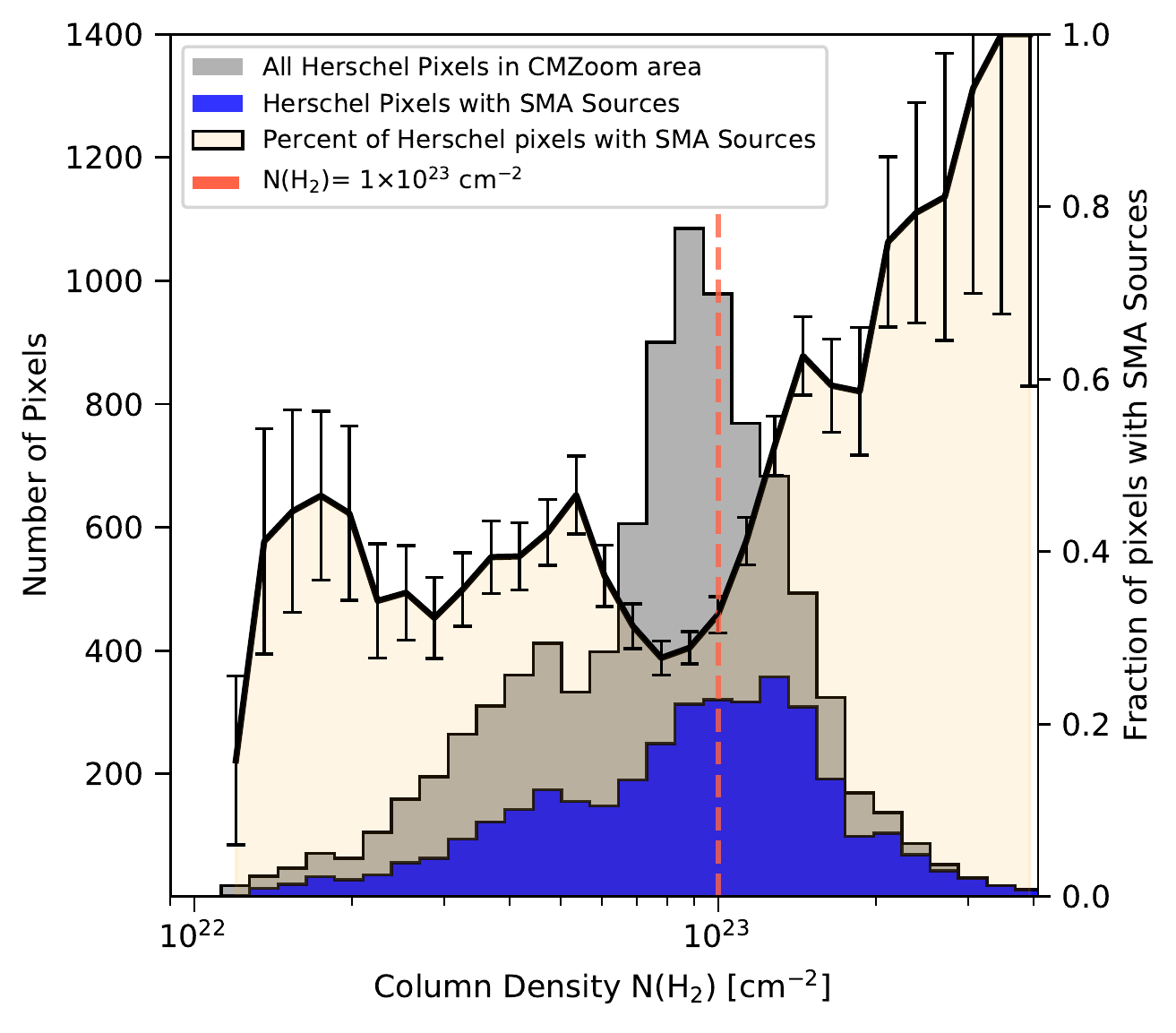}
\end{center}
\caption{A histogram of the {\it Herschel} column density (derived using the procedure from \cite{battersby_characterizing_2011}, from Battersby et al in prep) associated with every pixel in the \textit{CMZoom} survey. The gray histogram represents the column density associated with every pixel covered by the SMA survey, while the blue histogram represents only the pixels within one {\it Herschel} beam (36\arcsec) of a catalog leaf centroid in the high-completeness version of the catalog. The ratio of pixels within one {\it Herschel} beam of a source in the SMA catalog to the total number of pixels in the SMA's map for a given column density is shown as a solid line, which experiences a sharp uptick around a column density of N = 2$\times 10^{23}$ cm$^{-2}$. Pixels in the Sgr B2 region have been excluded from this analysis due to their much higher column densities.}
\label{fig:herschel_hist_complete}
\end{figure}

\begin{figure}
\begin{center}
\includegraphics[trim = 0mm 0mm 0mm 0mm, clip, width = 0.6\textwidth]{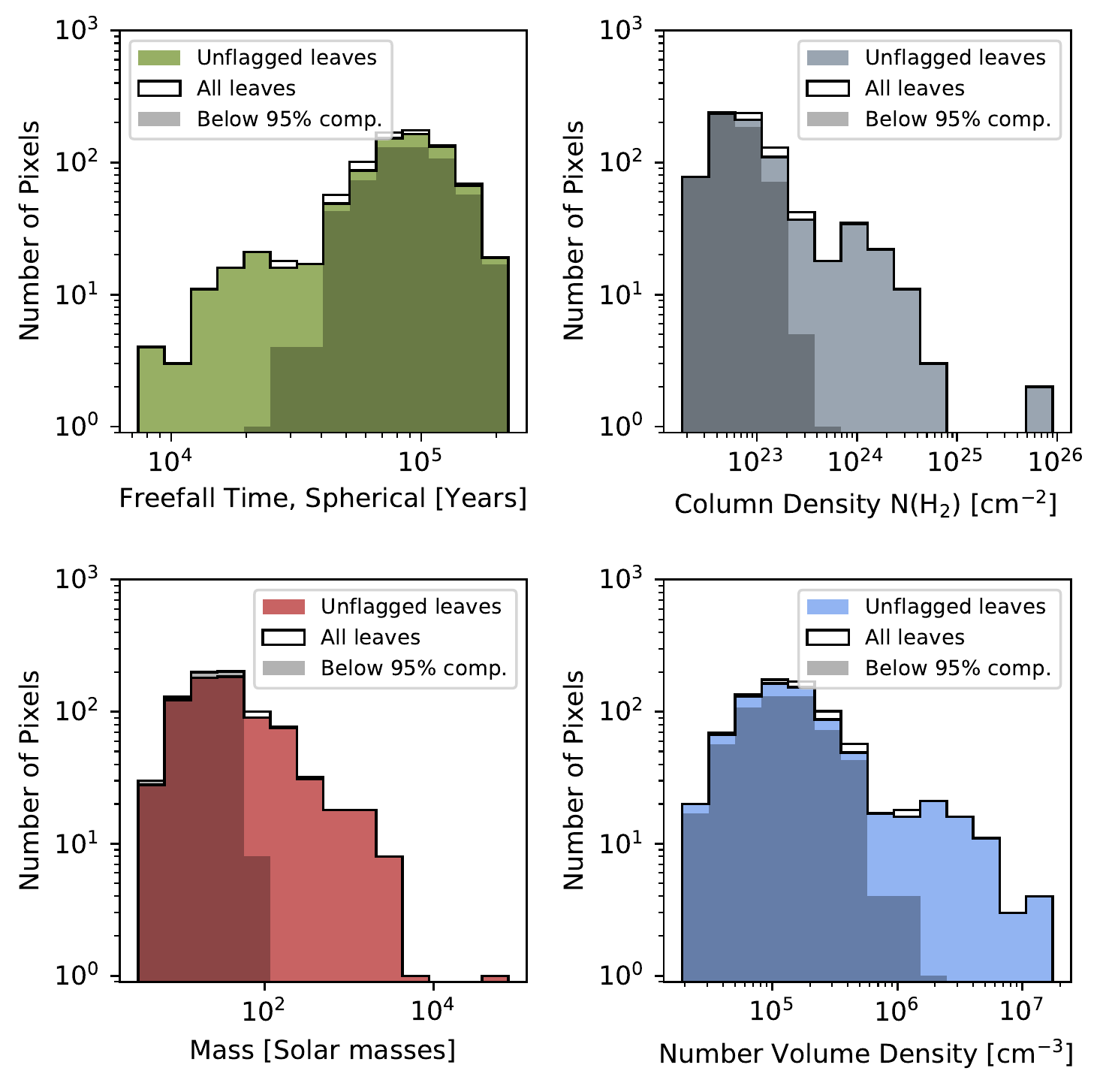}
\end{center}
\caption{Histograms of the physical properties for the leaves contained in the high-completeness version of the catalog. These properties are calculated according to the methods described in section \ref{sec:phys_prop}. The upper left panel shows the distribution of freefall times calculated by assuming a spherical distribution of gas according to the leaf's effective radius. The upper left panel shows the distribution of column densities for each leaf. The distribution of leaf masses is shown in the bottom left panel, and the volume density distribution is shown in the bottom right which is calculated again by assuming a spherical volume distribution according to the leaf's effective radius. The darker shaded regions correspond to the portion of leaves below the catalog's 95\% mass completeness limit of ~60 M$_\odot$. Leaves flagged as suspicious for being within 15 pixels of the map's edge are included as a non-filled histogram, though they largely follow the same trends as the more robust leaves.}
\label{fig:physprop_histo_radius_complete}
\end{figure}

\begin{figure}
\begin{center}
\includegraphics[trim = 0mm 0mm 0mm 0mm, clip, width = 0.45\textwidth]{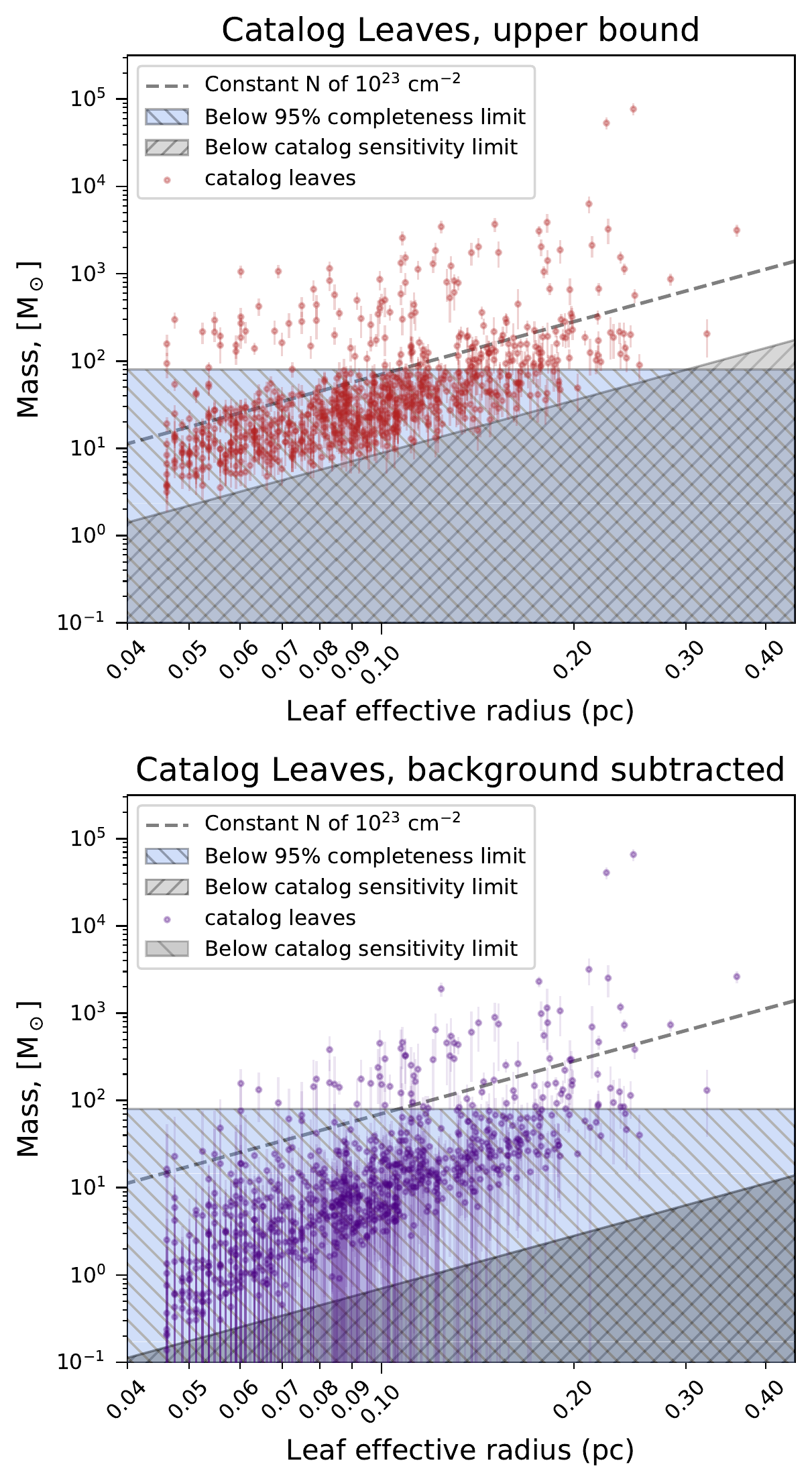}
\end{center}
\caption{Two representations of the mass ($M_\odot$) vs. radius (parsecs) for each leaf in the high-completeness version of the catalog. The upper panel shows the mass calculated using the flux achieved by integrating over the entire leaf structure, acting as an upper bound for the true structure mass. The bottom panel displays the background subtracted leaf masses, calculated by subtracting the minimum value of the leaf and integrated over the entire leaf area. This background subtracted mass is likely an underestimate for the true structure mass. Both panels show a line of constant column density N = $10^{23}$ cm$^{-2}$. The range of masses below 95\% completeness is hatched right and shaded blue, and the region below minimum flux possible for a cataloged leaf is hatched left and shaded gray.}
\label{fig:mr2fig_complete}
\end{figure}

Here we present versions of figures \ref{fig:herschel_hist}, \ref{fig:physprop_histo}, and \ref{fig:mr2fig} as well as tables \ref{tab:base}, \ref{tab:prop}, and \ref{tab:unc} for the high-completeness version of the catalog. Figure \ref{fig:herschel_hist_complete} presents the histogram of Herschel column densities and fraction of pixels with a leaf for the complete catalog. Similar to Figure \ref{fig:herschel_hist}, there is a sudden uptick in the frequency of Herschels pixel containing at least one leaf around N=$10^{23}$ cm$^{-2}$, though the baseline frequency is higher, as would be expected from a more complete catalog with a greater number of leaves. Figure \ref{fig:physprop_histo_radius_complete} shows the histograms of freefall time, column density, mass, and volume density. The distributions are largely similar to Figure \ref{fig:physprop_histo}, with a larger number of leaves flagged as suspicious due to their proximity to the \textit{CMZoom} map edges. Figure \ref{fig:mr2fig_complete} displays the mass-radius distribution of complete catalog leaves, for their upper limit masses and background subtracted masses, following largely the same trends as Figure \ref{fig:mr2fig} and discussed in Section \ref{sec:sgrb2}. Tables \ref{tab:comp_base}, \ref{tab:comp_prop}, and \ref{tab:comp_unc} have also been made available online at \href{https://doi.org/10.7910/DVN/RDE1CH}{https://doi.org/10.7910/DVN/RDE1CH}.

\begin{table*}
\centering
\caption{Leaf Properties, High-Completeness}
\begin{tabular}{lllllllll}
\hline\hline
 Leaf ID & Area & $l$ & $b$ & R$_{\rm eff}$ & Integrated Flux & Peak Cont. Flux & Mean Cont. Flux & RMS \\
 & (as$^2)$ & (deg.) & (deg.) & (pc) & (Jy) & (Jy sr$^{-1}$) & (Jy sr$^{-1}$) & (Jy sr$^{-1}$) \\
\hline
G359.615-0.243a & 259.25 & -0.39 & -0.24 & 0.36 & 2.09e+00 & 1.73e+09 & 3.43e+08 & 9.88e+06 \\
G359.615-0.243b & 34.75 & -0.38 & -0.24 & 0.13 & 7.52e-02 & 1.70e+08 & 9.20e+07 & 1.25e+07 \\
G359.615-0.243c & 45.00 & -0.38 & -0.25 & 0.15 & 6.71e-02 & 1.39e+08 & 6.35e+07 & 1.68e+07 \\
G359.615-0.243d & 43.75 & -0.39 & -0.25 & 0.15 & 3.35e-02 & 7.06e+07 & 3.26e+07 & 1.44e+07 \\
G359.615-0.243e & 8.00 & -0.40 & -0.24 & 0.06 & 2.24e-02 & 1.40e+08 & 1.19e+08 & 3.42e+07 \\
G359.615-0.243f & 6.75 & -0.40 & -0.24 & 0.06 & 1.95e-02 & 1.48e+08 & 1.23e+08 & 3.13e+07 \\
G359.615-0.243g & 14.00 & -0.38 & -0.25 & 0.08 & 1.64e-02 & 6.51e+07 & 4.97e+07 & 1.53e+07 \\
G359.615-0.243h & 15.50 & -0.39 & -0.24 & 0.09 & 1.43e-02 & 5.91e+07 & 3.92e+07 & 1.30e+07 \\
G359.615-0.243i & 4.75 & -0.38 & -0.25 & 0.05 & 6.50e-03 & 6.35e+07 & 5.82e+07 & 1.17e+07 \\
G0.316-0.201a & 115.25 & 0.32 & -0.20 & 0.24 & 9.12e-01 & 1.09e+09 & 3.37e+08 & 9.56e+06 \\
... & ... & ... & ... & ... & ... & ... & ... & ...\\
\hline\hline
\end{tabular}
\label{tab:comp_base}
\end{table*}

\begin{table*}
\caption{Leaf Properties, High-Completeness (continued)}
\centering
\begin{tabular}{llllllll}
\hline\hline
Leaf ID & N$_{\rm Herschel}$ & N$_{\rm SMA}$ & Mass & Mass (bg. sub.) & n & $\rho$ & t$_{\rm ff}$ \\
 & (cm$^{-2})$ & (cm$^{-2})$ & (M$_\odot)$ & (M$_\odot$) & (cm$^{-3})$ & (g cm$^{-3})$ & (yr) \\
\hline
G359.615-0.243a & 1.46e+23 & 1.73e+24& 3.15e+03 & 2.62e+03 & 2.33e+05& 1.09e-15 & 6.37e+04 \\
G359.615-0.243b & 8.59e+22 & 1.73e+23& 1.14e+02 & 4.28e+01 & 1.72e+05& 8.07e-16 & 7.41e+04 \\
G359.615-0.243c & 6.81e+22 & 1.42e+23& 1.03e+02 & 4.16e+01 & 1.05e+05& 4.92e-16 & 9.49e+04 \\
G359.615-0.243d & 8.04e+22 & 6.36e+22& 4.53e+01 & 2.36e+01 & 4.82e+04& 2.26e-16 & 1.40e+05 \\
G359.615-0.243e & 5.33e+22 & 1.36e+23& 3.25e+01 & 4.31e+00 & 4.42e+05& 2.07e-15 & 4.63e+04 \\
G359.615-0.243f & 5.33e+22 & 1.43e+23& 2.83e+01 & 4.76e+00 & 4.98e+05& 2.33e-15 & 4.36e+04 \\
G359.615-0.243g & 6.81e+22 & 6.65e+22& 2.51e+01 & 5.94e+00 & 1.48e+05& 6.91e-16 & 8.01e+04 \\
G359.615-0.243h & 1.31e+23 & 5.57e+22& 2.02e+01 & 7.52e+00 & 1.02e+05& 4.79e-16 & 9.62e+04 \\
G359.615-0.243i & 9.96e+22 & 6.20e+22& 9.52e+00 & 8.45e-01 & 2.83e+05& 1.33e-15 & 5.78e+04 \\
G0.316-0.201a & 5.74e+22 & 9.00e+23& 1.13e+03 & 7.31e+02 & 2.82e+05& 1.32e-15 & 5.79e+04 \\
... & ... & ... & ... & ... & ... & ... & ... \\
\hline\hline
\end{tabular}
\label{tab:comp_prop}
\end{table*}

\begin{table*}
\caption{Leaf Property Uncertainties, High-Completeness)}
\centering
\begin{tabular}{llllll}
\hline\hline
Leaf ID & Mass Unc. & N$_{\rm SMA}$ unc. & n unc. & $\rho$ unc. & t$_{\rm ff}$ unc. \\
 & (M$_\odot)$ & (cm$^{-2})$ & (cm$^{-3})$ & (g cm$^{-3})$ (yr) \\
\hline
G359.615-0.243a & 4.92e+02 & 3.48e+24& 7.85e+04 & 3.68e-16 & 2.15e+04 \\
G359.615-0.243b & 2.34e+01 & 3.47e+23& 6.20e+04 & 2.90e-16 & 2.67e+04 \\
G359.615-0.243c & 3.15e+01 & 2.85e+23& 4.47e+04 & 2.09e-16 & 4.04e+04 \\
G359.615-0.243d & 2.11e+01 & 1.28e+23& 3.04e+04 & 1.42e-16 & 8.82e+04 \\
G359.615-0.243e & 1.06e+01 & 2.74e+23& 2.06e+05 & 9.63e-16 & 2.15e+04 \\
G359.615-0.243f & 8.40e+00 & 2.88e+23& 2.20e+05 & 1.03e-15 & 1.93e+04 \\
G359.615-0.243g & 8.62e+00 & 1.34e+23& 6.69e+04 & 3.14e-16 & 3.63e+04 \\
G359.615-0.243h & 7.37e+00 & 1.12e+23& 5.21e+04 & 2.44e-16 & 4.90e+04 \\
G359.615-0.243i & 2.40e+00 & 1.25e+23& 1.15e+05 & 5.39e-16 & 2.35e+04 \\
G0.316-0.201a & 1.72e+02 & 1.80e+24& 1.15e+05 & 5.41e-16 & 2.37e+04 \\
... & ... & ... & ... & ... & ... \\
\hline\hline
\end{tabular}
\label{tab:comp_unc}
\end{table*}

\section{Catalog Parameter Studies}\label{sec:param_study}

The output tree of a dendrogram is highly dependent on the choice of parameters given to the dendrogram algorithm. Given that our cataloging algorithm is a collection of dendrogram leaves pruned according to local noise, it is necessary to understand how our catalog changes with the choices of both the parameters of the dendrogram, and also our method of estimating local noise. In this section, we perform simple parameter studies to explore the variability of the final catalog with small changes in the initial dendrogram parameters as well as the smoothing kernel for generating our RMS noise maps.

\begin{figure*}
\begin{center}
\includegraphics[trim = 0mm 0mm 0mm 0mm, clip, width=0.5 \textwidth]{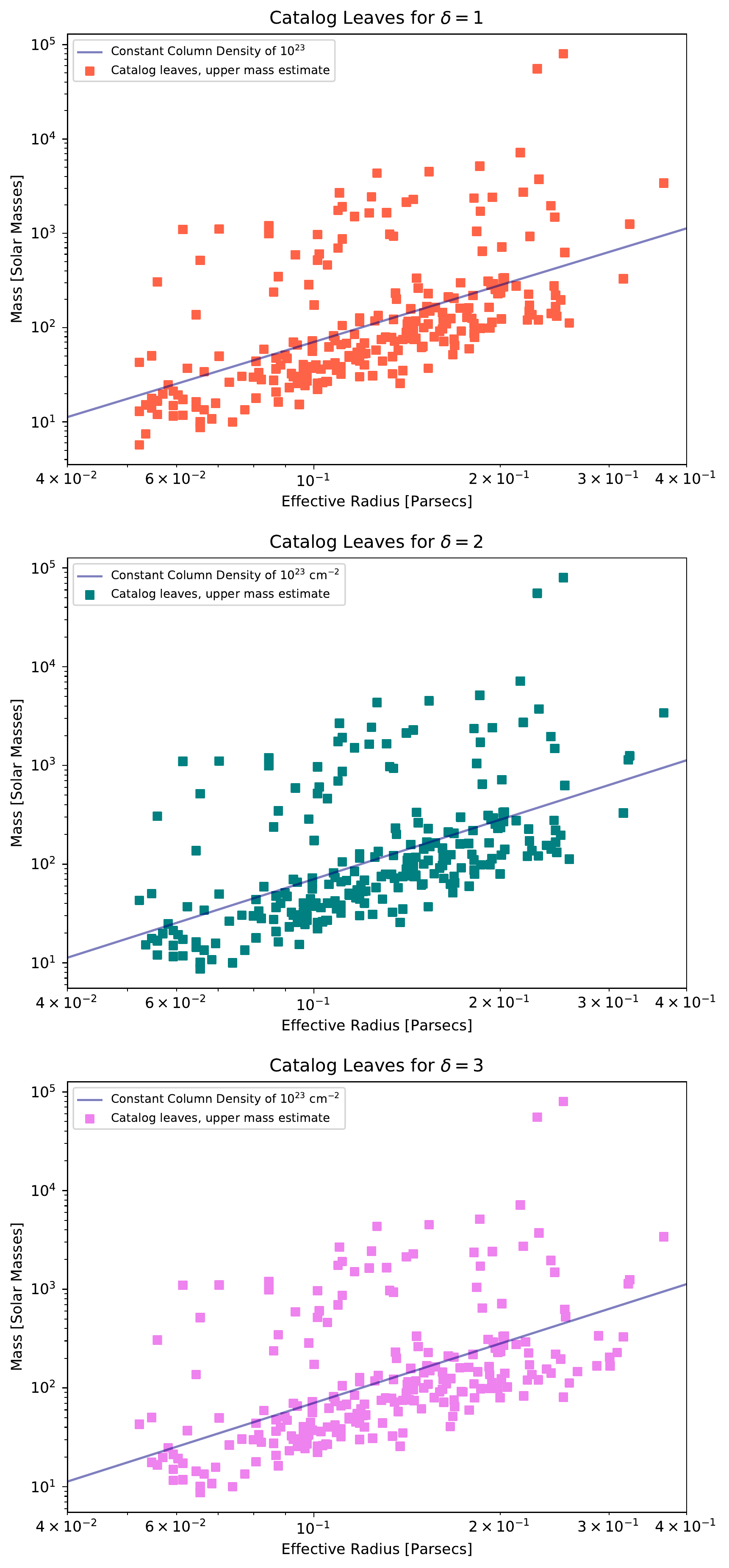}
\end{center}
\caption{Three versions of the catalog, calculated for different choices of the initial dendrogram parameter for the minimum separation of structures, $\delta$. From top to bottom, these catalogs were constructed using $\delta=2$, $\delta=3$, and $\delta=4$. The middle panel is identical to the actual catalog described in the rest of this work. The majority of leaves are robust with respect to changes in $\delta$, and the objects most affected seem to be those with a large effective radius and a small mass.}
\label{fig:mr_full_parameterstudy}
\end{figure*}

\begin{figure*}
\begin{center}
\includegraphics[trim = 0mm 0mm 0mm 0mm, clip, width=0.5 \textwidth]{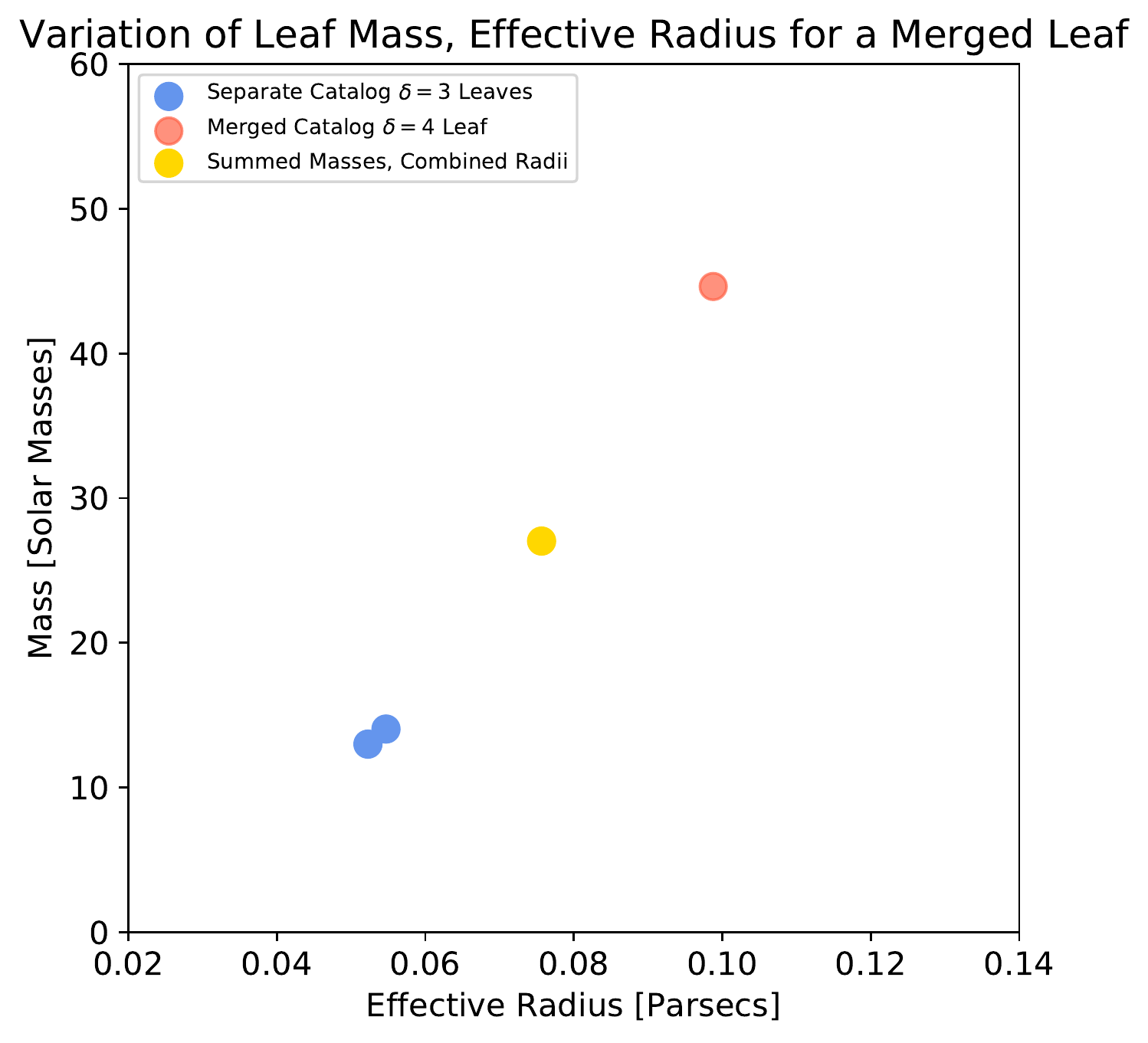}
\end{center}
\caption{The behavior of a pair of nearby local flux peaks included as separate leaves in the catalog for a low $\delta=1,23$, but which merge together for $\delta=3$. This behavior is uncommon in the catalog but has a significant effect on the total amount of flux included in the structure. The amount of flux enclosed by the merged structure is more than a factor of 2 larger than the sum of the individual leaves' mass.}
\label{fig:mr_merge_parameterstudy}
\end{figure*}

\begin{figure*}
\begin{center}
\includegraphics[trim = 0mm 0mm 0mm 0mm, clip, width=0.5 \textwidth]{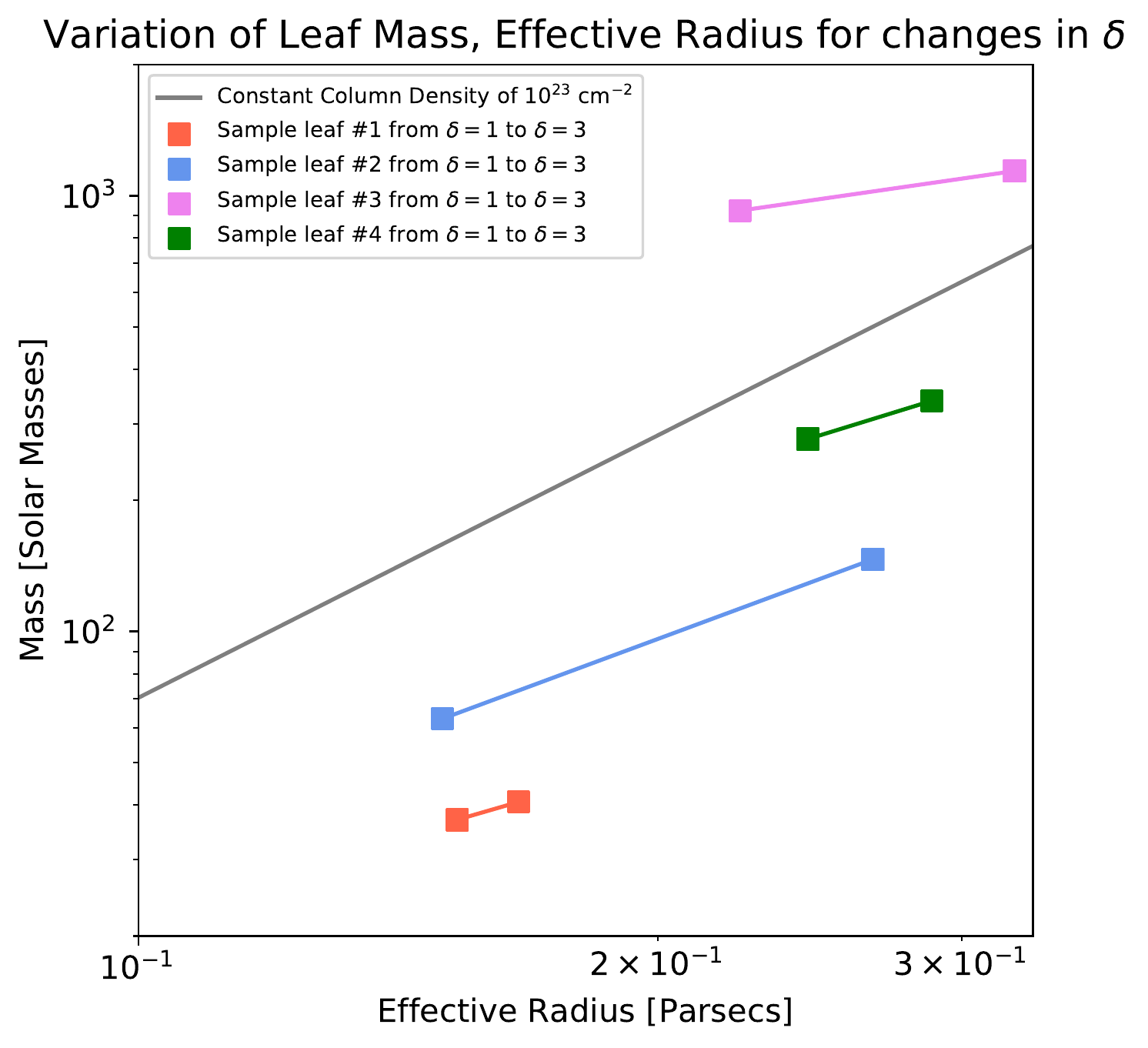}
\end{center}
\caption{The variation of four leaves that change mass and effective radius for varying the initial dendrogram minimum difference parameter $\delta$. A line of constant column density $10^{23}$ is shown for comparison to Figure \ref{fig:mr2fig}. }
\label{fig:mr_vary_parameterstudy}
\end{figure*}

\begin{figure*}
\label{fig:simobs_kernel}
\begin{center}
\includegraphics[trim = 0mm 0mm 0mm 0mm, clip, width=0.5 \textwidth]{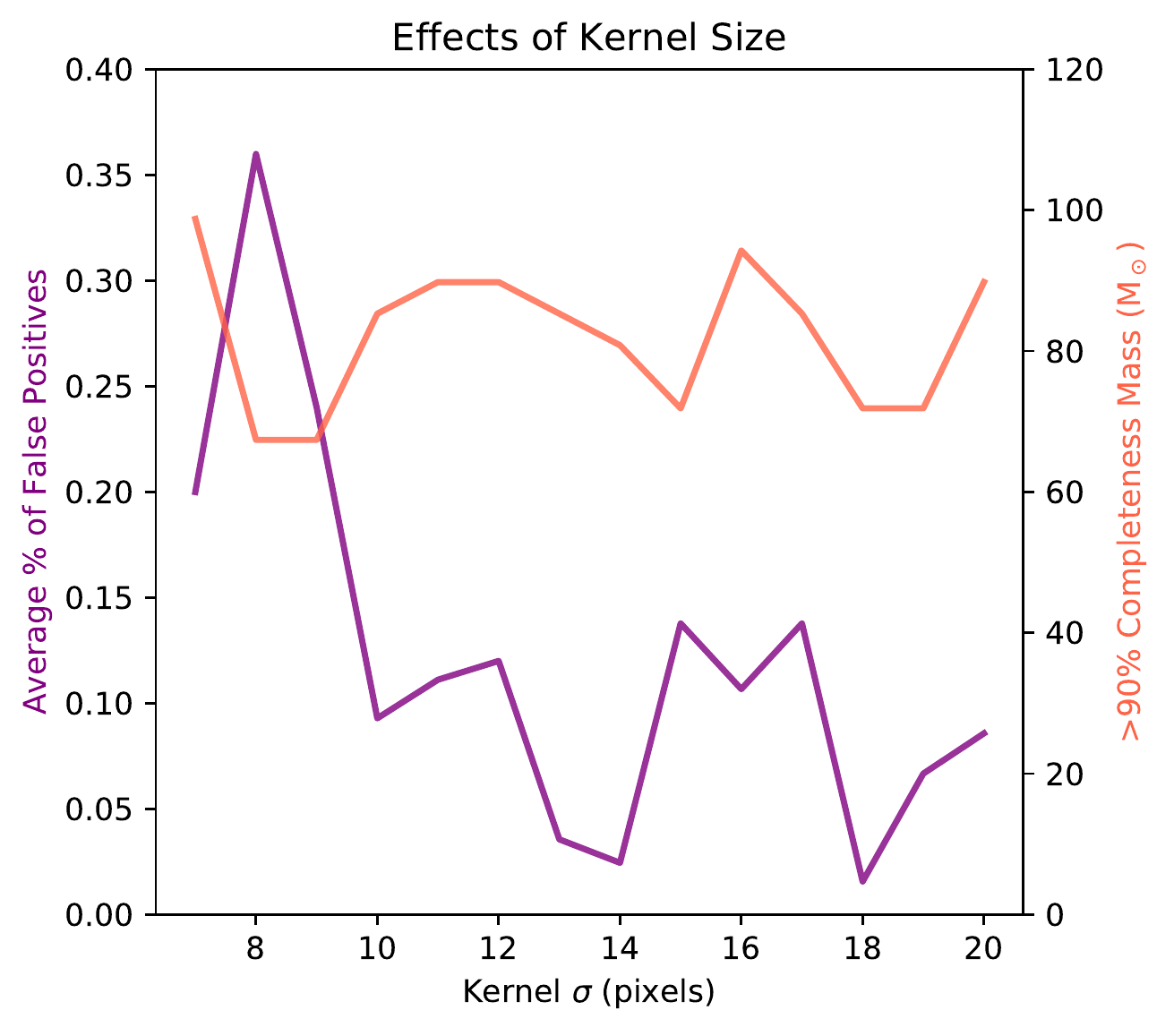}
\end{center}
\caption{The effects of altering the smoothing kernel used in the convolutions for generating the local noise map used for pruning. Certain choices of kernel size lead to a much higher percentage of false positive source detections for points objects above the completeness threshold of 70M$_\odot$ used in the high reliability version of the catalog (see section\,\ref{sec:simobs})}.
\end{figure*}

Small changes in the minimum value parameter for the original dendrogram (see \ref{sec:catalog}) have no effect on the total dendrogram, as all of the RMS noise values are by design greater than the low noise estimate used to scale the initial dendrogram parameters. Therefore, we neglect an experiment varying this quantity. The initial dendrogram, however, is sensitive to changes in the minimum significance parameter $\delta$. It is possible, by changing $\delta$, to cause nearby leaves of similar flux distributions to merge together or split apart into separate leaves. This phenomenon is rare in the case of our catalogs, with $\lessapprox3\%$ of leaves contained in the robust catalog displaying this behavior. In Figure \ref{fig:mr_merge_parameterstudy} we show the relationship between cataloged leaf masses before and after a leaf merger. The merged leaf mass and radius is larger than would be naively expected by summing the radii and masses of the separate leaves, though this is not surprising given the clustering procedure at work in the dendrogram algorithm.

 Similarly, leaves with particularly shallow distributions might change slightly in size with changes in $\delta$. These changes can affect the inferred physical properties of the leaves in a non-trivial way. Figure \ref{fig:mr_vary_parameterstudy} shows a selection of leaves identified in the robust catalog that vary with changes in $\delta$. The leaf size and inferred mass appear to vary as a roughly consistent power law, with leaves varying in mass (radius) less than a factor of $\sim$2 ($\sim$3). Even with these rare cases where leaves merge or change shape with varying $\delta$, we find that the mass and radius distribution of leaf properties in the catalog is well preserved (see Figure \ref{fig:mr_full_parameterstudy}), with the leaves associated with Sgr B2 and those elsewhere still clearly separated. 

We also consider the effect of changes to the local noise map generation on the catalog. The only scalable parameter affecting the contents of the catalog is the convolution kernel size used in smoothing the RMS noise map. We explore the sensitivity of the catalog to changes in the smoothing kernel size using a similar simulated observation method as section \ref{sec:simobs}. For each iteration of the simulated observations, and for a given kernel size, we record the false positive percentage for point-source like emission. In order to get a single number for the false positive rate over a number of trials, we average together all simulated observation runs for source masses above the 95\% completeness for that run. It is worth noting that the number of false positives is higher for larger mass point-sources, as the imaging artifacts have correspondingly higher fluxes and are less likely to be pruned away by the cataloging algorithm. The results from this process are displayed in Figure \ref{fig:simobs_kernel}. It appears that a kernel size of 14 pixels or 18 pixels minimizes the average number of false positives for objects above our 80\% completeness limit. We chose to use the 14 pixel kernel as this choice also has a slightly lower average completeness mass. The variations from trial to trial of this experiment are large enough that there is no ideal choice of kernel for any given configuration of point-sources, and we suspect the results would be further complicated by non-point-source-like emission. However, the variation in completeness is small typically smaller for choices in the range of 10-20 pixels, and resulting effects on the catalog varying between these choices are small relative to the overall trends in completeness in Figure \ref{fig:simobs_percent}.

\section{Zoom-in Images of Key Regions}\label{sec:zoomins}
\begin{figure*}
\begin{center}
\subfigure{
\includegraphics[width=0.48\textwidth]{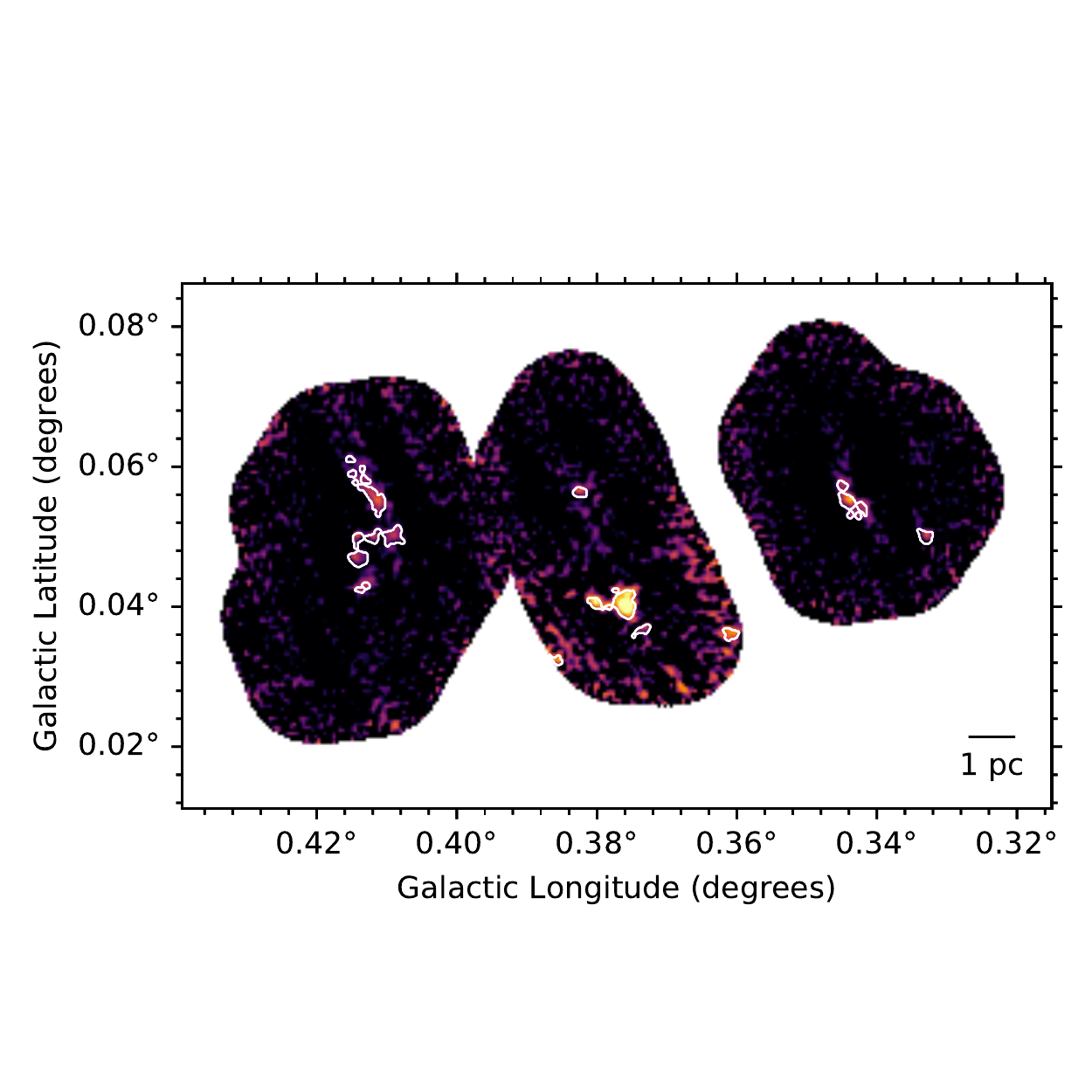}}
\subfigure{
\includegraphics[width=0.48\textwidth]{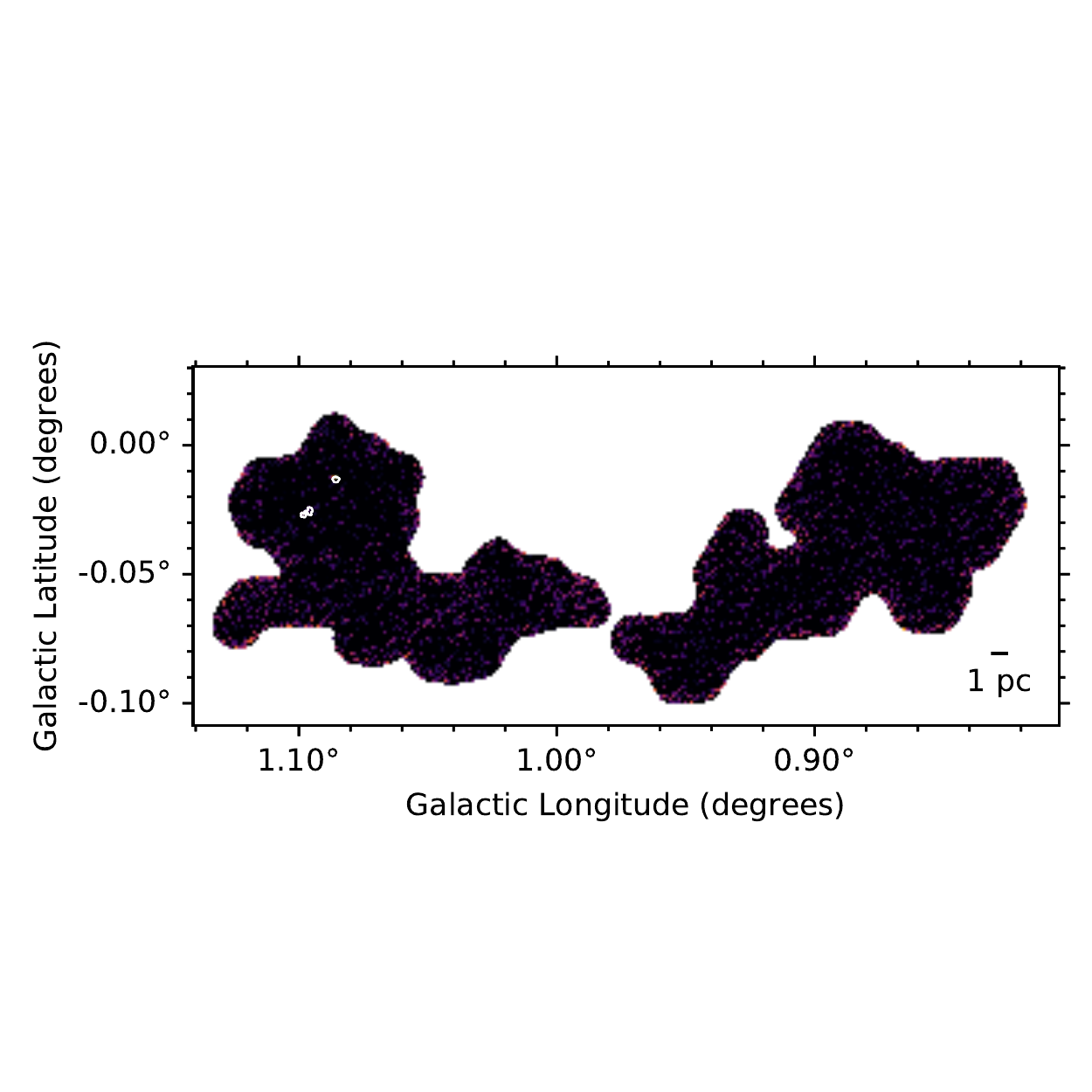}}
\subfigure{
\includegraphics[width=0.48\textwidth]{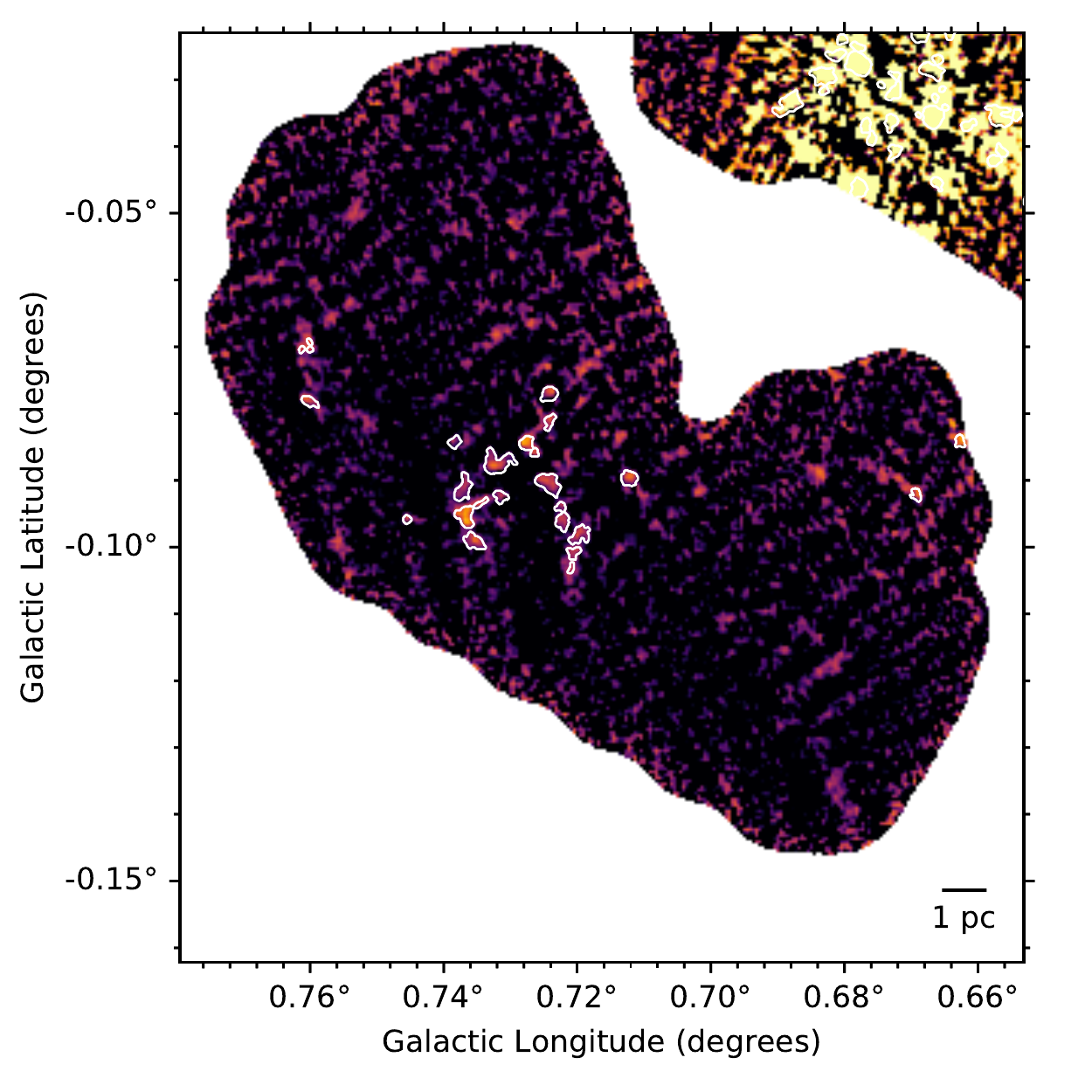}}
\subfigure{
\includegraphics[width=0.48\textwidth]{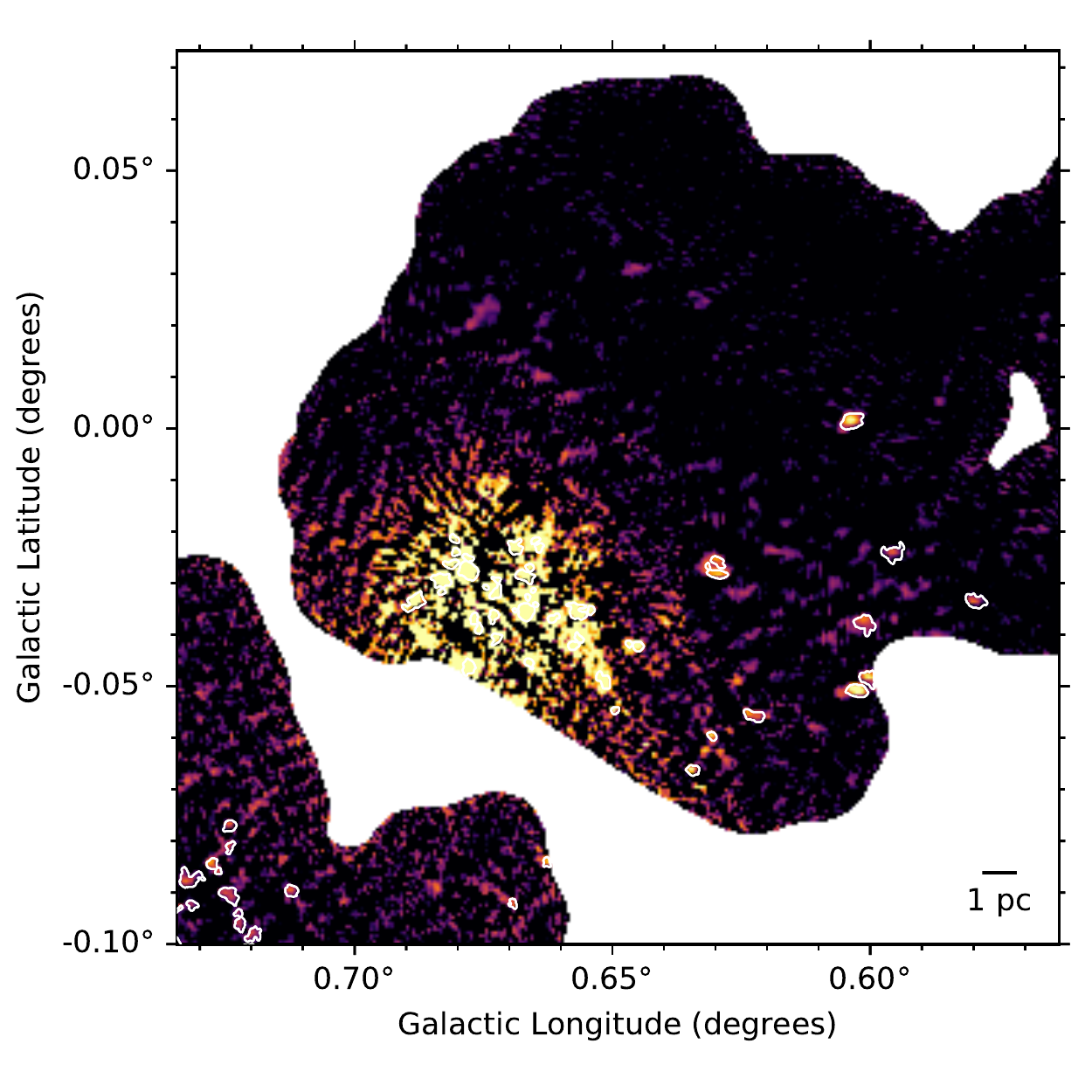}}
\singlespace\caption{Robust catalog leaf contours superimposed over the 1.3 mm dust continuum from the \textit{CMZoom} Survey with a 1 pc scale-bar. All images are displayed on a log scale with limits between 1$\times 10^7$ Jy sr$^{-1}$ and 3$\times 10^8$ Jy sr$^{-1}$} 
\end{center}
\label{fig:cutout_1}
\end{figure*}

\begin{figure*}
\begin{center}
\subfigure{
\includegraphics[width=0.48\textwidth]{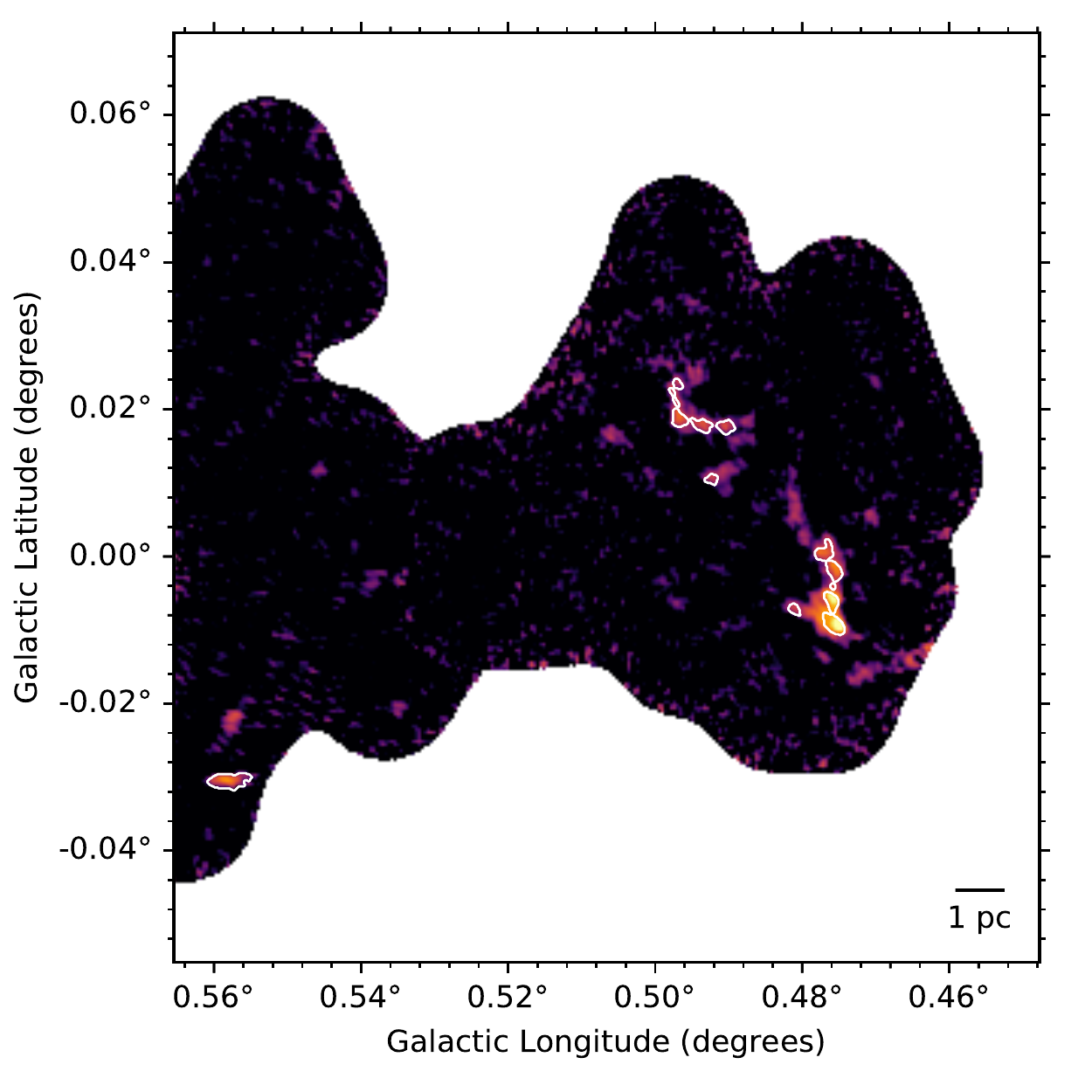}}
\subfigure{
\includegraphics[width=0.48\textwidth]{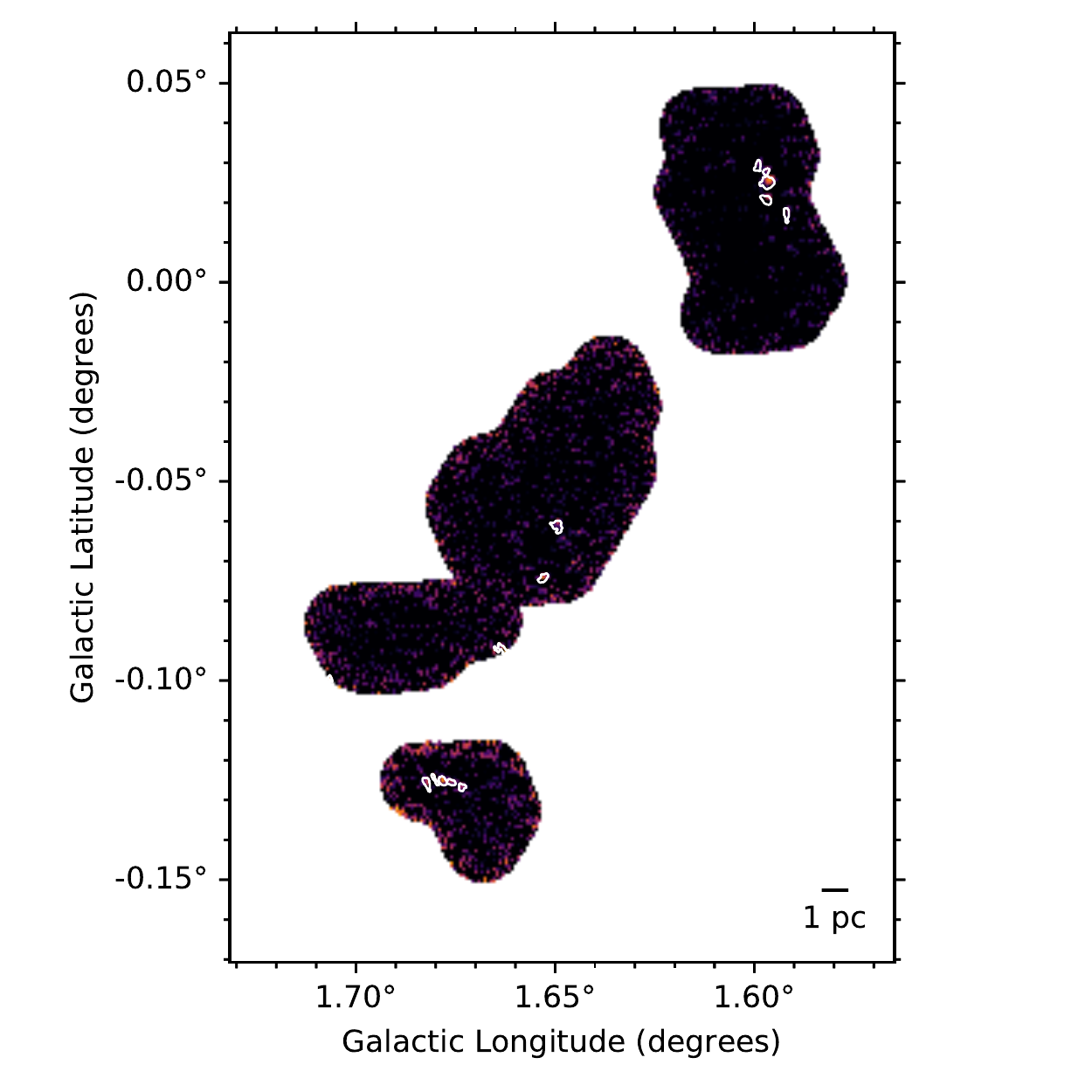}}
\subfigure{
\includegraphics[width=0.48\textwidth]{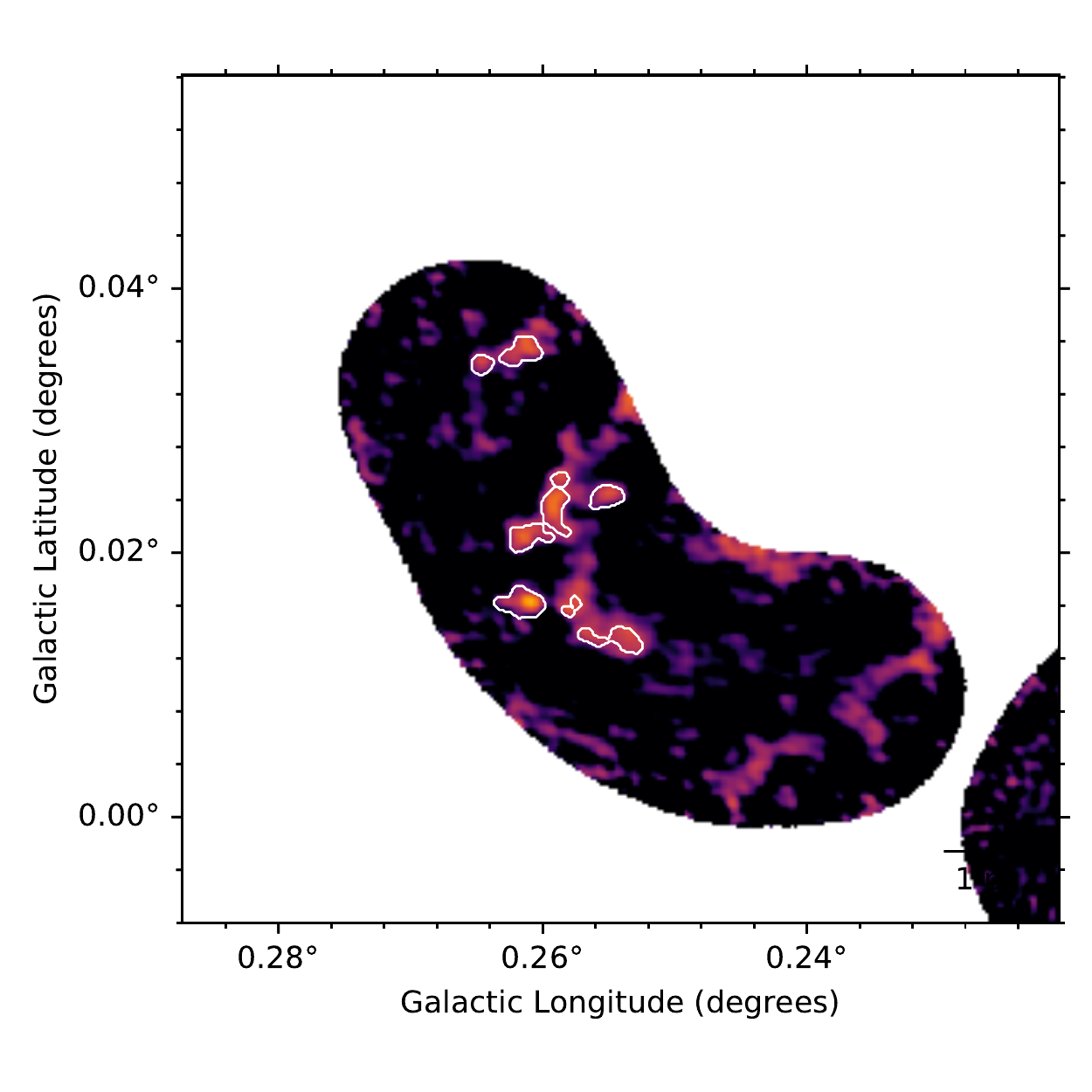}}
\subfigure{
\includegraphics[width=0.48\textwidth]{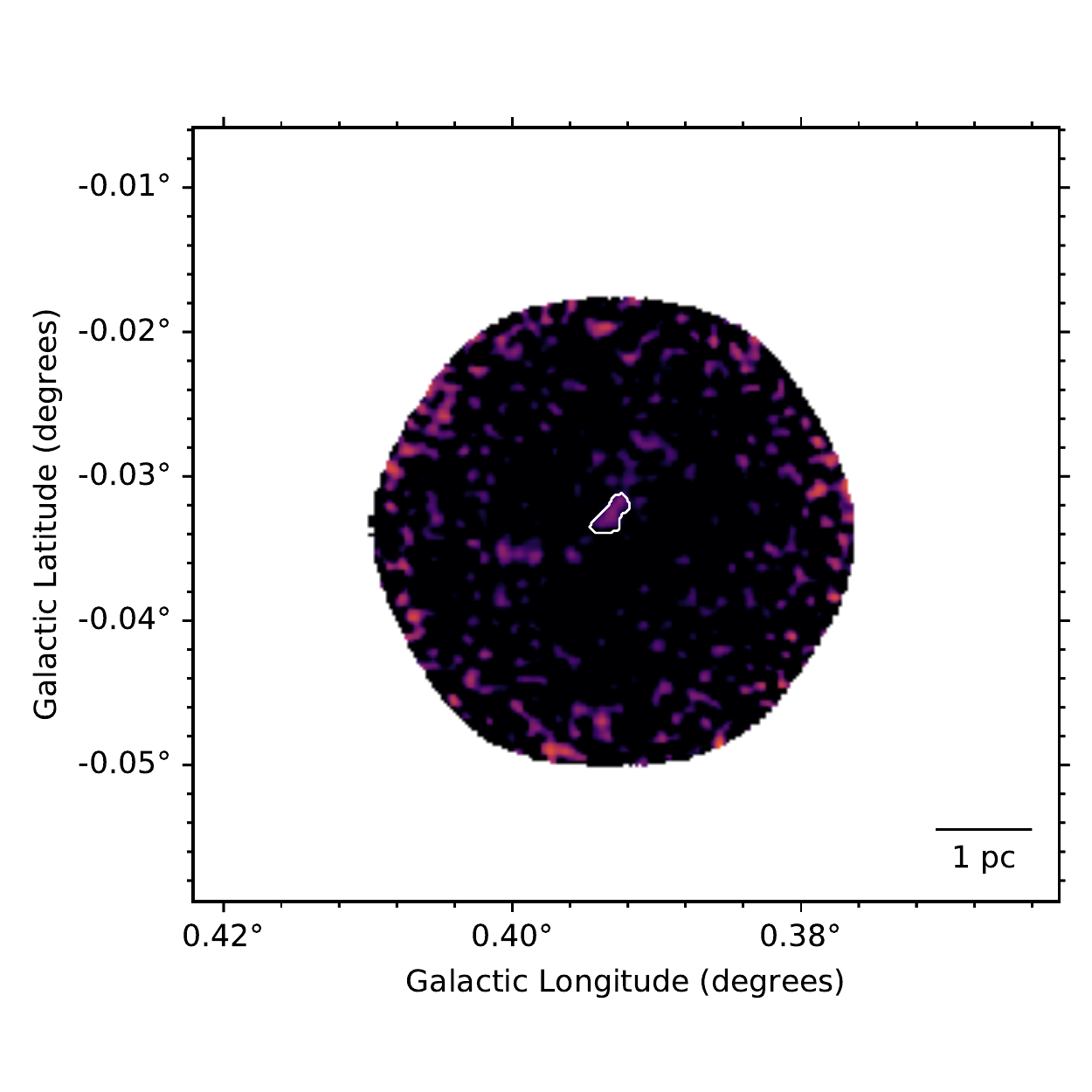}}
\singlespace\caption{Robust catalog leaf contours superimposed over the 1.3 mm dust continuum from the \textit{CMZoom} Survey with a 1 pc scale-bar. All images are displayed on a log scale with limits between 1$\times 10^7$ Jy sr$^{-1}$ and 3$\times 10^8$ Jy sr$^{-1}$} 
\end{center}
\label{fig:cutout_2}
\end{figure*}

\begin{figure*}
\begin{center}
\subfigure{
\includegraphics[width=0.48\textwidth]{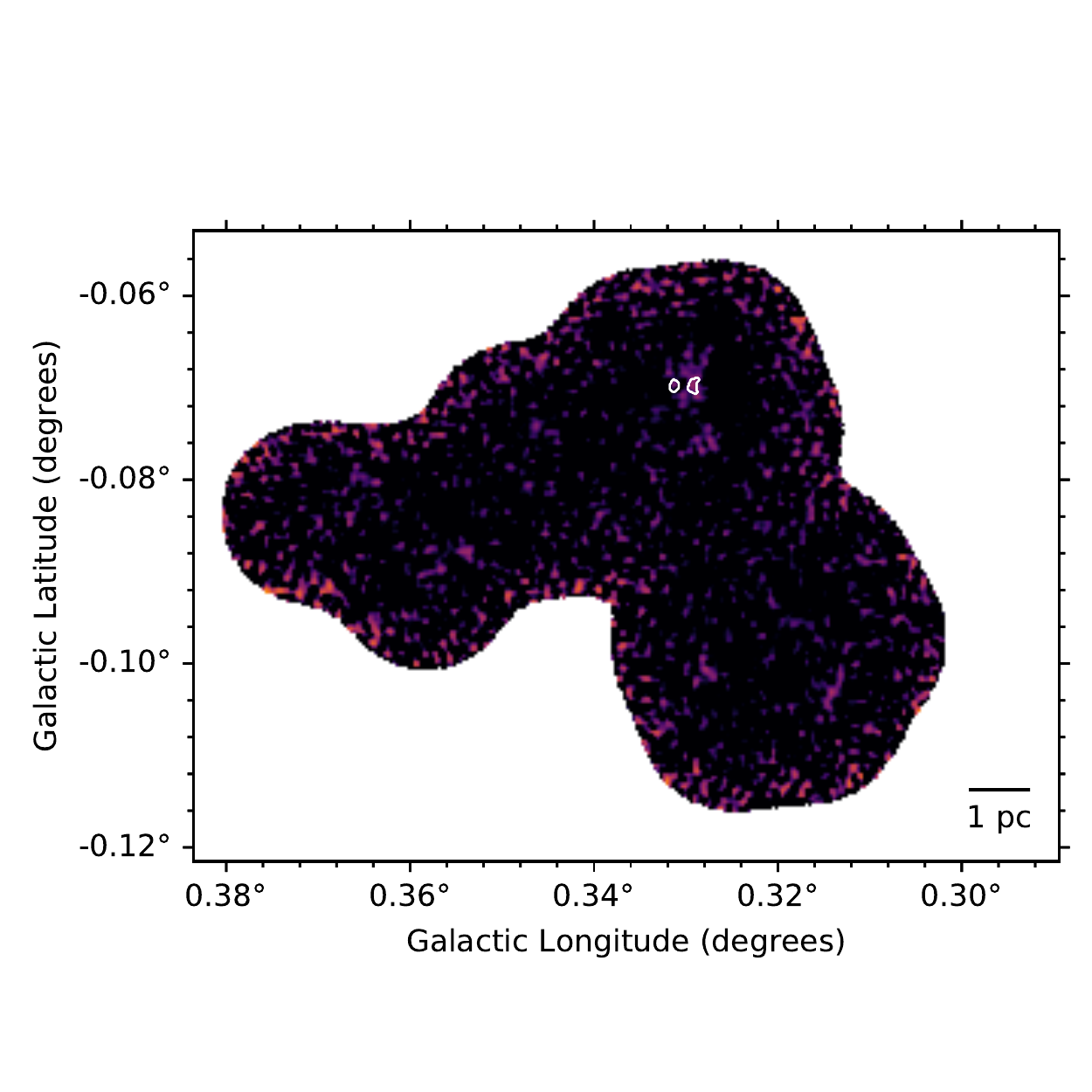}}
\subfigure{
\includegraphics[width=0.48\textwidth]{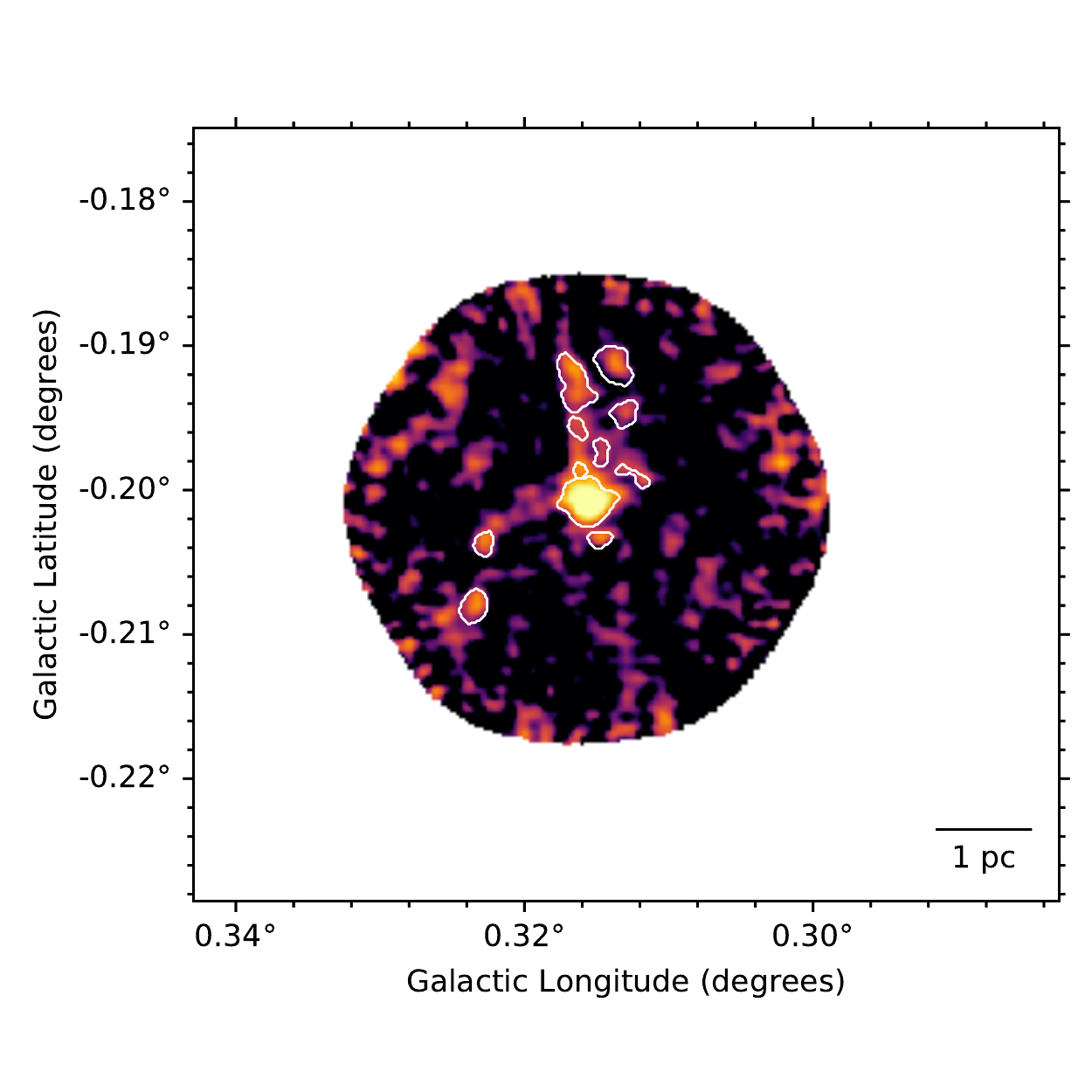}}
\subfigure{
\includegraphics[width=0.48\textwidth]{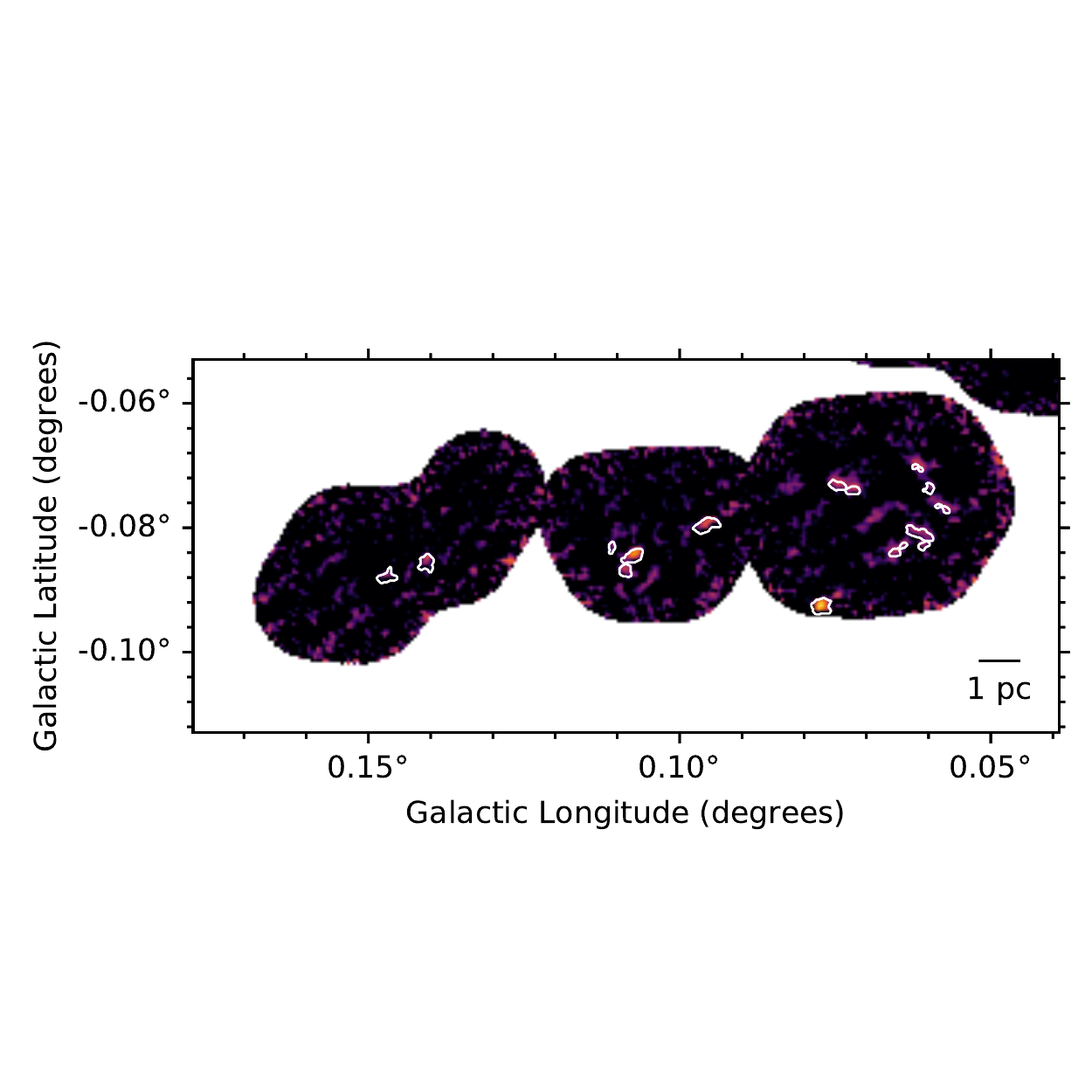}}
\subfigure{
\includegraphics[width=0.48\textwidth]{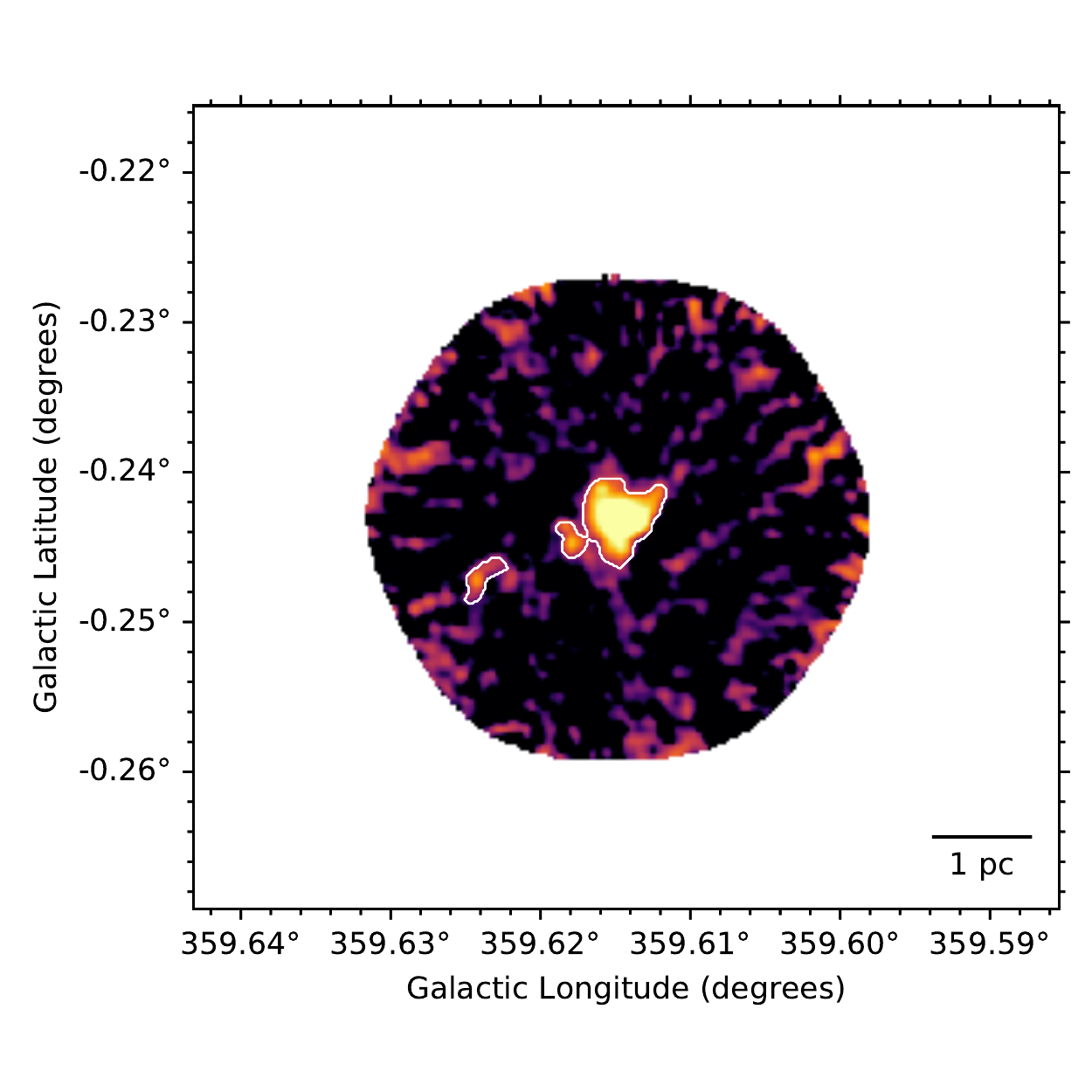}}
\singlespace\caption{Robust catalog leaf contours superimposed over the 1.3 mm dust continuum from the \textit{CMZoom} Survey with a 1 pc scale-bar. All images are displayed on a log scale with limits between 1$\times 10^7$ Jy sr$^{-1}$ and 3$\times 10^8$ Jy sr$^{-1}$. } 
\end{center}
\label{fig:cutout_3}
\end{figure*}

\begin{figure*}
\begin{center}
\subfigure{
\includegraphics[width=0.48\textwidth]{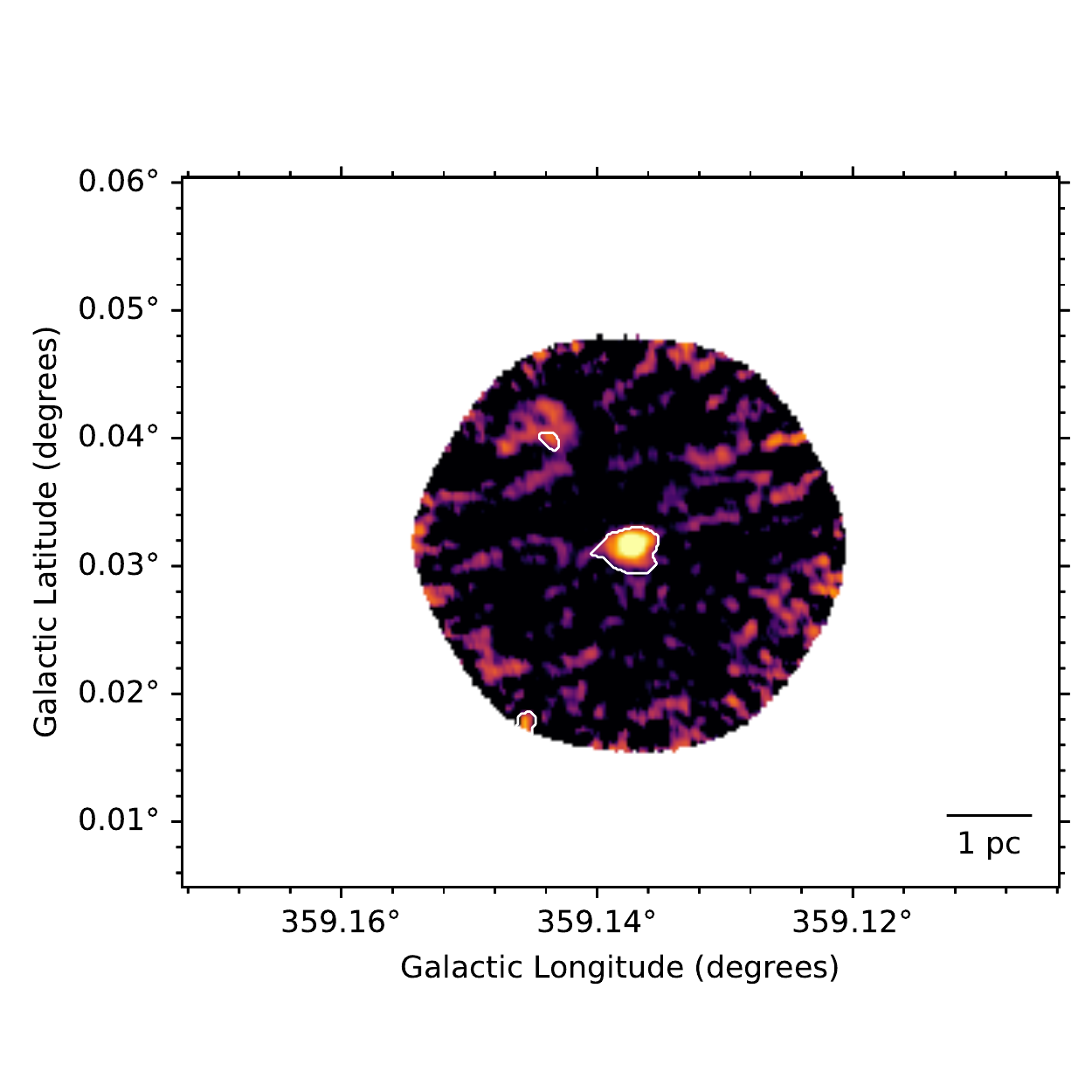}}
\subfigure{
\includegraphics[width=0.48\textwidth]{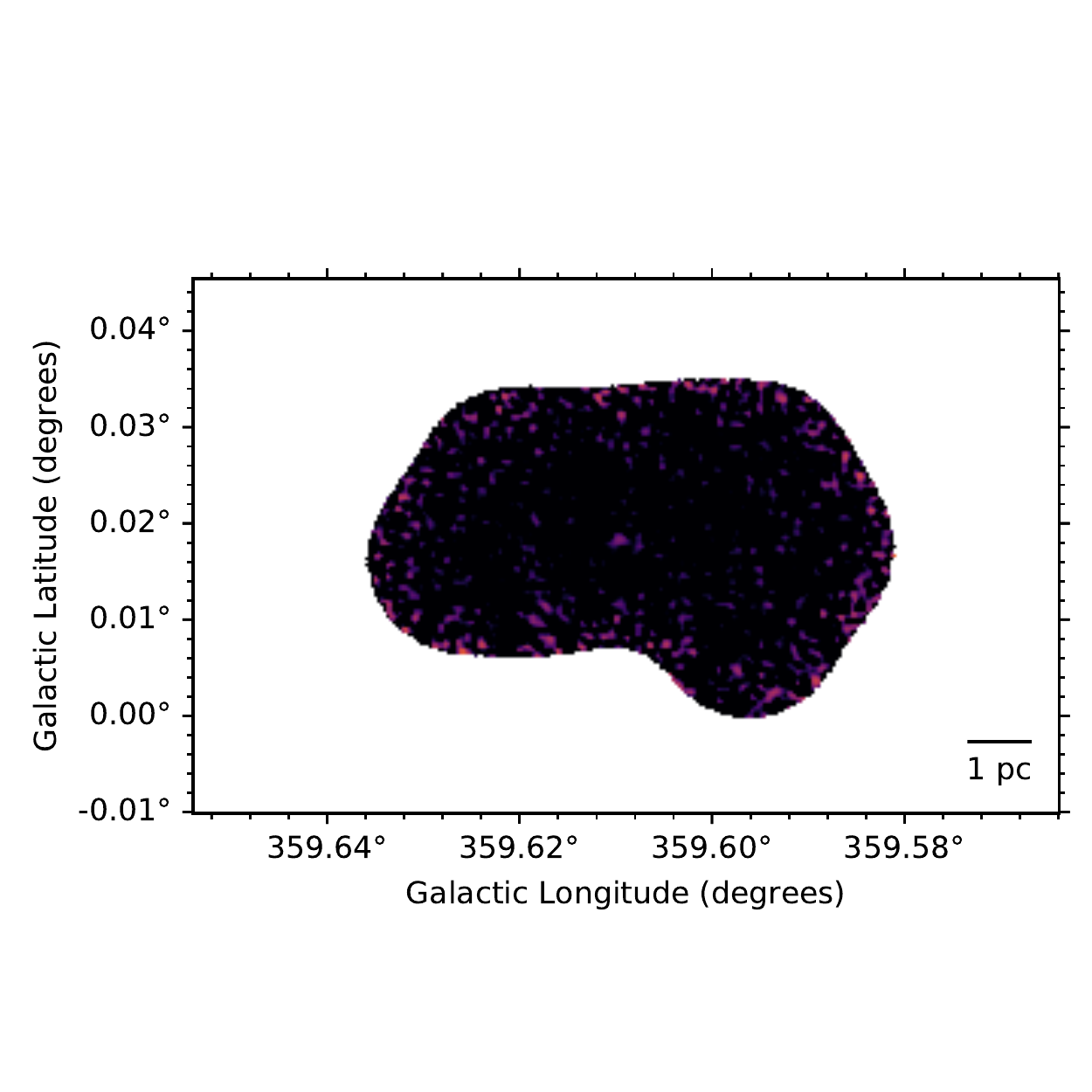}}
\subfigure{
\includegraphics[width=0.48\textwidth]{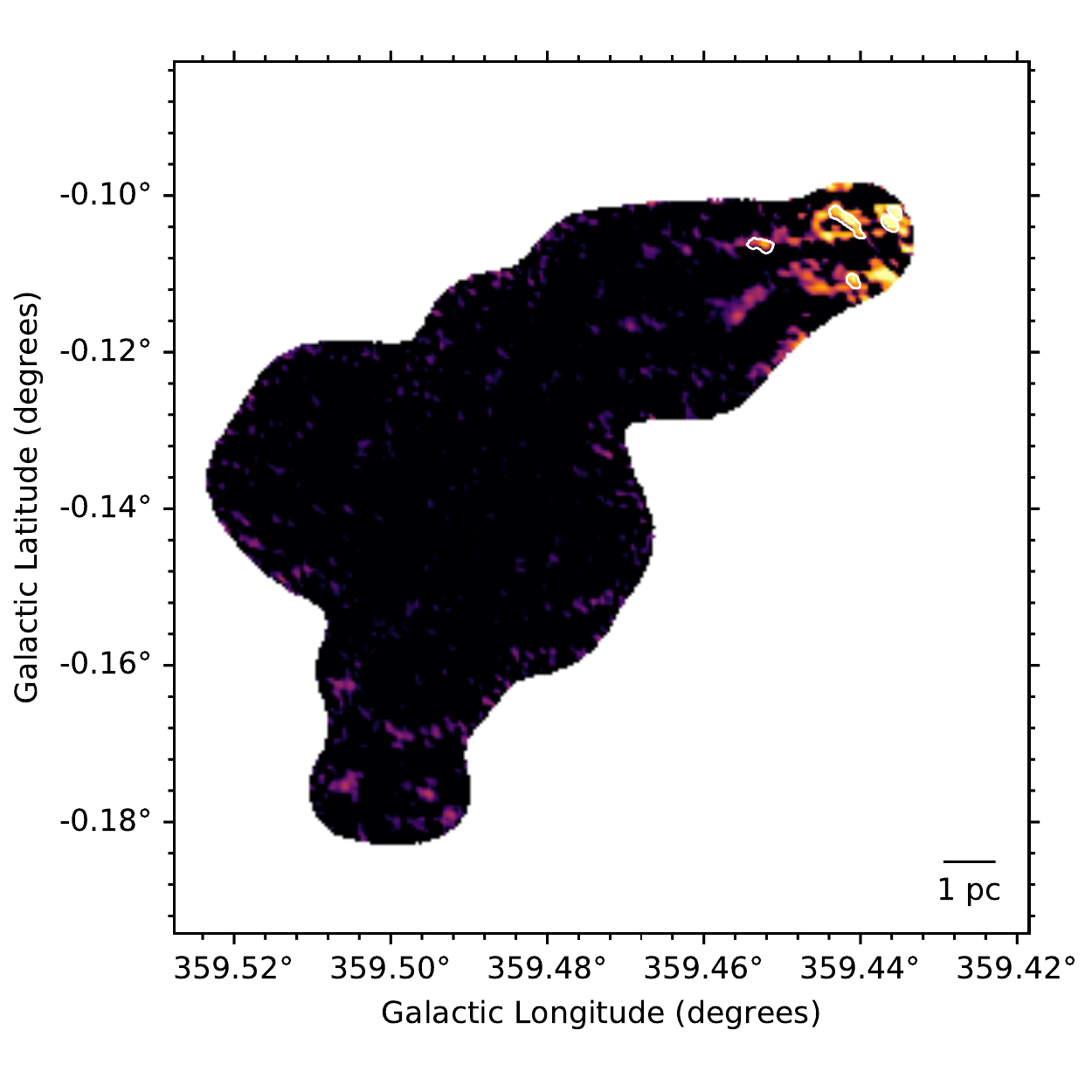}}
\subfigure{
\includegraphics[width=0.48\textwidth]{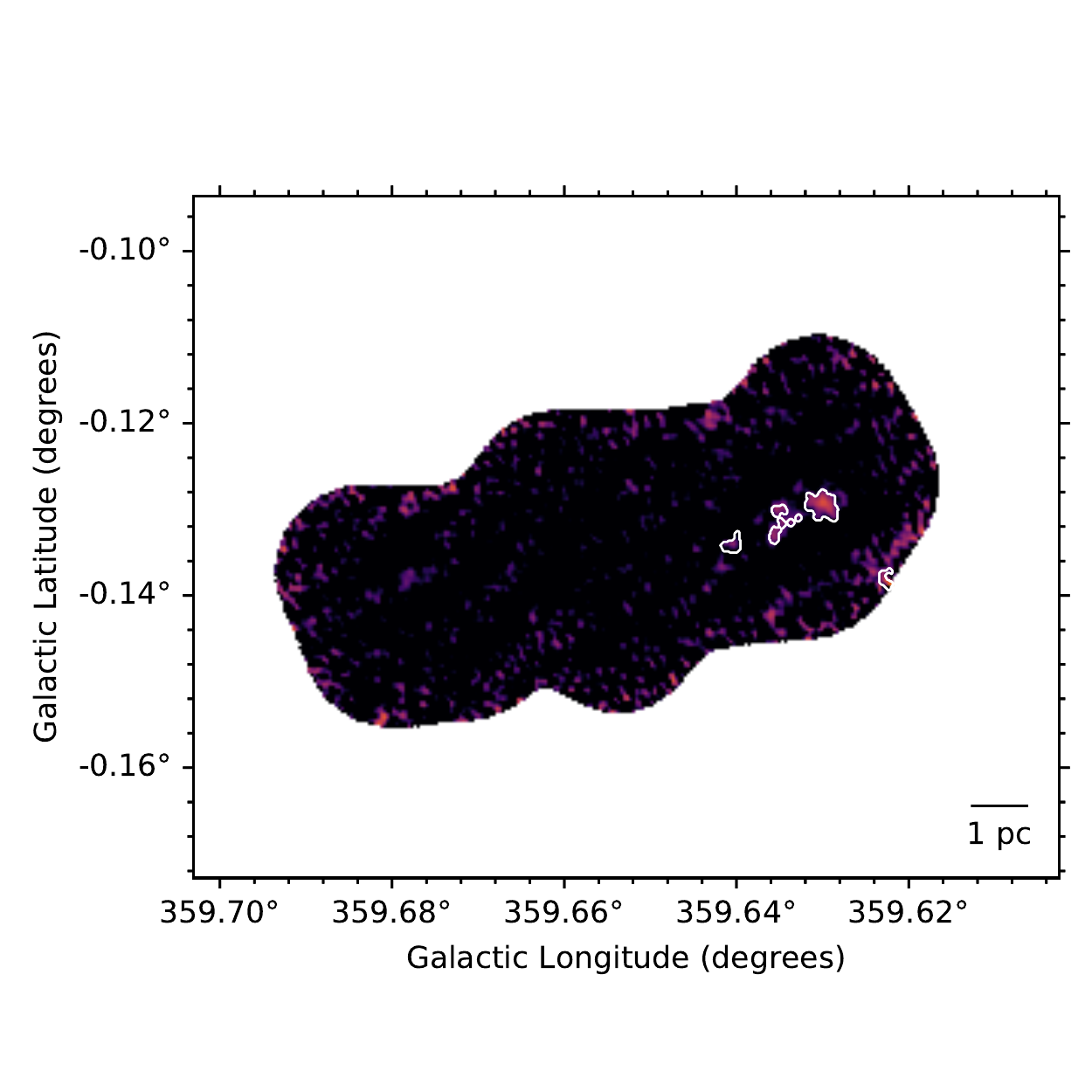}}
\singlespace\caption{Robust catalog leaf contours superimposed over the 1.3 mm dust continuum from the \textit{CMZoom} Survey with a 1 pc scale-bar. All images are displayed on a log scale with limits between 1$\times 10^7$ Jy sr$^{-1}$ and 3$\times 10^8$ Jy sr$^{-1}$. } 
\end{center}
\label{fig:cutout_4}
\end{figure*}

\begin{figure*}
\begin{center}
\subfigure{
\includegraphics[width=0.48\textwidth]{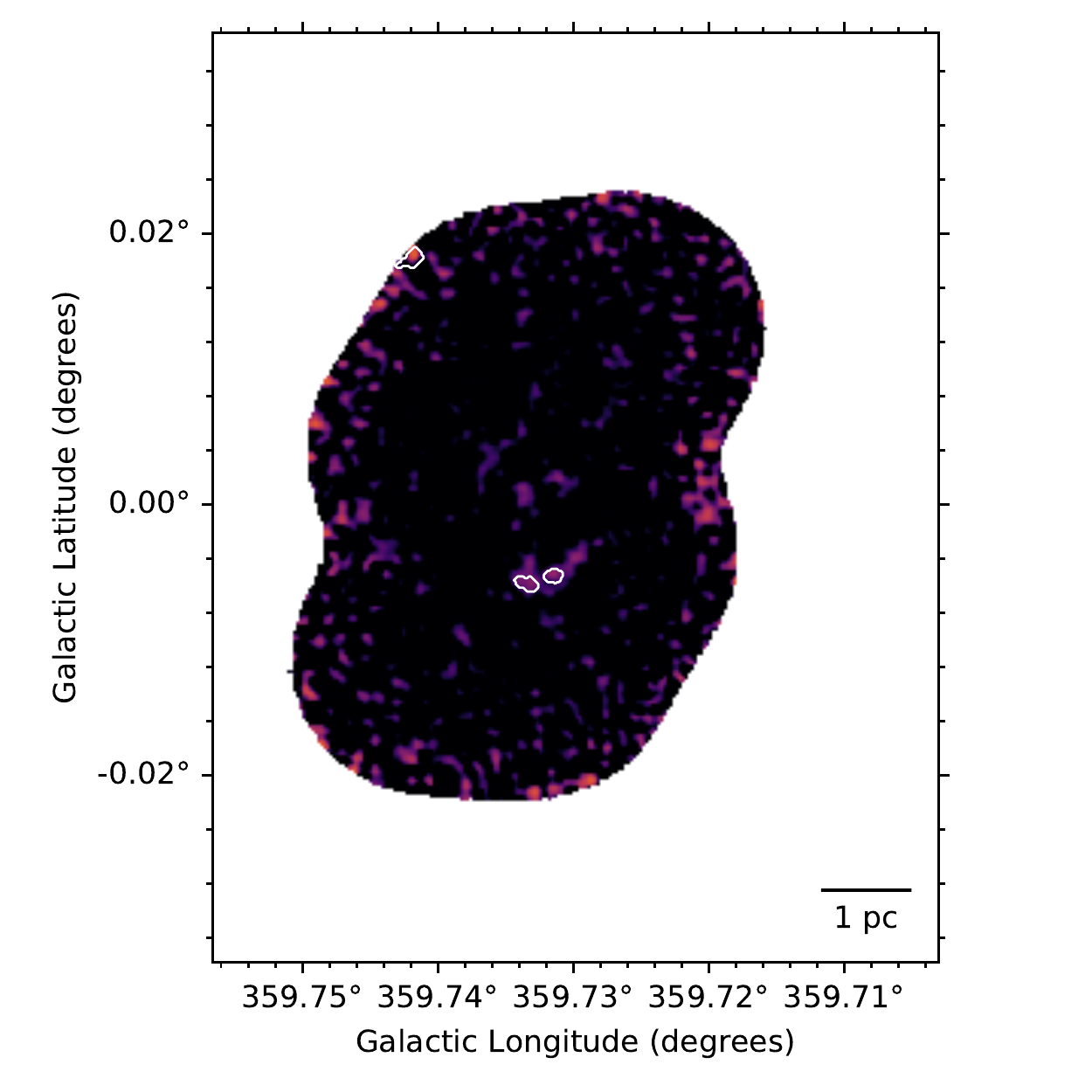}}
\subfigure{
\includegraphics[width=0.48\textwidth]{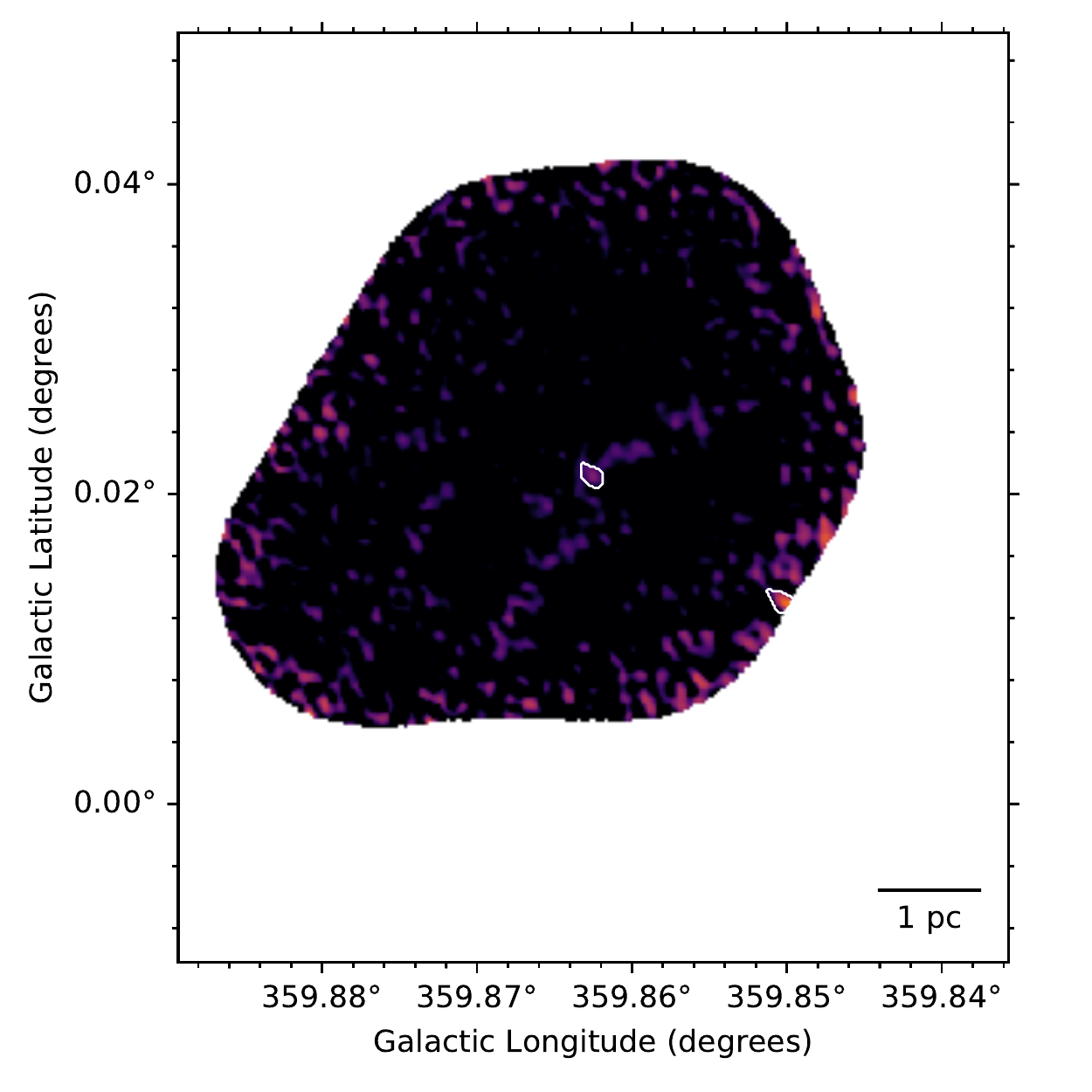}}
\subfigure{
\includegraphics[width=0.48\textwidth]{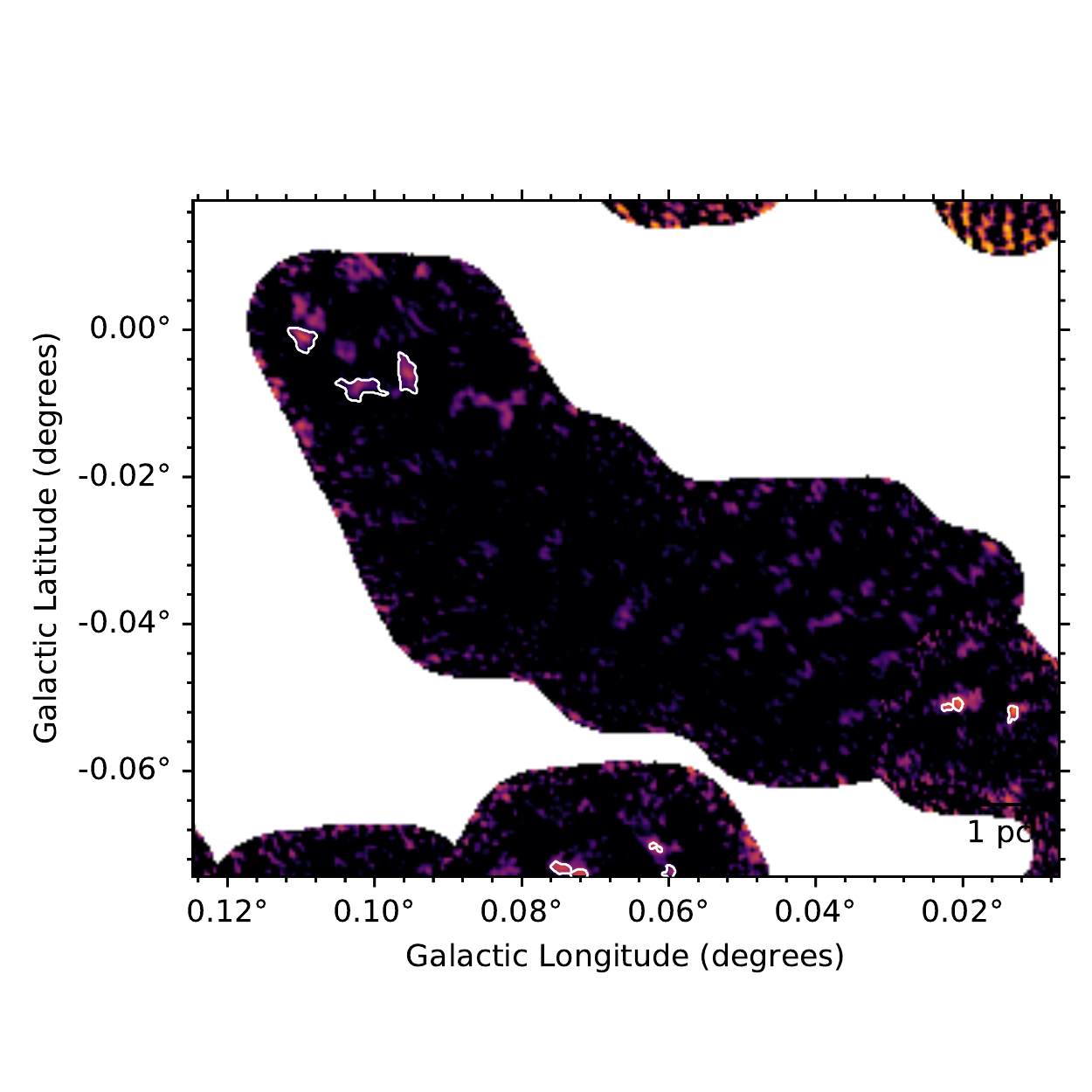}}
\subfigure{
\includegraphics[width=0.48\textwidth]{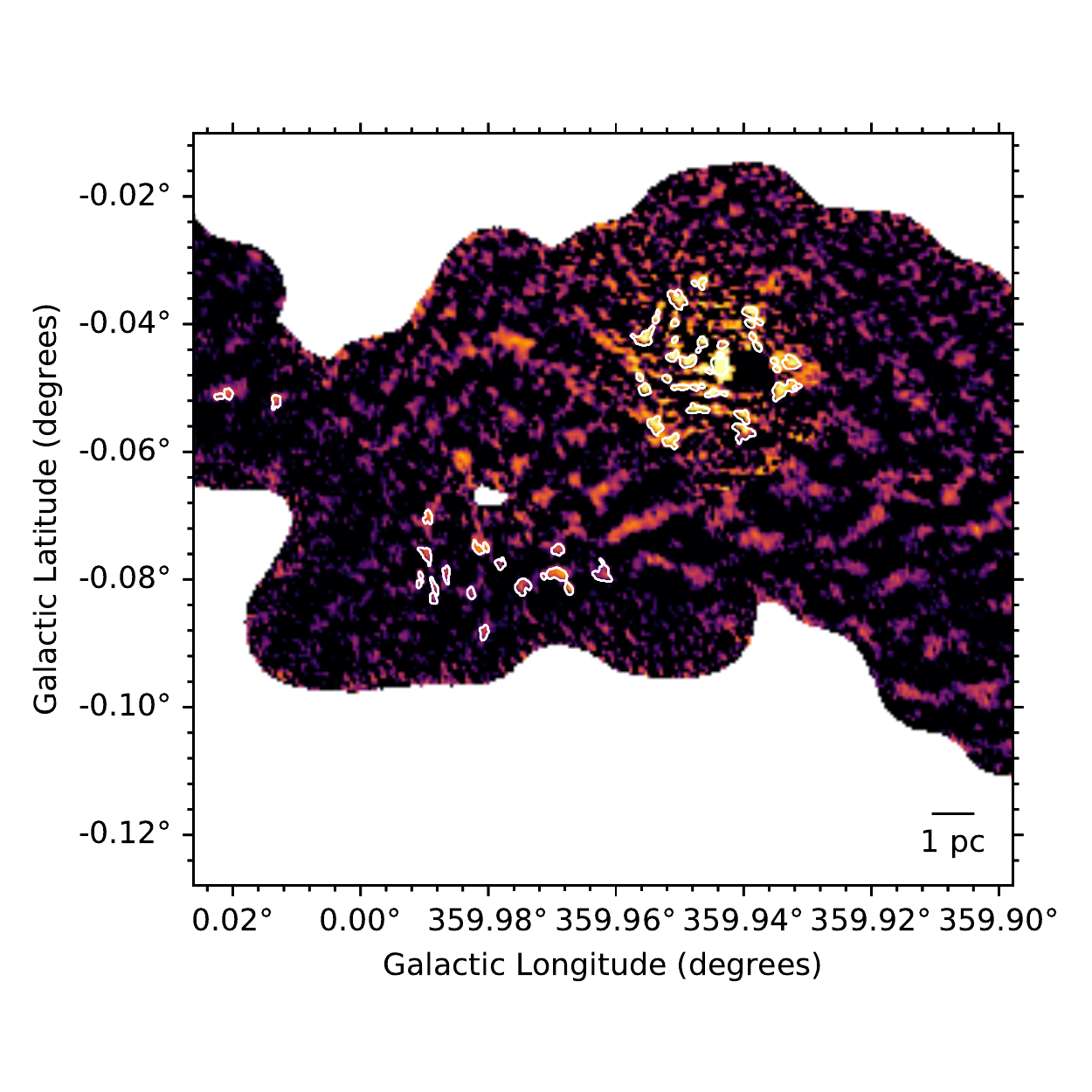}}
\singlespace\caption{Robust catalog leaf contours superimposed over the 1.3 mm dust continuum from the \textit{CMZoom} Survey with a 1 pc scale-bar. All images are displayed on a log scale with limits between 1$\times 10^7$ Jy sr$^{-1}$ and 3$\times 10^8$ Jy sr$^{-1}$. } 
\end{center}
\label{fig:cutout_5}
\end{figure*}

\begin{figure*}
\begin{center}
\subfigure{
\includegraphics[width=0.48\textwidth]{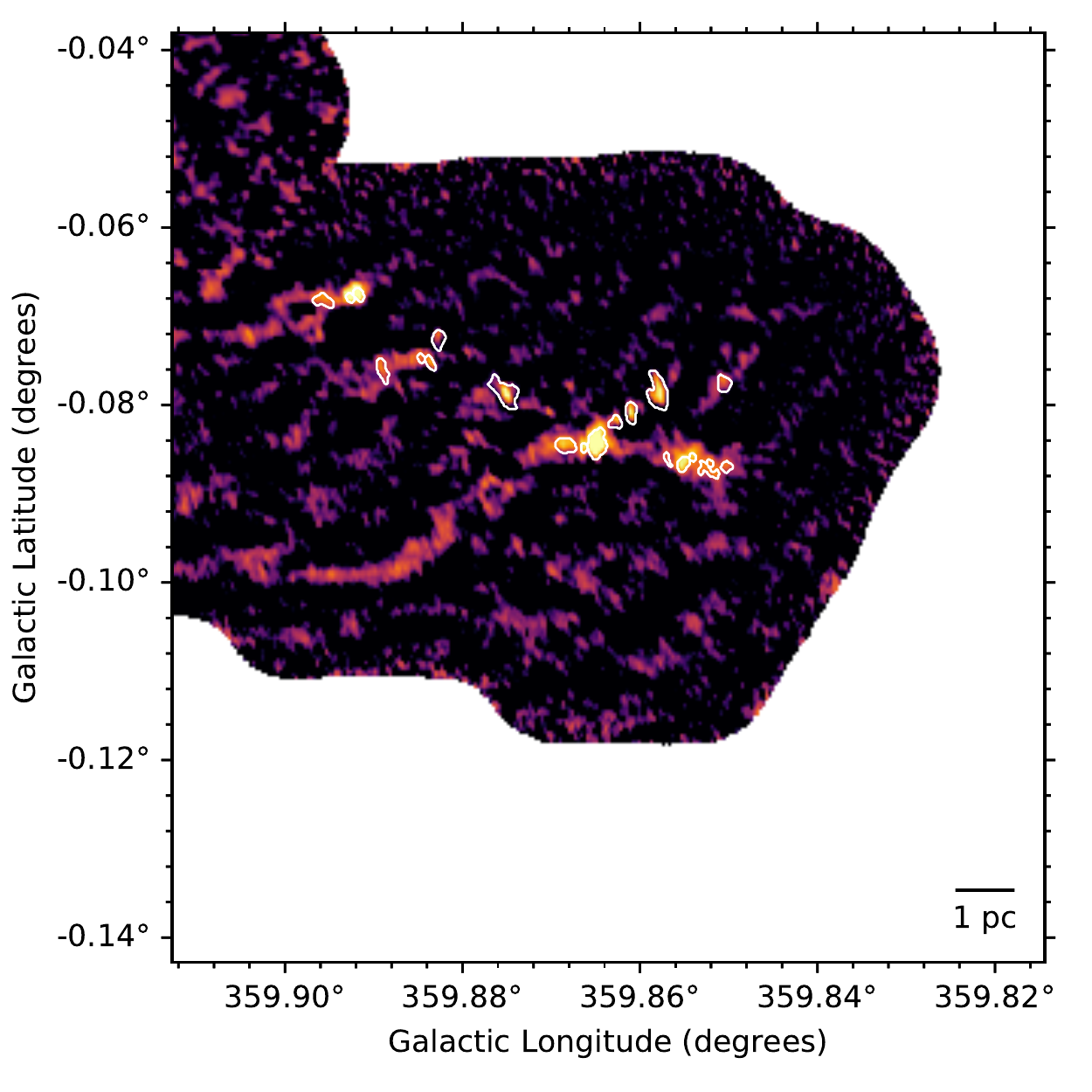}}
\subfigure{
\includegraphics[width=0.48\textwidth]{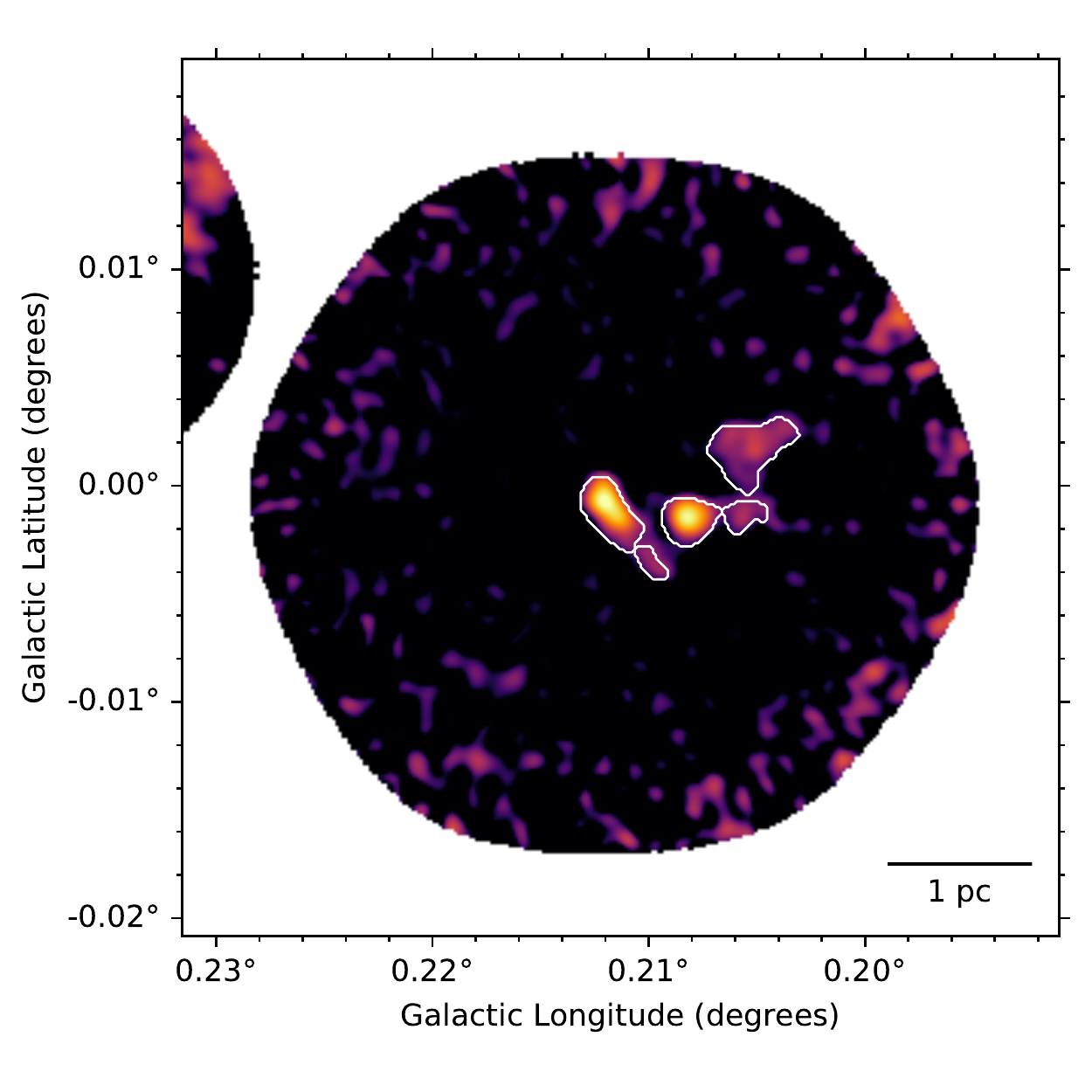}}
\subfigure{
\includegraphics[width=0.48\textwidth]{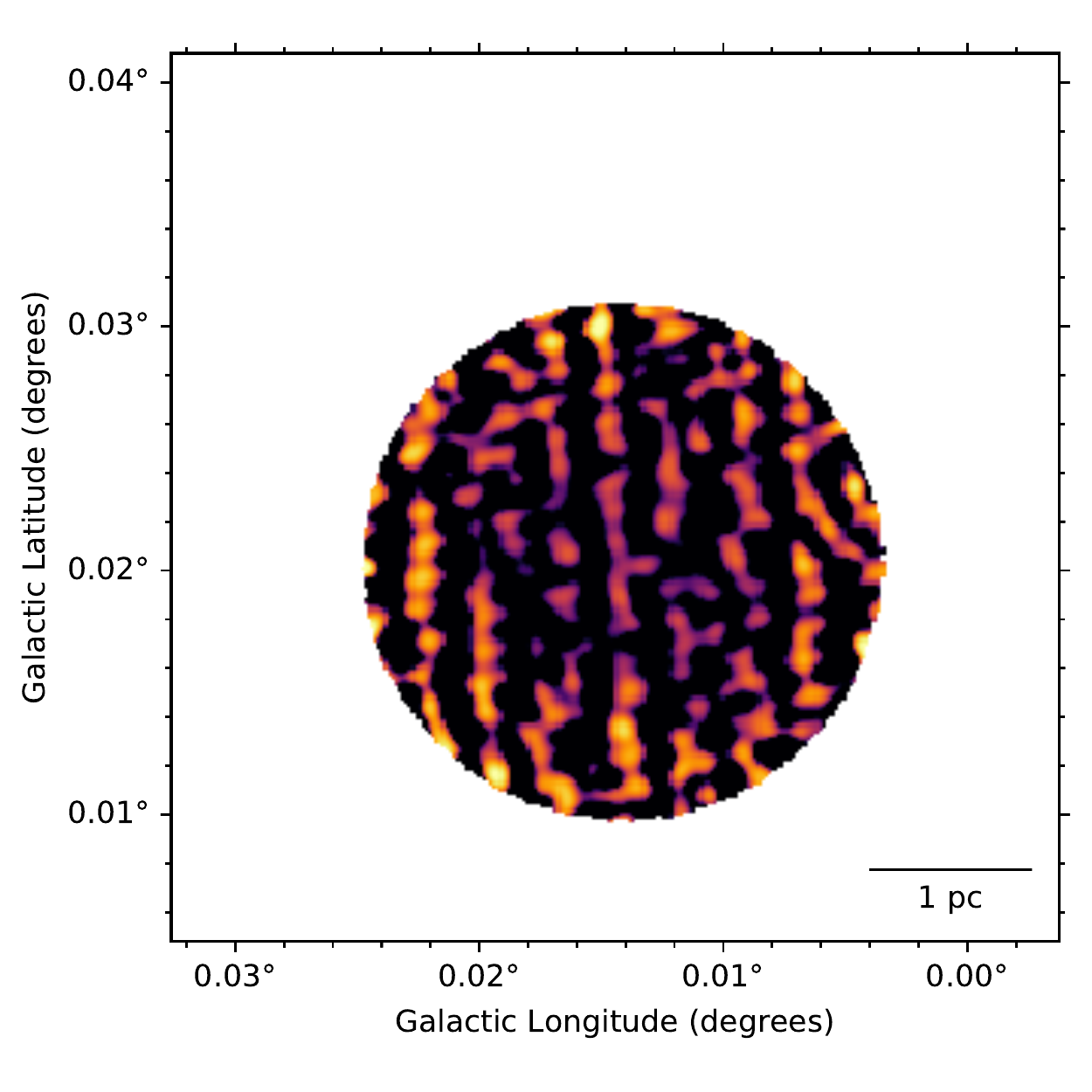}}
\subfigure{
\includegraphics[width=0.48\textwidth]{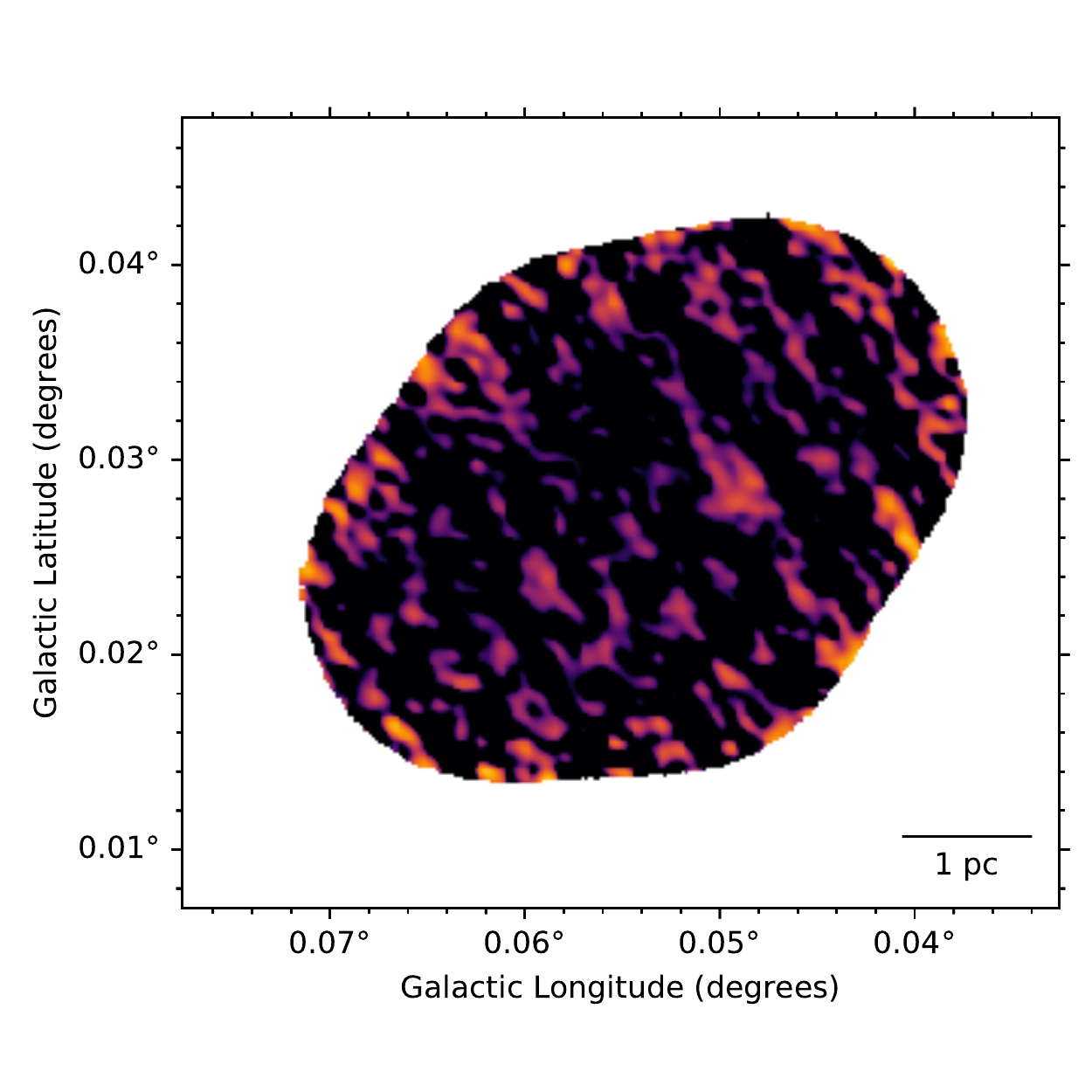}}
\singlespace\caption{Robust catalog leaf contours superimposed over the 1.3 mm dust continuum from the \textit{CMZoom} Survey with a 1 pc scale-bar. All images are displayed on a log scale with limits between 1$\times 10^7$ Jy sr$^{-1}$ and 3$\times 10^8$ Jy sr$^{-1}$. } 
\end{center}
\label{fig:cutout_6}
\end{figure*}

In this section, we show a gallery of the regions included in the \textit{CMZoom} survey, with the high reliability catalog leaf contours overplotted in white. These cutouts are presented in Figures \ref{fig:cutout_1}, \ref{fig:cutout_2}, \ref{fig:cutout_3}, \ref{fig:cutout_4}, \ref{fig:cutout_5}, and \ref{fig:cutout_6}. The SMA 1.3mm continuum maps, as well as the leaf contour mask used to generate this gallery, have been made available at \hyperlink{https://dataverse.harvard.edu/dataverse/cmzoom}{https://dataverse.harvard.edu/dataverse/cmzoom}. All images are displayed with a 1 pc scale-bar and are on a log scale with limits between 1$\times 10^7$ Jy sr$^{-1}$ and 3$\times 10^8$ Jy sr$^{-1}$.

\section{Comparison with Sgr B2 ALMA Data}\label{sec:alma_comp}

In order to further probe the trustworthiness of the catalog contents, the leaves in several key regions of our SMA continuum maps were compared with results from recent observations from ALMA of the same regions. In particular, we used 1.1mm  continuum maps from Dust Ridge clouds D, E, and F \cite{barnes_young_2019}, as well as the Sgr B2 complex at 3mm \citep{ginsburg_distributed_2018}. These maps from ALMA are compared side by side in Figure \ref{fig:cloud_d_alma}, \ref{fig:clouds_ef_alma}, and \ref{fig:sgrb2_alma} below. Dust Ridge clouds D, E and F appear by eye to have a very good agreement, with all Catalog leaves having a clear counterpart in the ALMA maps. Some of the ALMA sources do not appear to have been significant enough detections for inclusion in the catalog, but this is reasonable considering the higher sensitivity given by ALMA. The SMA Sgr B2 map, however, contains some compact sources that do not appear to be present in the ALMA map. While it is possible this might be due to the fact that this region contains the highest noise and highest intensity emission contained in the \textit{CMZoom} survey coverage, it is also likely that differences in emission between the 1.3mm and 3mm maps, particularly given the widespread compact and extended HII regions known to populate the area could also account for the discrepancy. We also note that the leaves, highlighted in Figure \ref{fig:sgrb2_alma}, lie next to a deep negative bowl in the SMA map, supporting the explanation that these leaves in particular are an artifact of the imaging process. These SMA artifact leaves do not appear to have properties remarkably different from nearby leaves which are in agreement with the ALMA maps, and therefore do not significantly influence our interpretation of Sgr B2 leaves in section \ref{sec:sgrb2}. 
\cite{}
\begin{figure*}
\begin{center}
\includegraphics[trim = 0mm 0mm 0mm 0mm, clip, width = 0.5\textwidth,angle=270]{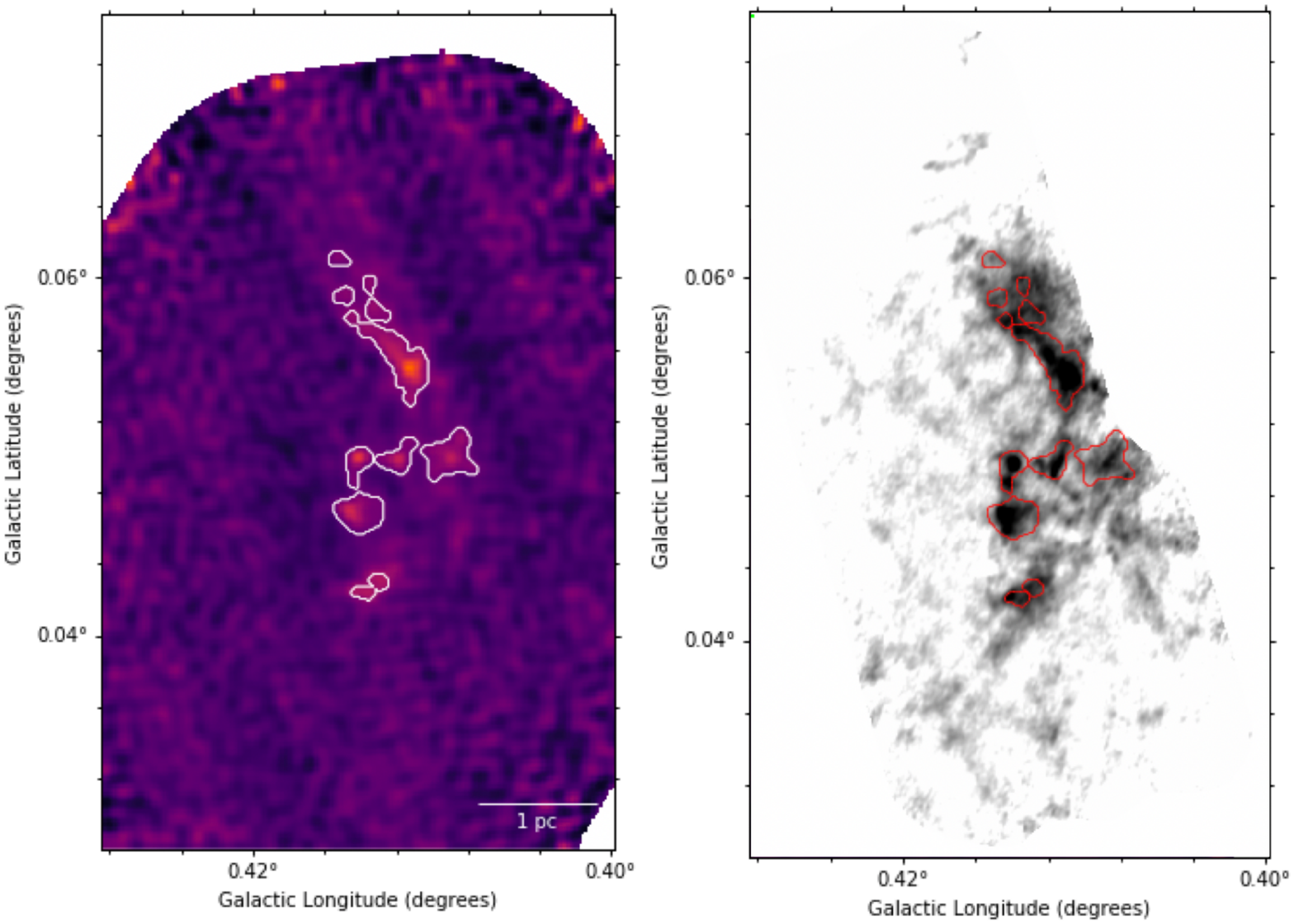}
\end{center}
\caption{A side by side comparison of dust ridge cloud D in the ALMA 1.1mm continuum combined with Bolocam single dish map (right) from \citealt{barnes_young_2019}, and the 1.3mm SMA continuum maps from the \textit{CMZoom} survey (left), both with contours of the catalog leaves in red. All of the catalog leaves have corresponding sources in the ALMA map.  The SMA map is scaled between $10^7$ and 3$\times 10^8$ Jy sr$^{-1}$. The ALMA map is scaled between -0.0026 and 1.1502 Jy beam$^{-1}$. }
\label{fig:cloud_d_alma}
\end{figure*}

\begin{figure*}
\begin{center}
\includegraphics[trim = 0mm 0mm 0mm 0mm, clip, width = 0.5\textwidth,angle=270]{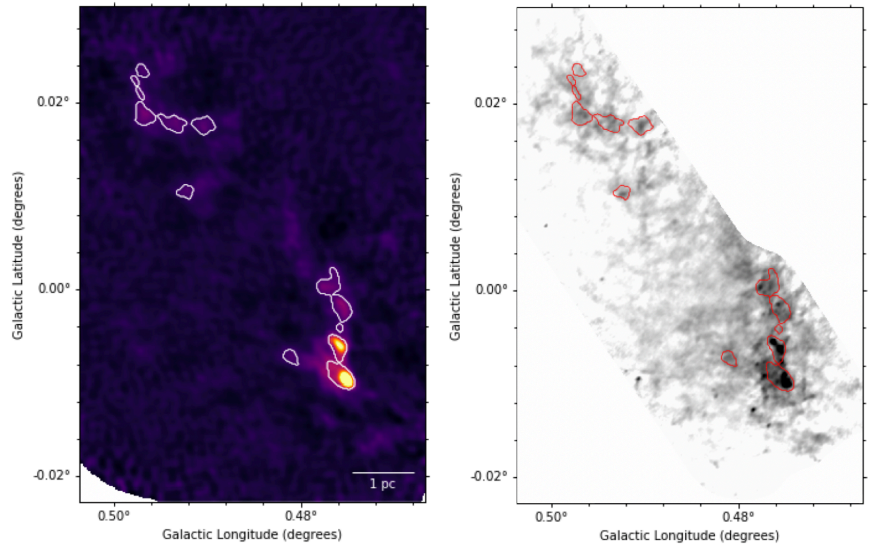}
\end{center}
\caption{A side by side comparison of Dust Ridge Clouds E and F in the ALMA 1.1mm continuum combined with Bolocam single dish map from \citealt{barnes_young_2019} (right), the 1.3mm SMA continuum maps from the \textit{CMZoom} survey (left). both with contours of the catalog leaves in red and white respectively. All of the catalog leaves have corresponding sources in the ALMA map. The 1.3 mm SMA map seems to more clearly resemble a filamentary structure of the emission in these clouds that fragments into more discrete substructure on the scales revealed by ALMA. The SMA map is scaled between $10^7$ and 1.6$\times 10^8$ Jy sr$^{-1}$. The ALMA map is scaled between -0.0026 and 1.1502 Jy beam$^{-1}$.}
\label{fig:clouds_ef_alma}
\end{figure*}

\begin{figure*}
\begin{center}
\includegraphics[trim = 0mm 0mm 0mm 0mm, clip, width = .5\textwidth,angle=270]{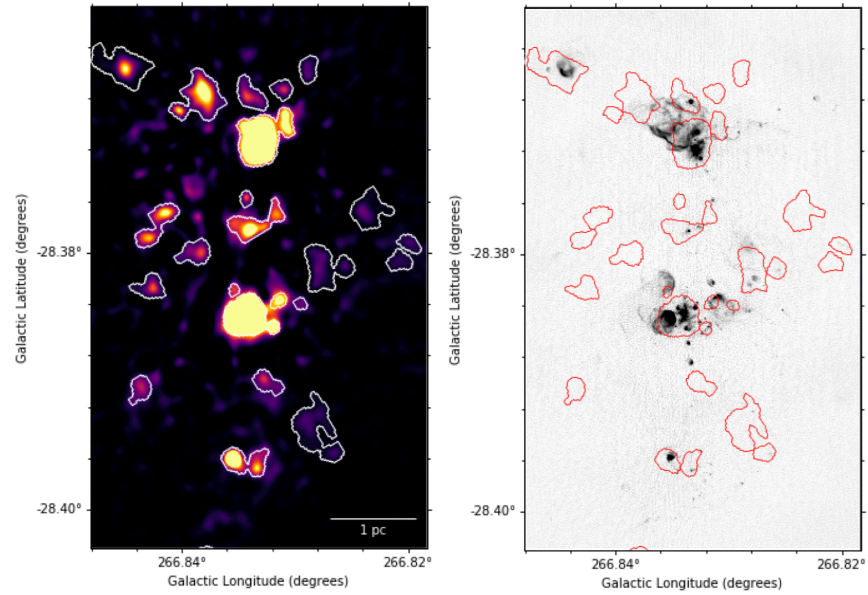}
\end{center}
\caption{A side by side comparison of the Sgr B2 complex in the 1.3mm SMA continuum map (left) from the \textit{CMZoom} survey and the ALMA 3mm continuum map (right) from \citealt{ginsburg_distributed_2018}, both with contours of the catalog leaves in white and red respectively. This figure highlights several catalog leaves which do not appear to have corresponding emission in the ALMA map. We suspect that these sources and the ones that appear symmetrically on the other side of the SgrB2 complex might not be real emission, but instead imaging artifacts. Some of the extended emission visible in the ALMA 3mm map are optically thin HII regions which would not be significant at 1.3mm. The SMA map is scaled between $10^7$ and 3$\times 10^8$ Jy sr$^{-1}$. The ALMA map is scaled between -0.0026 and 1.1502 Jy beam$^{-1}$.}
\label{fig:sgrb2_alma}
\end{figure*}

\end{document}